\documentclass[11pt]{article}

\usepackage{amssymb,amsmath}
\usepackage[dvips]{graphicx}
\usepackage{verbatim}
\usepackage{ifthen}
\usepackage{pstricks}

\newboolean{stocformat}
\setboolean{stocformat}{false}

\ifthenelse{\boolean{stocformat}}{
}{
 \addtolength{\voffset}{-0.7in}
 \addtolength{\textheight}{1.4in}
 \addtolength{\hoffset}{-0.8in}
 \addtolength{\textwidth}{1.6in}
}

\ifthenelse{\boolean{stocformat}}{
 \newcommand{\vskipline}{\vskip 7pt}
}{
 \newcommand{\vskipline}{\vskip 11pt}
}

\newboolean{sec2}
\newboolean{sec3}
\newboolean{sec4}
\newboolean{sec5}
\newboolean{sec6}
\newboolean{sec7}
\setboolean{sec2}{true}
\setboolean{sec3}{true}
\setboolean{sec4}{true}
\setboolean{sec5}{true}
\setboolean{sec6}{true}
\setboolean{sec7}{true}

\newboolean{app}
\setboolean{app}{true}

\DeclareMathOperator{\poly}{poly}

\newtheorem{thm}{Theorem}

\newtheorem{lem}[thm]{Lemma}

\newcommand{\set}[1]{\lbrace #1 \rbrace}

\newcommand{\ket}[1]{| #1 \rangle}

\newcommand{\norm}[1]{\lVert #1 \rVert}

\newcommand{\RR}{\mathbb{R}}
\newcommand{\CC}{\mathbb{C}}
\newcommand{\ZZ}{\mathbb{Z}}
\newcommand{\vecx}{\vec{x}}
\newcommand{\veck}{\vec{k}}
\newcommand{\vecb}{\vec{b}}
\newcommand{\vectheta}{\vec{\theta}}
\newcommand{\vecphi}{\vec{\phi}}
\newcommand{\vecc}{\vec{c}}
\newcommand{\betatilde}{\tilde{\beta}}
\newcommand{\calI}{\cal{I}}
\newcommand{\calA}{\cal{A}}

\newcommand{\undef}{\text{``undef''}}
\newcommand{\coarse}{\text{``coarse''}}
\newcommand{\fine}{\text{``fine''}}
\newcommand{\ceil}[1]{\lceil #1 \rceil}
\newcommand{\floor}[1]{\lfloor #1 \rfloor}

\begin{document}

\ifthenelse{\boolean{stocformat}}{
 \conferenceinfo{STOC'09,}{May 31--June 2, 2009, Bethesda, Maryland, USA.}
 \CopyrightYear{2009}
 \crdata{978-1-60558-506-2/09/05}
}{
 %nothing
}

\title{Quantum Algorithms Using the Curvelet Transform}

\ifthenelse{\boolean{stocformat}}{
 \numberofauthors{1}
 \author{
 \alignauthor Yi-Kai Liu\\
 \affaddr{Institute for Quantum Information}\\
 \affaddr{California Institute of Technology}\\
 \affaddr{Pasadena, CA, USA}\\
 \email{yikailiu@caltech.edu}
 }
}{
 \author{Yi-Kai Liu\\
 Institute for Quantum Information\\
 California Institute of Technology\\
 Pasadena, CA, USA\\
 \texttt{yikailiu@caltech.edu}}
}

\date{Mar. 25, 2009}

\maketitle

%%%%%%%%%%%%%%%%%%%%%%%%%%%%%%%%%%%%%%%%%%%%%%%%%%%%%%%%%%%%%%%%%%%%%%%%%%%%%%%

\begin{abstract}
The curvelet transform is a directional wavelet transform over $\RR^n$, which is used to analyze functions that have singularities along smooth surfaces (Cand\`es and Donoho, 2002).  I demonstrate how this can lead to new quantum algorithms.  I give an efficient implementation of a quantum curvelet transform, together with two applications:  a single-shot measurement procedure for approximately finding the center of a ball in $\RR^n$, given a quantum-sample over the ball; and, a quantum algorithm for finding the center of a radial function over $\RR^n$, given oracle access to the function.  I conjecture that these algorithms succeed with constant probability, using one quantum-sample and $O(1)$ oracle queries, respectively, independent of the dimension $n$ --- this can be interpreted as a quantum speed-up.  To support this conjecture, I prove rigorous bounds on the distribution of probability mass for the continuous curvelet transform.  This shows that the above algorithms work in an idealized ``continuous'' model.
\end{abstract}

\ifthenelse{\boolean{stocformat}}{
}{
 %nothing
}

%\newpage

\ifthenelse{\boolean{stocformat}}{
 \sloppy
}{
 %nothing
}

%%%%%%%%%%%%%%%%%%%%%%%%%%%%%%%%%%%%%%%%%%%%%%%%%%%%%%%%%%%%%%%%%%%%%%%%%%%%%%%

\section{Introduction}

% NOTE:  Give the broader context of this paper -- 
% motivation, our contributions, related work

% Start with the big picture -- what kind of paper is this? what did we do?

One of the most remarkable demonstrations of the power of a quantum computer is Shor's algorithm for factoring and discrete logarithms \cite{shor-alg}.  This has motivated many researchers to try to generalize its key components---the quantum Fourier transform over $\mathbb{Z}_N$, and the algorithm for period-finding---to solve other problems \cite{childs-vandam-survey}.  In particular, this motivated the study of the quantum Fourier transform and the hidden subgroup problem (HSP) on non-Abelian groups, as a route to solving certain lattice problems and the graph isomorphism problem \cite{MRS-graph-isom, BCD-semidirect-prod, regev-dhsp}.  

In this paper we study a different generalization of the Fourier transform, namely the curvelet transform over $\RR^n$ \cite{curvelets-2002}.  This is a kind of ``directional'' wavelet transform, which can resolve features over the spatial and frequency domains simultaneously.  A curvelet basis function resembles a wave-packet, with high-frequency oscillations in one direction (like a plane wave $e^{i\veck\cdot\vecx}$, as in the Fourier transform on $\RR^n$), but which is also supported on a small region of space (unlike the plane wave).  We show that this leads to fast quantum algorithms for some new classes of problems, outside the framework of the HSP.  To the best of our knowledge, this is the first attempt to design quantum algorithms based on the curvelet transform.  

Intuitively, the curvelet transform is helpful in analyzing functions on $\RR^n$ that are discontinuous along $(n-1)$-dimensional surfaces.  If a function $f$ is discontinuous along a surface $S$, then its curvelet transform $\Gamma_f$ will be ``large'' at those locations $(\vecb,\vectheta)$, where $\vecb$ is a point on $S$ and $\vectheta$ is the vector normal to $S$ at $\vecb$.  The set of all such pairs $(\vecb,\vectheta)$ is called the ``wavefront set'' of $f$.

The basic model for a quantum algorithm using the curvelet transform is as follows:  first prepare a quantum state $\sum_{\vecx \in \RR^n}$ $f(\vecx) \ket{\vecx}$, which is a weighted superposition of points in $\RR^n$; then apply the quantum curvelet transform, to get the state $\sum_{\vecb,\vectheta} \Gamma_f(\vecb,\vectheta) \ket{\vecb,\vectheta}$; and finally measure $\vecb$ and $\vectheta$.  For the time being, we ignore implementation issues, such as how to discretize $\RR^n$ and how to compute the quantum curvelet transform efficiently.  The more basic question is whether we can find functions $f$ such that we can prepare the initial state efficiently, and such that a measurement in the ``curvelet basis'' yields useful information.  

One example consists of letting $f$ be the indicator function of a ball in $\RR^n$.  We can efficiently prepare a uniform superposition over points in a ball, using the techniques of \cite{aharonov-tashma, grover-rudolph}.  (This is called ``quantum-sampling,'' and it can be done more generally, e.g., for convex bodies.)  Then, measuring in the curvelet basis extracts information about the center of the ball.

Another example consists of choosing $f$ to be the indicator function of a spherical shell in $\RR^n$.  This is motivated by the problem of finding the center of a radial function on $\RR^n$.  Let $G$ be a radial function, centered around some unknown point $\vecc$.  Prepare a uniform superposition over a large region of space, then compute $G(\vecx)$ and measure it; this produces a uniform superposition over one of the level sets of $G$, which is a spherical shell centered at $\vecc$.  Then, measuring in the curvelet basis extracts information about the location of $\vecc$.

The goal of this paper is to make this intuition precise.  We can interpret $\Gamma_f$ as a wavefunction, and we want to show that its probability mass $|\Gamma_f|^2$ is concentrated near the wavefront set.  
% This is a strong claim---in particular, it is significantly stronger than the results shown in \cite{cont-curvelet-trans-1}.  Our claim is similar in spirit to \cite{curvelets-2002}, though that paper only considers functions on $\RR^2$, not $\RR^n$.  
We can prove this for the continuous curvelet transform, for the two cases of interest, where $f$ is the indicator function of a ball or a spherical shell in $\RR^n$.  In these cases, $(\vecb,\vectheta)$ is near the wavefront set with high probability.  This implies that the line $\set{\vecb+\lambda\vectheta \;|\; \lambda\in\RR}$ passes near the center of the ball or spherical shell.
%(see Figure \ref{fig-ball}) 
%\begin{figure}
%\input{ball.pst}
%\caption{A point $\vecb$ and vector $\vectheta$ that lie near the wavefront set of the ball / spherical shell.}
%\label{fig-ball}
%\end{figure}

Next, we give an efficient implementation of a quantum curvelet transform.  (This is a discrete version of the transform described above, acting on superposition states.)  
% for a certain class of curvelets.  This uses ideas from the classical setting \cite{fast-curvelet-trans}, but the curvelets must be specially designed so that certain quantum superposition states can be prepared efficiently.  
Then we propose polynomial-time quantum algorithms for two problems:  (1) given a single quantum-sample over a ball in $\RR^n$, find the center of the ball, with accuracy $\pm\Delta$ where $\Delta$ is a constant fraction of the radius of the ball; (2) given oracle access to a radial function $f$ that is centered around some unknown point $\vecc \in \RR^n$, find the point $\vecc$ exactly (i.e., with accuracy $\pm\Delta$ in time $\poly(\log\frac{1}{\Delta})$, assuming that the function $f$ fluctuates on sufficiently small scales).  

For the first problem, we conjecture that our quantum procedure succeeds with constant probability, while the best classical procedure succeeds with probability that is exponentially small in $n$.  Classically, this problem is hard because the volume of a ball in $\RR^n$ is concentrated near its surface.  But this same fact is helpful to the quantum curvelet transform, which works by finding a line normal to the surface of the ball.

For the second problem, we conjecture that our quantum algorithm uses only a constant number of queries, whereas any classical algorithm requires $\Omega(n)$ queries.  Intuitively, this is because the curvelet transform uses constructive interference to find a direction in $\RR^n$ from just one query.

We then prove that these algorithms work in an idealized ``continuous'' model --- this follows from our rigorous results on the continuous curvelet transform.  However, we do not have a rigorous analysis of the effects caused by discretization, though we can argue that these should be small.

These examples demonstrate that one can use the curvelet transform to obtain a quantum speed-up.  These examples are artificially simple, in order to allow a rigorous analysis.  But the underlying idea---using the curvelet transform to find normal vectors to a surface---should work on more complicated geometric objects.

\vskipline
%\subsection{Technical Contributions}

% Methods - how did we do it?

\noindent \textbf{1.1 Technical Contributions:}
First, in Section 2, we define the continuous curvelet transform over $\RR^n$.  This generalizes the definition over $\RR^2$ given in \cite{cont-curvelet-trans-2}.  Given a function $f(\vecx)$, the continuous curvelet transform returns a function $\Gamma_f(a,\vecb,\vectheta)$.  Here, $\vecx \in \RR^n$ represents a ``location,'' while $0<a<1$ is a ``scale'' (smaller values denote finer scales, larger values denote coarser scales), $\vecb \in \RR^n$ is a ``location,'' and $\vectheta \in S^{n-1}$ (the unit sphere in $\RR^n$) is a ``direction.''

Next, we study the distribution of probability mass $|\Gamma_f|^2$ over different values of $(a,\vecb,\vectheta)$.  This is technically quite difficult.  $\Gamma_f(a,\vecb,\vectheta)$ is defined by an oscillatory integral, and while there are various methods for bounding the asymptotic decay rates of such quantities \cite{stein-murphy, cont-curvelet-trans-1}, we need non-asymptotic bounds on the total probability mass in a given region.  In Section 3 we develop some tools for proving such bounds, in the case where $f$ is a radial function.  Then, in Sections 4 and 5, we specialize to the case where $f$ is the indicator function of a ball or a spherical shell.  Here, the analysis relies on powerful classical results that bound the oscillation and decay of Bessel functions \cite{abram-stegun, watson}.

% In section 3 we specialize to the case where $f$ is a radial function, $f(\vecx) = f_0(|\vecx|)$.  This implies that $\hat{f}$ is also radial, $\hat{f}(\veck) = F_0(|\veck|)$.  We show that the probability of observing a fine-scale element $a \leq \eta$ is essentially given by the amount of probability mass of $\hat{f}$ at frequencies above $1/\eta$.  Also, by symmetry, the direction $\vectheta$ is uniformly random, and the location $\vecb$ has expectation value $\vec{0}$.  Finally, we bound the variance of $\vecb$.  Our argument is somewhat indirect, since even for a radial function $f$, we do not have an exact closed-form expression for $\Gamma_f$.  Instead, we use Plancherel's theorem to convert the integral over $\vecb$ into an integral over $\veck$.  This is helpful because $\Gamma_f$ has a simpler description over frequency space, while multiplication by $b_j$ turns into differentiation with respect to $k_j$.  We can then bound this integral in a (relatively) straightforward way.

In Section 4 we let $f$ be the indicator function of a ball in $\RR^n$, with radius $\beta$, centered at the origin.  We expect that, after applying the curvelet transform, $\vecb$ and $\vectheta$ will be concentrated near the wavefront set of $f$:  that is, $\vecb$ will be concentrated near the line $\set{\lambda\vectheta \;|\; \lambda\in\RR}$, at about distance $\beta$ from the origin.  Furthermore, we expect that $\vecb$ will become more tightly concentrated, the smaller the value of $a$.
We show that this is essentially what happens.  In particular, with constant probability, the distance from $\vecb$ to the line $\set{\lambda\vectheta \;|\; \lambda\in\RR}$ will be at most a constant fraction of $\beta$; remarkably, this holds independent of the dimension $n$.  

% We show that this is essentially what happens.  We show that the probability of observing $a \leq \eta$ is at least $\Omega(\eta)$.  This is due to the fact that $\hat{f}$ has a ``heavy tail,'' which is caused by the discontinuity of $f$ along the surface of the ball.  (For comparison, one would not observe this behavior if $f$ were, say, a Gaussian.)  Next, we show that the variance of $\vecb$, conditioned on $a \leq \eta$, is at most $O(\beta^2)$ in the $\vectheta$ direction, and at most $O(\eta\beta^2)$ in the subspace orthogonal to $\vectheta$.  We interpret this as follows:  $\vecb$ and $\vectheta$ define a line that passes near the center of the ball; and, with constant probability, the error (i.e., the displacement away from the center) is at most a constant fraction of the radius of the ball.  

% As mentioned previously, it is remarkable that these bounds do not depend on the dimension $n$.  

In Section 5, we let $f$ be supported on a thin spherical shell, having radius $\beta$ and thickness $\delta \ll \beta$.  Here, after applying the curvelet transform, we get a qualitatively similar behavior of $a$, $\vecb$ and $\vectheta$.  Quantitatively, however, we find that we can observe much smaller scales $a$, on the order of $\delta/\beta$; and thus we can find the center of the shell with much greater precision.  Essentially, by making the shell extremely thin, we can find its center with arbitrarily high precision.  

% We show the following.  Let $\varepsilon = \delta/\beta$, and let $\eta_c = \varepsilon(n-2)/e$.  Then we observe $a \leq \eta_c$ with at least constant probability.  Furthermore, conditioned on this event, the variance of $\vecb$ is at most $O(\beta^2)$ in the $\vectheta$ direction, and at most $O(\varepsilon n \beta^2)$ in the subspace orthogonal to $\vectheta$.  Here, $\varepsilon$ is the dominant factor.  In the context of a quantum algorithm for finding the center of a radial function, $\varepsilon$ can be extremely small:  when the oracle computes the function to $m$ bits of precision, $\varepsilon$ can be of size $2^{-\Omega(m)}$.

% \vskipline

Finally, we turn to the discrete curvelet transform, and quantum algorithms.  In Section 6 we give an efficient implementation of a quantum curvelet transform.  This uses ideas from the fast classical curvelet transform \cite{fast-curvelet-trans}.  However, there is a new complication in the quantum case:  we need to prepare certain states which are superpositions of different scales and directions $(a,\vectheta)$.  We design families of curvelets that allow this step to be performed efficiently, and that have similar analytic properties to the curvelets used in Sections 3-5.

% In Section 6 we give an efficient implementation of a quantum curvelet transform.  At a high level, this consists of a quantum Fourier transform, followed by an operation $\cal{X}$ that prepares certain superpositions over $a$ and $\vectheta$, followed by an inverse quantum Fourier transform:  
% \begin{equation}
% \begin{split}
% \sum_{\vecx} f(\vecx) \ket{\vecx} \ket{0,\vec{0}}
%  &\mapsto \sum_{\veck} \hat{f}(\veck) \ket{\veck} \ket{0,\vec{0}} \\
%  &\mapsto \sum_{\veck} \hat{f}(\veck) \ket{\veck} \sum_{a,\vectheta} \chi_{a,\vectheta}(\veck) \ket{a,\vectheta} \\
%  &\mapsto \sum_{\vecb,a,\vectheta} \Gamma_f(a,\vecb,\vectheta) \ket{\vecb} \ket{a,\vectheta}.
% \end{split}
% \end{equation}
% This is the same basic approach used to calculate the classical curvelet transform \cite{fast-curvelet-trans}.  However, in the quantum case, the operation $\cal{X}$ is tricky to implement, since the superposition can involve $2^{\Theta(n)}$ different terms, when we are working over $\RR^n$.  This depends on the precise form of the window functions $\chi_{a,\vectheta}$.  We show how to implement $\cal{X}$ when the window functions are constructed in a particular way, using spherical coordinates in $\RR^n$.  (It is not clear how to do this for the window functions used in \cite{fast-curvelet-trans}.)  We argue that this quantum curvelet transform is a reasonable approximation to the continuous curvelet transform that we studied previously.

In Section 7, we formally define the two problems mentioned earlier:  estimating the center of a ball, given a single quantum-sample state; and finding the center of a radial function, given oracle access.  We present quantum algorithms for these problems, and use our results from Sections 4 and 5 to prove that the algorithms work in a continuous model.  We also sketch a classical lower bound for finding the center of a radial function.  % For the first problem, we conjecture that our quantum procedure succeeds with constant probability, while the best classical procedure succeeds with probability that is exponentially small in $n$.  For the second problem, we conjecture that our quantum algorithm uses only a constant number of queries, independent of the dimension $n$.  These conjectures are supported by our (rigorous) results on the continuous curvelet transform; and we can argue (heuristically) that the effects caused by discretization should be small.  

This paper omits most of the proofs, due to lack of space.  The proofs can be found in the full version \cite{Liu-curvelets}.  Also, note that this paper contains some additional results there were not present in the first version of \cite{Liu-curvelets}.  This paper contains an improved algorithm for finding the center of a radial function, and a classical lower bound for that problem.  

\vskipline
%\subsection{Related Work}

\noindent \textbf{1.2 Related Work:}
Curvelets over $\RR^2$ and $\RR^3$ have been studied as a tool for image processing and simulating wave propagation \cite{curvelets-2002, curvelets-wavepropagators, fast-curvelet-trans, fast-curvelet-trans-3d}.  The curvelet transform is also related to older ideas from harmonic analysis; see, e.g., Smith \cite{smith-fio}.  It can be viewed as an algorithmic implementation of a technique known as second dyadic decomposition \cite{stein-murphy}.

There are a few rigorous results on the behavior of the curvelet transform which are similar in spirit to our work \cite{curvelets-2002, cont-curvelet-trans-1, cont-curvelet-trans-2, Blanco-Silva-curvelets}.  
% For instance, in \cite{curvelets-2002} it is shown that, if $f$ is a function on $\RR^2$, and $f$ is $C^2$-smooth away from a $C^2$-smooth edge, then the curvelet coefficients decay at a certain rate.  From the proof, it is clear that the large curvelet coefficients are concentrated around the wavefront set.  
% Also, in \cite{cont-curvelet-trans-1}, it is shown that, for a function $f$ on $\RR^2$, the wavefront set of $f$ consists of points $(\vecb,\vectheta)$ where $\Gamma_f(a,\vecb,\vectheta)$ does not decay rapidly as $a \rightarrow 0$.  
These results apply to much broader classes of functions, but they are only known to hold over $\RR^2$ (or $\RR^3$ in some cases).  Although one would expect them to generalize in some fashion to $\RR^n$, it is perhaps surprising that the scaling with $n$ is as favorable as we find here.  
% (In general, curvelets over $\RR^n$ have received less attention, as most practical applications involve $\RR^2$ and $\RR^3$, and the analysis becomes increasingly complicated in higher dimensions.)

In connection with quantum algorithms, there has been some work on the quantum wavelet transform \cite{hoyer-wavelets, fijany-wavelets, hoyer-ordsearch, freedman-wavelets}.  
% Essentially, the quantum wavelet transform (using, say, Daubechies wavelets) can be computed efficiently, but few applications are known.  
But the curvelet transform on $\RR^n$ is quite different from the ``ordinary'' wavelet transform on $\RR^n$, which consists of a product of 1-D transforms.  The ordinary wavelet transform on $\RR^n$ can detect the locations of discontinuities, but it cannot resolve directions.

The geometric problems studied in this paper are reminiscent of some recent work on finding hidden nonlinear structures, although the details are different.  Shifted subset problems \cite{childs-nonlin, montanaro-shifted} involve translational invariance, so the natural tool for solving them is the Fourier transform, rather than curvelets.  Hidden polynomial problems \cite{childs-nonlin, wocjan-hpp} resemble our problem of finding the center of a radial function.  However, they are much more general (and thus much harder), and they are set over a finite field rather than $\RR^n$.  

These problems can also be studied from the perspective of quantum state discrimination \cite{chefles}, e.g., what is the optimal quantum measurement for estimating the center of a ball?  However, we stress that our algorithms using the quantum curvelet transform are computationally efficient.

Finally, we recently became aware of a quantum algorithm for estimating the gradient of a function $f$ on $\RR^n$, using only $O(1)$ queries \cite{jordan-gradient}.  
This is quite similar to our algorithm for finding the center of a radial function --- it is like applying the curvelet transform at a single location $\vecb$.  Viewing this as a curvelet transform has the advantage of providing a more general framework, where one can do this procedure on an arbitrary input state.  Also, note that our emphasis in this paper is on functions $f$ that are not smooth---in this case, the gradient is not well-defined.

\ifthenelse{\boolean{stocformat}}{
 \section{The Continuous Curvelet \\ Transform}
}{
 \section{The Continuous Curvelet Transform}
}

\ifthenelse{\boolean{sec2}}{

We begin by defining the continuous curvelet transform over $\RR^n$.  This generalizes the definition of \cite{cont-curvelet-trans-2} over $\RR^2$.  Given a function $f(\vecx)$, the continuous curvelet transform returns a function $\Gamma_f(a,\vecb,\vectheta)$.  Here, $\vecx \in \RR^n$ represents a ``location,'' while $0<a<1$ is a ``scale'' (smaller values denote finer scales, larger values denote coarser scales), $\vecb \in \RR^n$ is a ``location,'' and $\vectheta \in S^{n-1}$ (the unit sphere in $\RR^n$) is a ``direction.''  All functions return values in $\CC$.  

% Intuitively, if $f$ is discontinuous along a smooth surface $S$ of dimension $n-1$, then $\Gamma_f$ will be ``large'' near points $(a,\vecb,\vectheta)$ that satisfy the following conditions:  $\vecb$ lies on the surface $S$, $a$ has an appropriate value that matches the ``sharpness'' of the discontinuity at $\vecb$, and $\vectheta$ points in the direction normal to $S$ at $\vecb$.  

Intuitively, the curvelet transform decomposes $f$ into pieces corresponding to different scales $a$ and directions $\vectheta$; we can view $\Gamma_f(a,\vecb,\vectheta)$ as a family of functions, indexed by $a$ and $\vectheta$, each representing some ``piece'' of $f(\vecx)$ (note that the variables $\vecb$ and $\vecx$ both represent locations in space).  To get one such piece of $f$, we will take its Fourier transform, and then multiply by a ``window function'' $\chi_{a,\vectheta}$ which is defined over the frequency domain.

More precisely, the curvelet transform is defined to be 
\begin{equation}
\Gamma_f(a,\vecb,\vectheta) := \int_{\RR^n} \hat{f}(\veck) \chi_{a,\vectheta}(\veck) e^{2\pi i \veck \cdot \vecb} d\veck.  
\label{eqn-cct-def1}
\end{equation}
Here, $\hat{f}$ is the Fourier transform of $f$, and $\chi_{a,\vectheta}$ is a function that is smooth, real, non-negative, and supported on a ``sector'' of frequency space $S_{a,\vectheta} \subset \RR^n$.  We will describe these sectors below.  Before doing so, we remark that the curvelet transform consists of (1) taking the Fourier transform of $f$, (2) separating $\hat{f}$ into pieces corresponding to different scales and directions, and (3) taking the inverse Fourier transform.  This description suggests how to compute the curvelet transform efficiently.

The ``sector'' $S_{a,\vectheta}$ is roughly given by the intersection of the cone centered around the vector $\vectheta$ with angular width $\sqrt{a}$, and the annulus with inner radius $1/a$ and outer radius $2/a$.  Thus, the ``piece'' of $f$ at scale $a$ and direction $\vectheta$ is somewhat like the restriction of $\hat{f}$ to frequencies $\veck \approx (1/a)\vectheta$ (which represent oscillations in direction $\vectheta$, at higher frequencies when the scale $a$ is small).  Note, however, that the sector has dimensions $1/a \times (1/\sqrt{a})^{n-1}$, so its shape is not constant --- the sector becomes longer and narrower when the scale $a$ is small.

This construction can also be understood from a second perspective.  We can define a family of curvelet basis functions $\gamma_{a,\vecb,\vectheta}$ as follows.  The curvelet at location $\vecb$ is obtained by translation from the curvelet at location $\vec{0}$, that is, $\gamma_{a,\vecb,\vectheta}(\vecx) := \gamma_{a,\vec{0},\vectheta}(\vecx-\vecb)$.  The curvelet at location $\vec{0}$ is defined in terms of its Fourier transform, which is simply the window function $\chi_{a,\vectheta}$, that is, $\hat{\gamma}_{a,\vec{0},\vectheta}(\veck) := \chi_{a,\vectheta}(\veck)$.  It is easy to check that the curvelet transform defined earlier is equivalent to taking inner products with this family of curvelet basis functions:  $\Gamma_f(a,\vecb,\vectheta) = \langle \gamma_{a,\vecb,\vectheta}, f \rangle$.  

Now we can see how our choice of the window function $\chi_{a,\vectheta}$ implies (and is motivated by) certain properties of the curvelet basis functions $\gamma_{a,\vecb,\vectheta}$.  Since $\chi_{a,\vectheta}$ is smooth, the $\gamma_{a,\vecb,\vectheta}$ are rapidly decaying.  Also, each $\gamma_{a,\vecb,\vectheta}$ has high-frequency oscillations in the $\vectheta$ direction, and is essentially supported on a plate-like region, centered at location $\vecb$, orthogonal to $\vectheta$, with dimensions $a \times (\sqrt{a})^{n-1}$.  Intuitively, $\gamma_{a,\vecb,\vectheta}$ resembles a plane-wave in direction $\vectheta$, localized around the point $\vecb$.

% NOTE:  see our notes from 5/5/08, 6/3/08

Finally, we define the window function $\chi_{a,\vectheta}$ as follows.  Write $\veck$ using spherical coordinates $(r,\phi_1,\ldots,\phi_{n-1})$, centered around the direction $\vectheta$, so that $\phi_1$ is the angle between $\veck$ and $\vectheta$.  Then define 
%\begin{equation}
$\chi_{a,\vectheta}(\veck) := W(\lambda ar) V(\phi_1/\sqrt{a}) \Lambda_a(\phi_1)$.  
%\end{equation}
Here $\lambda$ is a constant, which can be chosen freely; we will explain how to set it later.  $W$ is a radial window function, real, nonnegative, supported on the interval $[1/e,1]$, and satisfying the admissibility condition 
%\begin{equation}
$\int_0^\infty W(r)^2 \frac{dr}{r} = 1$.
%\end{equation}
$V$ is an angular window function, real, nonnegative, supported on the interval $[0,\pi/2]$, and satisfying the admissibility condition 
%\begin{equation}
$\int_{S^{n-1}} V(\phi_1)^2 d\sigma(\phi_1,\ldots,\phi_{n-1}) = 1$, 
%\end{equation}
where $d\sigma$ denotes integration over the unit sphere $S^{n-1}$ in $\RR^n$.  

$\Lambda_a$ is a normalization and adjustment factor:
%\begin{equation}
$\Lambda_a(\phi_1) := a^{(n+1)/4} \Bigl( \frac{\sin(\phi_1/\sqrt{a}) \sqrt{a}}{\sin(\phi_1)} \Bigr)^{(n-2)/2}$.
%\end{equation}
This is needed because the volume of the sector $S_{a,\vectheta}$ (on which $\chi_{a,\vectheta}$ is supported) changes with $a$.  Note that, when $\phi_1$ is small, $\Lambda_a(\phi_1) \approx a^{(n+1)/4}$.  This is the main point where defining curvelets over $\RR^n$ is more complicated than over $\RR^2$; note that in dimension $n=2$, $\Lambda_a(\phi_1) = a^{(n+1)/4} = a^{3/4}$ exactly.  We remark that a simpler approach would be to use a constant normalization factor that only depends on $a$ and not $\phi_1$; however, our more complicated construction will be more convenient in the later sections of this paper.

We now state some basic properties of the curvelet transform.  First, note that the curvelet transform works primarily on the high-frequency components of $f$, which correspond to fine-scale elements ($a$ small).  The constant factor $\lambda$, mentioned above, sets the low-frequency cutoff value, which corresponds to the coarsest scale ($a=1$).  For convenience, here we assume that $f$ has no low-frequency components below the cutoff value.  In practice, when $f$ has such low-frequency components, the curvelet transform leaves them unchanged, and simply returns them as a residual function $f_{\text{res}}$.

Next, we define the reference measure 
%\begin{equation}
$d\mu(a,\vecb,\vectheta) := \frac{da \: d\vecb \: d\sigma(\vectheta)}{a^{n+1}}$.
%\end{equation}
This weights the contributions of $\Gamma_f(a,\vecb,\vectheta)$ differently according to the scale $a$.  Intuitively, this is needed because the sectors $S_{a,\vectheta}$ do not cover the frequency domain uniformly, and the translations of a curvelet $\gamma_{a,\vec{0},\vectheta}$ (for fixed $a$ and $\vectheta$) to different locations $\vecb$ do not cover the spatial domain uniformly.  Rather, $(a,\vecb,\vectheta)$ should be ``sampled'' in a certain way.  To see this, write 
%\begin{equation}
$d\mu(a,\vecb,\vectheta) = \frac{da}{a} \frac{d\vecb}{a^{(n+1)/2}} \frac{d\sigma(\vectheta)}{a^{(n-1)/2}}$.
%\end{equation}
Note that $\frac{da}{a} = d(\log a)$, suggesting that we should sample $\log(a)$ at uniform intervals, i.e., we should set $a$ equal to powers of 2; we should sample $\vecb$ on a grid in $\RR^n$ whose cells have size $a \times (\sqrt{a})^{n-1}$; and we should sample $\vectheta$ on a mesh on $S^{n-1}$ whose cells have size $(\sqrt{a})^{n-1}$.  Later, when we construct the discrete curvelet transform, we will use this sampling trick for $a$ and $\vectheta$, in place of the reference measure.

Then we have the following theorems:
\begin{thm}
\label{thm-cct-1}
Suppose that $\hat{f}(\veck) = 0$ for all $|\veck| < 1/\lambda$.  Then we can recover $f$ from its curvelet transform $\Gamma_f$: \ifthenelse{\boolean{stocformat}}{\\}{}
%\begin{equation}
$f(\vecx) = \int_{a<1} \Gamma_f(a,\vecb,\vectheta) \gamma_{a,\vecb,\vectheta}(\vecx) d\mu(a,\vecb,\vectheta)$.
%\end{equation}
\end{thm}
\begin{thm}
\label{thm-cct-2}
Suppose that $\hat{f}(\veck) = 0$ for all $|\veck| < 1/\lambda$.  Then the curvelet transform preserves the $L^2$ norm: \ifthenelse{\boolean{stocformat}}{\\}{}
%\begin{equation}
$\int_{\RR^n} |f(\vecx)|^2 d\vecx = \int_{a<1} |\Gamma_f(a,\vecb,\vectheta)|^2 d\mu(a,\vecb,\vectheta)$.
%\end{equation}
\end{thm}
These are straightforward generalizations (to the case of $\RR^n$) of results in \cite{cont-curvelet-trans-2}.  We sketch the proofs in \ifthenelse{\boolean{app}}{Appendix A}{\cite{Liu-curvelets}}.

}{}

%%%%%%%%%%%%%%%%%%%%%%%%%%%%%%%%%%%%%%%%%%%%%%%%%%%%%%%%%%%%%%%%%%%%%%%%%%%%%%%

\section{The Curvelet Transform of a Radial Function}

\ifthenelse{\boolean{sec3}}{

$|\Gamma_f(a,\vecb,\vectheta)|^2 d\mu(a,\vecb,\vectheta)$ can be interpreted as a probability density over the different scales, locations and directions $(a,\vecb,\vectheta)$.  In this section we will develop some tools for understanding where this probability mass is concentrated.  
% This is technically quite difficult.  $\Gamma_f(a,\vecb,\vectheta)$ is defined by an oscillatory integral, and while there are various methods for bounding the asymptotic decay rates of such quantities \cite{stein-murphy, cont-curvelet-trans-1}, we need non-asymptotic bounds.  We cannot solve this problem in general, but we can handle some special cases.
We will consider the case where $f$ has rotational symmetry.  Though there is no simple analytic expression for $\Gamma_f$, we can deduce certain properties from symmetry, and we can upper-bound the variance of $\vecb$ (this latter point is our main result).

Let $f$ be a radial function, $f(\vecx) = f_0(|\vecx|)$.  Its Fourier transform is also radial, $\hat{f}(\veck) = F_0(|\veck|)$, where 
%\begin{equation}
$F_0(\rho) = \frac{2\pi}{\rho^{(n-2)/2}} \int_0^\infty J_{(n-2)/2}(2\pi\rho r) f_0(r) r^{n/2} dr$, 
%\label{eqn-ft-radial}
%\end{equation}
and $J$ is a Bessel function (see, e.g., \cite{stein-shak}).  We assume that $f$ is normalized so that $\int_{\RR^n} |f(\vecx)|^2 d\vecx = 1$.  

%\sloppy
When $f$ is radial, $\Gamma_f$ has the following symmetries:  
%\begin{equation}
$\Gamma_f(a,\vecb,\vectheta) = \Gamma_f(a,-\vecb,-\vectheta)$,
%\label{eqn-rf-reflection-symm}
%\end{equation}
and for any rotation $R$, 
%\begin{equation}
$\Gamma_f(a,\vecb,\vectheta) = \Gamma_f(a,R(\vecb),R(\vectheta))$.
%\label{eqn-rf-rotation-symm}
%\end{equation}

%\fussy
We make a particular choice for the radial and angular windows $W$ and $V$.  These windows are $C^1$ smooth, which is necessary in our analysis of the variance of $\vecb$.  We let 
$W(r)$ = [$C_w \sin(\pi\log r)^2$ if $1/e \leq r \leq 1$; 0 otherwise] 
and 
$V(t)$ = [$C_v \cos(t)^2$ if $0 \leq t \leq \pi/2$; 0 otherwise], 
%\begin{equation}
%W(r) = \begin{cases}
%  C_w \sin(\pi\log r)^2, & 1/e \leq r \leq 1, \\
%  0, & \text{otherwise},
%\end{cases}
%\end{equation}
%%\quad \text{and} \quad
%\begin{equation}
%V(t) = \begin{cases}
%  C_v \cos(t)^2, & 0 \leq t \leq \pi/2, \\
%  0, & \text{otherwise},
%\end{cases}
%\end{equation}
where 
%\begin{equation}
$C_w = \sqrt{8/3}$,
%\end{equation}
and 
%\begin{equation}
$C_v = \sqrt{\frac{2(n+2)n}{3S_0}}$.
%\end{equation}

\vskipline
%\subsection{The probability of observing a scale $a$}

% NOTE:  see our notes from 6/3/08

\noindent \textbf{3.1 The probability of observing a scale $a$:}
First, we claim that the probability of observing a fine-scale element $a \leq \eta$ is essentially given by the amount of probability mass of $\hat{f}$ at frequencies above $1/(\lambda\eta)$:  
%\begin{equation}
$\Pr[a \leq \eta]
 = \int_{a \leq \eta} |\Gamma_f(a,\vecb,\vectheta)|^2 d\mu(a,\vecb,\vectheta)
 \geq \int_{|\veck| \geq 1/(\lambda\eta)} |\hat{f}(\veck)|^2 d\veck$.
%\end{equation}
This follows from the same argument used to prove Theorem \ref{thm-cct-2}.  In the case of a radial function, we write this as:
\begin{equation}
\Pr[a \leq \eta]
 \geq S_0 \int_{1/(\lambda\eta)}^\infty F_0(\rho)^2 \rho^{n-1} d\rho, 
\label{eqn-rf-praleqeta}
\end{equation}
where $S_0 = \frac{2\pi^{n/2}}{\Gamma(n/2)}$ is the surface area of the sphere $S^{n-1} \subset \RR^n$.  

\vskipline
%\subsection{The location $\vecb$ and direction $\vectheta$}

% NOTE:  see our notes from 8/8/08, 10/23/08

\noindent \textbf{3.2 The location $\vecb$ and direction $\vectheta$:}
We claim that the location $\vecb$ has expectation value $\vec{0}$.  %To see this, write:
%\begin{equation*}
%\begin{split}
%\text{E}(b_j)
% &= \int b_j |\Gamma_f(a,\vecb,\vectheta)|^2 d\mu(a,\vecb,\vectheta) \\
% &= \int_{S^{n-1}} \int_0^1 \int_{\RR^n} b_j |\Gamma_f(a,\vecb,\vectheta)|^2 d\vecb \frac{da}{a^{n+1}} d\sigma(\vectheta) \\
% &= \int_{S^{n-1}} \int_0^1 \int_{\RR^n} -b_j |\Gamma_f(a,\vecb,\vectheta)|^2 d\vecb \frac{da}{a^{n+1}} d\sigma(\vectheta) \\
% &= \text{E}(-b_j),
%\end{split}
%\end{equation*}
%using equation (\ref{eqn-rf-reflection-symm}) and a change of variables; thus $\text{E}(b_j) = 0$.  
To see this, observe that $\text{E}(b_j) = -\text{E}(b_j)$, due to the reflection symmetry of $\Gamma_f$; thus $\text{E}(b_j) = 0$.  Note that this remains true when we condition on the value of $a$.

Also, we claim that the direction $\vectheta$ is uniformly distributed.  %To see this, let $\vectheta_0, \vectheta_1 \in S^{n-1}$, and write:
%\begin{equation*}
%\begin{split}
%\text{P}(\vectheta = \vectheta_0)
% &= \int_{\vectheta = \vectheta_0} |\Gamma_f(a,\vecb,\vectheta)|^2 d\mu(a,\vecb,\vectheta) \\
% &= \int_0^1 \int_{\RR^n} |\Gamma_f(a,\vecb,\vectheta_0)|^2 d\vecb \frac{da}{a^{n+1}} \\
% &= \int_0^1 \int_{\RR^n} |\Gamma_f(a,\vecb,\vectheta_1)|^2 d\vecb \frac{da}{a^{n+1}} \\
% &= \text{P}(\vectheta = \vectheta_1),
%\end{split}
%\end{equation*}
%using equation (\ref{eqn-rf-rotation-symm}) and a change of variables; thus the distribution is uniform.  
This follows from the rotational symmetry of $\Gamma_f$.  Note that this remains true when we condition on the value of $a$, and when we condition on the value of $\vecb \cdot \vectheta$ (since this preserves the rotational symmetry).

%%%%%%%%%%%%%%%%%%%%%%%%%%%%%%%%%%%%%%%%%%%%%%%%%%%%%%%%%%%%%%%%%%%%%%%

\vskipline
%\subsection{The variance of $\vecb$ perpendicular to $\vectheta$}

% NOTE:  see our notes from 5/19/08, 5/26/08 (and following days), 7/15/08

\noindent \textbf{3.3 The variance of $\vecb$ perpendicular to $\vectheta$:}
Finally, we seek to upper-bound the variance of $\vecb$, in the directions perpendicular to $\vectheta$, as well as parallel to $\vectheta$.  These results are rather complicated, so we defer most of the details to \ifthenelse{\boolean{app}}{Appendix B.1}{\cite{Liu-curvelets}}.  However, these results are a basic component of our proofs in Sections 4 and 5, so we will sketch some of the calculations.

First, the variance of $\vecb$ perpendicular to $\vectheta$ is:
\ifthenelse{\boolean{stocformat}}{ % STOC format
 \begin{multline}
 \text{E}(\vecb^T (I-\vectheta\vectheta^T) \vecb) = 
 \int_{S^{n-1}} \int_0^1 \int_{\RR^n} (\vecb^T (I-\vectheta\vectheta^T) \vecb) \\
 |\Gamma_f(a,\vecb,\vectheta)|^2 d\vecb \frac{da}{a^{n+1}} d\sigma(\vectheta).
 \end{multline}
}{ % Regular
 \begin{equation}
 \text{E}(\vecb^T (I-\vectheta\vectheta^T) \vecb) = 
 \int_{S^{n-1}} \int_0^1 \int_{\RR^n} (\vecb^T (I-\vectheta\vectheta^T) \vecb) 
 |\Gamma_f(a,\vecb,\vectheta)|^2 d\vecb \frac{da}{a^{n+1}} d\sigma(\vectheta).
 \end{equation}
}
Note that a similar formula holds when we condition on observing $a \leq \eta$.

We can take advantage of rotational symmetry to do the $\vectheta$ integral.  Fix a vector $\vec{u} = (1,0,\ldots,0)$, and for each $\vectheta$, let $R$ be a rotation that maps $\vectheta$ to $\vec{u}$.  Then we can replace the expression inside the integral with $(R(\vecb)^T (I-\vec{u}\vec{u}^T) R(\vecb)) |\Gamma_f(a,R(\vecb),\vec{u})|^2$.  
%\ifthenelse{\boolean{stocformat}}{ % STOC format
% \begin{multline}
% \text{E}(\vecb^T (I-\vectheta\vectheta^T) \vecb) = 
% \int_{S^{n-1}} \int_0^1 \int_{\RR^n} (R(\vecb)^T (I-\vec{u}\vec{u}^T) R(\vecb)) \\
% |\Gamma_f(a,R(\vecb),\vec{u})|^2 d\vecb \frac{da}{a^{n+1}} d\sigma(\vectheta).
% \end{multline}
%}{ % Regular
% \begin{equation}
% \text{E}(\vecb^T (I-\vectheta\vectheta^T) \vecb) = 
% \int_{S^{n-1}} \int_0^1 \int_{\RR^n} (R(\vecb)^T (I-\vec{u}\vec{u}^T) R(\vecb)) 
% |\Gamma_f(a,R(\vecb),\vec{u})|^2 d\vecb \frac{da}{a^{n+1}} d\sigma(\vectheta).
% \end{equation}
%}
Then change variables $\vecb \mapsto R^{-1}(\vecb)$.  The integrand is now independent of $\vectheta$, so we can do the $\vectheta$ integral.  We get:
\ifthenelse{\boolean{stocformat}}{ % STOC format
 \begin{multline}
 \text{E}(\vecb^T (I-\vectheta\vectheta^T) \vecb) = 
 S_0 \int_0^1 \int_{\RR^n} (\vecb^T (I-\vec{u}\vec{u}^T) \vecb) \\
 |\Gamma_f(a,\vecb,\vec{u})|^2 d\vecb \frac{da}{a^{n+1}}.
 \label{eqn-georgia-okeeffe}
 \end{multline}
}{ % Regular
 \begin{equation}
 \text{E}(\vecb^T (I-\vectheta\vectheta^T) \vecb) = 
 S_0 \int_0^1 \int_{\RR^n} (\vecb^T (I-\vec{u}\vec{u}^T) \vecb) 
 |\Gamma_f(a,\vecb,\vec{u})|^2 d\vecb \frac{da}{a^{n+1}}.
 \label{eqn-georgia-okeeffe}
 \end{equation}
}

Now the key idea is to replace integration over the spatial domain with integration over the frequency domain, via Plancherel's theorem.  (Recall that curvelets are defined more simply over the frequency domain.)  We introduce some new notation, 
%\begin{equation}
$\Phi_{a,\vectheta}(\vecb) := \Gamma_f(a,\vecb,\vectheta)$.  
%\end{equation}
%to emphasize that we view this as a function of $\vecb$.  
By equation (\ref{eqn-cct-def1}), the Fourier transform of $\Phi_{a,\vectheta}$ is given by 
%\begin{equation}
$\hat{\Phi}_{a,\vectheta}(\veck) = \hat{f}(\veck) \chi_{a,\vectheta}(\veck)$.  
%\end{equation}
%And we have:
%\begin{equation}
%\text{E}(\vecb^T (I-\vectheta\vectheta^T) \vecb)
% = S_0 \int_0^1 \int_{\RR^n} (\vecb^T (I-\vec{u}\vec{u}^T) \vecb) |\Phi_{a,\vec{u}}(\vecb)|^2 d\vecb \frac{da}{a^{n+1}}.  
%\end{equation}

Let $I_K$ denote the innermost integral in equation (\ref{eqn-georgia-okeeffe}).  Then 
%\begin{equation}
$I_K = \sum_{j=2}^n \int_{\RR^n} |b_j|^2 |\Phi_{a,\vec{u}}(\vecb)|^2 d\vecb$.  
%\end{equation}
Using Plancherel's theorem, and symmetry with respect to rotations around the $\vec{u}$ axis, we can write 
%\begin{equation}
%\begin{split}
$I_K = \sum_{j=2}^n \int_{\RR^n} |\tfrac{1}{2\pi i} \tfrac{\partial}{\partial k_j} \hat{\Phi}_{a,\vec{u}}(\veck) |^2 d\veck 
 = \tfrac{n-1}{(2\pi)^2} \int_{\RR^n} |\tfrac{\partial}{\partial k_2} \hat{\Phi}_{a,\vec{u}}(\veck) |^2 d\veck$.  
%\end{split}
%\end{equation}

We can expand out the integral on the right hand side, as follows.  Using spherical coordinates $\veck = (r,\phi_1,\ldots,\phi_{n-1})$, we write $\hat{\Phi}_{a,\vec{u}}(\veck)$ as a product of a radial part and an angular part:
%\begin{equation}
$\hat{\Phi}_{a,\vec{u}}(\veck) = L(r) M(\phi_1)$, 
%\end{equation}
where 
%\begin{equation}
$L(r) = F_0(r) W(\lambda ar)$, 
and %\quad
$M(\phi_1) = V(\phi_1/\sqrt{a}) \Lambda_a(\phi_1)$.  
%\end{equation}
Then we have 
\begin{equation}
\tfrac{\partial}{\partial k_2} \hat{\Phi}_{a,\vec{u}}(\veck)
 = L'(r) M(\phi_1) \tfrac{\partial r}{\partial k_2} + L(r) M'(\phi_1) \tfrac{\partial \phi_1}{\partial k_2}, 
\end{equation}
where 
%\begin{equation}
$\frac{\partial r}{\partial k_2} = \sin\phi_1 \cos\phi_2$, 
and %\quad
$\frac{\partial \phi_1}{\partial k_2} = \frac{\cos\phi_1 \cos\phi_2}{r}$.
%\end{equation}
So we get:  (note that $\hat{\Phi}_{a,\vec{u}}(\veck)$ is real)
\begin{equation}
\begin{split}
I_K &= \frac{n-1}{(2\pi)^2} \int_{S^{n-1}} \int_0^\infty \Bigl( L'(r) M(\phi_1) \sin\phi_1 \cos\phi_2 + \\
 &\qquad L(r) M'(\phi_1) r^{-1} \cos\phi_1 \cos\phi_2 \Bigr)^2 r^{n-1} dr d\sigma(\vec{\phi}).
\end{split}
\end{equation}
%\begin{equation}
%\begin{split}
%\int_{\RR^n} & |\tfrac{\partial}{\partial k_2} \hat{\Phi}_{a,\vec{u}}(\veck) |^2 d\veck \\
% &= \int_{S^{n-1}} \int_0^\infty \Bigl( L'(r) M(\phi_1) \sin\phi_1 \cos\phi_2 + \\
% &\qquad L(r) M'(\phi_1) r^{-1} \cos\phi_1 \cos\phi_2 \Bigr)^2 r^{n-1} dr d\sigma(\vec{\phi}).
%\end{split}
%\end{equation}

We can then upper-bound these integrals in terms of $F_0$ (the radial component of $\hat{f}$).  See \ifthenelse{\boolean{app}}{Appendix B.1}{\cite{Liu-curvelets}} for details.  The final result is:
\begin{equation}
\begin{split}
\text{E}&(\vecb^T (I-\vectheta\vectheta^T) \vecb \:|\: a \leq \eta) \\
 &\leq \frac{1}{\Pr[a \leq \eta]} \frac{n-1}{(2\pi)^2} S_0 \cdot 
 \Bigl[ \tfrac{1}{2}(n-2) \int_{1/(\lambda\eta e)}^\infty F_0(r)^2 r^{n-3} dr \\
 &+ \tfrac{5}{n-1} \Bigl( \tfrac{1}{\lambda} \int_{1/(\lambda\eta e)}^\infty F'_0(r)^2 r^{n-2} dr 
 + \tfrac{17}{\lambda} \int_{1/(\lambda\eta e)}^\infty F_0(r)^2 r^{n-4} dr \Bigr) \\
 &+ (2n+9+\tfrac{10}{n-3}) e\lambda \int_{1/(\lambda\eta e)}^\infty F_0(r)^2 r^{n-2} dr \Bigr].
\end{split}
\label{eqn-elephant-seal}
\end{equation}
In sections 4 and 5, we will explain how this is used.

%%%%%%%%%%%%%%%%%%%%%%%%%%%%%%%%%%%%%%%%%%%%%%%%%%%%%%%%%%%%%%%%%%%%%%%

\vskipline
%\subsection{The variance of $\vecb$ parallel to $\vectheta$}

\noindent \textbf{3.4 The variance of $\vecb$ parallel to $\vectheta$:}
In a similar way, we can upper-bound the variance of $\vecb \cdot \vectheta$.  See \ifthenelse{\boolean{app}}{Appendix B.2}{\cite{Liu-curvelets}} for details.

}{}

%%%%%%%%%%%%%%%%%%%%%%%%%%%%%%%%%%%%%%%%%%%%%%%%%%%%%%%%%%%%%%%%%%%%%%%%%%%%%%%

% Sections 4 and 5
% Now included directly in main file
% \input{curvelets-sec45.tex}

%%%%%%%%%%%%%%%%%%%%%%%%%%%%%%%%%%%%%%%%%%%%%%%%%%%%%%%%%%%%%%%%%%%%%%%%%%%%%%%

\section{The Ball in $\RR^n$}

\ifthenelse{\boolean{sec4}}{

Let $B$ be a ball in $\RR^n$, of radius $\beta$.  In this section we will analyze the curvelet transform of the function 
$f(\vecx)$ = [$1/\sqrt{\text{vol}(B)}$ if $\vecx \in B$, 0 otherwise].
%\begin{equation}
%f(\vecx) = \begin{cases}
%  1/\sqrt{\text{vol}(B)} & \text{if $\vecx \in B$}, \\
%  0 & \text{otherwise}.
%\end{cases}
%\end{equation}
This is the wavefunction one gets by quantum-sampling over $B$.  
% We will prove bounds on the distribution of probability mass resulting from the curvelet transform.  This will eventually lead to a single-shot measurement procedure for estimating the center of the ball (see Section 7).  

% We expect that, after applying the curvelet transform, $\vecb$ and $\vectheta$ will be concentrated near the wavefront set of $f$:  that is, $\vecb$ will be concentrated near the line with direction $\vectheta$ through the center, and $\vecb$ will lie at distance $\approx \beta$ from the center.  Furthermore, we expect that $\vecb$ will become more tightly concentrated, the smaller the value of $a$.

% NOTE:  see our notes from 5/26/08

\fussy
We assume $n \geq 4$, and we use the window functions $W$ and $V$ specified in Section 3.  We set the parameter $\lambda$ to lie in the range 
%\begin{equation}
$2\pi\beta e/n \leq \lambda \leq 2\cdot 2\pi\beta e/n$.
%\end{equation}
We show the following:
\begin{thm}
\label{thm-ball}
Almost all of the power in $\hat{f}$ is located at frequencies $|\veck| \geq 1/\lambda$:
%\begin{equation}
$\int_{|\veck| \leq 1/\lambda} |\hat{f}(\veck)|^2 d\veck < \frac{1}{\pi n}$.
%\end{equation}
For any $\eta \leq 1/e^2$, the probability of observing a fine-scale element $a \leq \eta$ is lower-bounded by:
%\begin{equation}
$\Pr[a\leq\eta] \geq \frac{e\eta}{14} (1-\tfrac{1}{n})$.
%\end{equation}
Furthermore, if $\eta \leq (1/2e^2) (1-\frac{2}{n+2})$, then the variance of $\vecb$, in the directions orthogonal / parallel to $\vectheta$, conditioned on $a \leq \eta$, is upper-bounded by:
\begin{equation}
E(\vecb^T(I-\vectheta\vectheta^T)\vecb \:|\: a\leq\eta)
 \leq \eta \beta^2 ( 14300 + O(\tfrac{1}{n}) ),
\end{equation}
%\quad 
\begin{equation}
E((\vecb\cdot\vectheta)^2 \:|\: a\leq\eta)
 \leq \beta^2 ( 242 + O(\tfrac{1}{n}) ).
\end{equation}
\end{thm}

The first claim shows that only an inverse-polynomial fraction of the probability mass lies below the low-frequency cutoff; this justifies our use of the curvelet transform and Theorems \ref{thm-cct-1} and \ref{thm-cct-2}.  The second claim shows that, for any sufficiently small constant $\eta$, we observe scale $a \leq \eta$ with constant probability.  This is due to the fact that $\hat{f}$ has a lot of power at high frequencies (a ``heavy tail''), which is caused by the discontinuity of $f$ along the surface of the ball.  (For comparison, one would not observe this behavior if $f$ were, say, a Gaussian.)  The third claim shows that, when $a \leq \eta$, $\vecb$ lies within distance $O(\sqrt{\eta} \beta)$ of the line that passes through the center of the ball with direction $\vectheta$.  Also, $\vecb$ lies within distance $O(\beta)$ of the center (and we expect, though we do not prove, that this distance is also lower bounded by $\Omega(\beta)$).

As mentioned previously, it is remarkable that these bounds do not depend on the dimension $n$.  
\ifthenelse{\boolean{stocformat}}{
%nothing
}{
It is also interesting that this concentration of probability mass can hold even when the window function $\chi_{a,\vectheta}$ is only $C^1$-smooth.  By contrast, in order for $\Gamma_f$ to be asymptotically rapidly decaying, $\chi_{a,\vectheta}$ must usually be $C^k$ or $C^\infty$-smooth.
}

We prove this using our results from Section 3.  Note that the curvelet transform behaves in a simple way when we translate the function $f$:  if $g(\vecx) = f(\vecx-\vec{z})$, then using equation (\ref{eqn-cct-def1}), we see that 
%\begin{equation}
$\Gamma_g(a,\vecb,\vectheta)
% = \int_{\RR^n} \hat{f}(\veck) e^{-2\pi i \veck \cdot \vec{z}} \chi_{a,\vectheta}(\veck) e^{2\pi i \veck \cdot \vecb} d\veck
 = \Gamma_f(a,\vecb-\vec{z},\vectheta)$.
%\end{equation}
Thus, without loss of generality, we can assume that the ball $B$ is centered at the origin.  In this case $f$ is a radial function.  

We will be interested in the Fourier transform of $f$.  We write $f(\vecx) = f_0(|\vecx|)$, where 
$f_0(r)$ = [$C$ if $r \leq \beta$, 0 otherwise], 
%\begin{equation}
%f_0(r) = \begin{cases}
%  C & \text{if $r \leq \beta$,} \\
%  0 & \text{otherwise,}
%\end{cases}
%\end{equation}
and $C = 1/\sqrt{\text{vol}(B)} = (S_0 \beta^n / n)^{-1/2}$, where $S_0$ is the surface area of the unit sphere $S^{n-1}$ in $\RR^n$.  
% NOTE:  see our notes from 4/7/08, 4/18/08
Then the Fourier transform of $f$ is given by $\hat{f}(\veck) = F_0(|\veck|)$, where 
\begin{equation}
\begin{split}
F_0(\rho)
% &= \tfrac{2\pi C}{\rho^{(n-2)/2}} \int_0^\beta J_{(n-2)/2}(2\pi\rho r) r^{n/2} dr \\
% &= \frac{2\pi C}{\rho^{(n-2)/2}} \int_0^{2\pi\rho\beta} J_{(n-2)/2}(s) s^{n/2} ds \frac{1}{(2\pi\rho)^{(n+2)/2}} \\
% &= \frac{2\pi C}{\rho^{(n-2)/2}} J_{n/2}(s) s^{n/2} \Big|_{s=0}^{s=2\pi\rho\beta} \frac{1}{(2\pi\rho)^{(n+2)/2}} \\
% &= \frac{C}{\rho^{n/2}} \beta^{n/2} J_{n/2}(2\pi\rho\beta) \\
 &= \sqrt{\tfrac{n}{S_0}} \tfrac{1}{\rho^{n/2}} J_{n/2}(2\pi\rho\beta), 
\end{split}
\end{equation}
% using equation (\ref{eqn-ft-radial}), 
using the definition from Section 3, 
\sloppy
and the identity $\frac{d}{dz} (z^\nu J_\nu(z)) = z^\nu J_{\nu-1}(z)$ (see \cite{abram-stegun}, eqn. (9.1.30)).  

\fussy
The behavior of $F_0(\rho)$ depends on the behavior of the Bessel function $J_\nu(z)$ when $\nu \geq 0$ and $z \geq 0$.  $J_\nu(z)$ is very small when $z \ll \nu$, it undergoes a transition near $z \approx \nu$, and it is approximately given by $\sqrt{2/(\pi z)} \cos(z - \frac{1}{2}\nu\pi - \frac{1}{4}\pi)$ when $z \gg \nu$ (see \cite{abram-stegun}).  Thus, our intuition is that $F_0(\rho) \approx 0$ when $\rho \lesssim (n/2) / (2\pi\beta)$, and $F_0(\rho) \approx (const) / \rho^{(n+1)/2}$ times an oscillating factor when $\rho \gtrsim (n/2) / (2\pi\beta)$.

We now sketch the proof, omitting the details.  First we choose $\lambda \approx (const) \cdot (2\pi\beta)/n$, so that very little power lies at frequencies below $1/\lambda$.  Substituting into (\ref{eqn-rf-praleqeta}), we get that $\Pr[a\leq\eta] \approx (const) \cdot \eta$.  

Then we bound the variance of $\vecb$ perpendicular to $\vectheta$, as follows.  (A similar argument holds for the variance of $\vecb\cdot\vectheta$.)  We start with (\ref{eqn-elephant-seal}), and use the fact that $r \geq 1/(\lambda\eta e)$ implies $r \cdot \lambda\eta e \geq 1$:
\begin{equation*}
\begin{split}
\text{E}&(\vecb^T (I-\vectheta\vectheta^T) \vecb \:|\: a \leq \eta) \\
 &\leq \frac{1}{\Pr[a \leq \eta]} \frac{n-1}{(2\pi)^2} S_0 \cdot 
 \Bigl[ \tfrac{1}{2}(n-2) (\lambda\eta e)^2 \int_{1/(\lambda\eta e)}^\infty F_0(r)^2 r^{n-1} dr \\
 &+ \tfrac{5}{n-1} \Bigl( \tfrac{\lambda\eta e}{\lambda} \int_{1/(\lambda\eta e)}^\infty F'_0(r)^2 r^{n-1} dr 
 \ifthenelse{\boolean{stocformat}}{\\ &}{} 
 + \tfrac{17 (\lambda\eta e)^3}{\lambda} \int_{1/(\lambda\eta e)}^\infty F_0(r)^2 r^{n-1} dr \Bigr) \\
 &+ (2n+9+\tfrac{10}{n-3}) e\lambda (\lambda\eta e) \int_{1/(\lambda\eta e)}^\infty F_0(r)^2 r^{n-1} dr \Bigr].
\end{split}
\end{equation*}
A standard calculation shows that $F'_0(r) = -\sqrt{\tfrac{n}{S_0}} r^{-n/2}$ $J_{(n/2)+1}(2\pi\beta r) (2\pi\beta)$, which behaves similarly to $F_0(r)$, except that it oscillates differently and is larger by a factor of $2\pi\beta$.  Thus we can get a reasonable estimate by replacing $F'_0(r)$ with $F_0(r)(2\pi\beta)$ in the integral:
\begin{equation*}
\begin{split}
\text{E}&(\vecb^T (I-\vectheta\vectheta^T) \vecb \:|\: a \leq \eta) \\
 &\lesssim \frac{1}{\Pr[a \leq \eta]} \frac{n-1}{(2\pi)^2} S_0 \cdot 
 \Bigl[ \tfrac{1}{2}(n-2) (\lambda\eta e)^2 \\
 &+ \tfrac{5}{n-1} \Bigl( \tfrac{\lambda\eta e}{\lambda} (2\pi\beta)^2 
 + \tfrac{17 (\lambda\eta e)^3}{\lambda} \Bigr) 
 + (2n+9+\tfrac{10}{n-3}) e\lambda (\lambda\eta e) \Bigr] \\
 &\cdot \int_{1/(\lambda\eta e)}^\infty F_0(r)^2 r^{n-1} dr \\
\end{split}
\end{equation*}
Using the definition of $\lambda$, 
\begin{equation*}
\begin{split}
\text{E}&(\vecb^T (I-\vectheta\vectheta^T) \vecb \:|\: a \leq \eta) \\
 &\lesssim \frac{1}{\Pr[a \leq \eta]} \frac{n-1}{(2\pi)^2} S_0 \cdot 
 \Bigl[ (const) \tfrac{(2\pi\beta)^2}{n} \eta^2 \\
 &+ \Bigl( (const) \tfrac{(2\pi\beta)^2}{n} \eta 
 + (const) \tfrac{(2\pi\beta)^2}{n^3} \eta^3 \Bigr) 
 + (const) \tfrac{(2\pi\beta)^2}{n} \eta \Bigr] \\
 &\cdot \int_{1/(\lambda\eta e)}^\infty F_0(r)^2 r^{n-1} dr.
\end{split}
\end{equation*}
Now recall the expression for $\Pr[a \leq \eta]$ given by (\ref{eqn-rf-praleqeta}).  We expect the two integrals to roughly cancel out, so we get 
\begin{equation}
\text{E}(\vecb^T (I-\vectheta\vectheta^T) \vecb \:|\: a \leq \eta)
 \lesssim O(\beta^2 \eta).
\end{equation}

This argument can be made rigorous, using known results about Bessel functions $J_\nu(z)$.  However, there is a technical obstacle:  our theorem concerns the case where $z$ is roughly proportional to $\nu$.  This is still in the transition regime, and the usual asymptotic expansions for $J_\nu(z)$ do not work here (they only work when $z \gtrsim \nu^2$, or when $z/\nu$ is some fixed ratio).  Fortunately, there are useful bounds on the quantity $M_\nu(z) := \sqrt{J_\nu(z)^2 + Y_\nu(z)^2}$, and representations of $J_\nu(z)$ and $Y_\nu(z)$ in terms of a modulus and phase, that do work in this regime \cite{abram-stegun, watson}.  This leads to a rigorous proof of our theorem---see \ifthenelse{\boolean{app}}{the Appendix}{\cite{Liu-curvelets}} for details.

}{}

%%%%%%%%%%%%%%%%%%%%%%%%%%%%%%%%%%%%%%%%%%%%%%%%%%%%%%%%%%%%%%%%%%%%%%%%%%%%%%%

\section{Spherical Shells}

\ifthenelse{\boolean{sec5}}{

We now consider the curvelet transform of a function supported on a thin spherical shell in $\RR^n$.  We will show results similar to the previous section, except that they now depend on the \textit{thickness} of the shell.  Intuitively, when the shell is very thin, we can measure very fine-scale elements ($a$ small) with significant probability, and $\vecb$ is tightly concentrated around the wavefront set.

% We will prove bounds on the distribution of probability mass, showing that the curvelet transform reveals information about the center of the spherical shell, just as it did with the ball, but with a significant improvement:  we can find the center with accuracy that scales with the \textit{thickness} of the shell.  Even if the shell has very large radius, if it is also very thin (i.e., its inner and outer surfaces are close together), then we can find its center very accurately.  

% We will use these results in Section 7, to design a quantum algorithm for an oracle problem, finding the center of a radial function.  We will give evidence that the quantum algorithm requires a constant number of queries and polynomial time, whereas known classical algorithms for the same problem require a polynomial number of queries and polynomial time.

% \vskipline

Without loss of generality, we can assume the shell is centered at the origin (see Section 4).  So consider the following function on $\RR^n$, 
$f(\vecx) = [C \; \text{if $\beta < |\vecx| \leq \beta+\delta$}, \; 0 \; \text{otherwise}]$, 
%\begin{equation}
%f(\vecx) = \begin{cases}
%  C, & \text{if $\beta < |\vecx| \leq \beta+\delta$}, \\
%  0, & \text{otherwise}, 
%\end{cases}
%\end{equation}
where $C = 1/\sqrt{(\beta+\delta)^n B_0 - \beta^n B_0}$, and $B_0$ is the volume of the unit ball in $\RR^n$.  This represents a uniform superposition over a spherical shell centered at the origin, with inner radius $\beta$ and thickness $\delta$.  We call this a spherical shell with ``square'' cross-section.

This is the exactly the kind of state that appears in our quantum algorithm.  However, it is difficult to analyze, as its Fourier transform involves a linear combination of two Bessel functions oscillating at different rates.  We are interested in the case where $\delta \ll \beta$.  In \ifthenelse{\boolean{app}}{Appendix D.1}{\cite{Liu-curvelets}}, we give a heuristic explanation of why the curvelet transform of this state will be tightly concentrated around the wavefront set.  (This holds when $\delta \lesssim \beta/n$.)

% \vskipline

Here, we give a more rigorous argument, for spherical shells that have ``Gaussian'' cross-sections---when $\delta \ll \beta$, these functions are similar to the above, but they are analytically tractable.  We define $f = C_f g*q$, where:  $C_f$ is a normalization factor; $g$ is a Gaussian of width $\delta$, that is, $g(\vecx) = \delta^{-n/2} \exp(-\pi |\vecx|^2/\delta^2)$; $q$ is the measure supported on the sphere of radius $\beta$ around the origin, which is obtained by restricting the usual volume measure on $\RR^n$; and the star denotes convolution.  Intuitively, $q$ represents a shell with infinitesimal thickness, and $f$ represents a ``smoothed'' shell with thickness $\delta$.

The Fourier transform of $f$ is given by $\hat{f} = C_f \hat{g} \cdot \hat{q}$, where:  $\hat{g}$ is a Gaussian of width $1/\delta$, $\hat{g}(\veck) = \delta^{n/2} \exp(-\pi \delta^2 |\veck|^2)$; and $\hat{q}$ is given by $\hat{q}(\veck) = Q_0(|\veck|)$, where $Q_0(\rho) = \frac{2\pi}{\rho^{(n/2)-1}}$ $J_{(n/2)-1}(2\pi\beta\rho) \beta^{n/2}$.  Intuitively, the Fourier transform of the spherical shell is somewhat like the Fourier transform of the ball, except that it decays more slowly (i.e., has more power at high frequencies), for frequencies up to roughly $1/\delta$; the power at frequencies above $1/\delta$ is suppressed by $\hat{g}$.

\ifthenelse{\boolean{app}}{
Note that this is quite similar to equation (\ref{eqn-ss-squarecs-F0}), describing a spherical shell with ``square'' cross-section, when we substitute in the upper and lower bounds on $C_f$ (to be given later in this section).  This suggests that the spherical shell with ``square'' cross-section can indeed be approximated by one with ``Gaussian'' cross-section, when $\delta \lesssim \beta/n$.
}{
We remark that this is quite similar to the Fourier transform of a spherical shell with ``square'' cross-section, when $\delta \lesssim \beta/n$.  See \cite{Liu-curvelets} for details.
}

% \vskipline

% We will prove bounds on the curvelet transform of a spherical shell with ``Gaussian'' cross-section.  
We will prove bounds on the continuous curvelet transform, with the same window functions as in Section 4.  We use a slightly different scaling parameter $\lambda$:  we set 
%\begin{equation}
$\lambda = \frac{2\pi\betatilde e}{n-2}$,
%\end{equation}
where $\betatilde$ is an estimate of the true radius of the shell, which satisfies $\beta \leq \betatilde \leq S\beta$, for some $S \geq 1$.  We assume that the dimension $n$ is at least 4, and we assume that the thickness of the shell is small compared to the radius: 
%\begin{equation}
$\delta = \varepsilon\beta$, 
where %\quad 
$\varepsilon \leq \min\Bigl( \frac{6}{(n-2)^2}, \frac{1}{n+2}, \frac{1}{en}, \frac{1}{5} \Bigr)$.  
%\end{equation}
(Note that for these values of $n$, $\frac{1}{en} \leq \frac{1}{n+2}$ and $\frac{1}{en} \leq \frac{1}{5}$, so the second and fourth conditions are actually redundant.)  
%Note that we can write $F_0(\rho)$ as follows, substituting in $\delta = \varepsilon\beta$:
%\begin{equation}
%F_0(\rho) = C_f \cdot \varepsilon^{n/2} \frac{2\pi}{\rho^{(n/2)-1}} \beta^n \cdot \exp(-\pi\varepsilon^2\beta^2\rho^2) \cdot J_{(n/2)-1}(2\pi\beta\rho).  
%\end{equation}
Under these assumptions, we prove the following:
\begin{thm}
\label{thm-sphericalshell}
Almost all of the power in $\hat{f}$ is located at frequencies $|\veck| \geq 1/\lambda$:
%\begin{equation}
$\int_{|\veck| \leq 1/\lambda} |\hat{f}(\veck)|^2 d\veck \leq \frac{\varepsilon}{5}$.
%\end{equation}
Let $\eta_c = (\delta/\betatilde)(n-2)/e$.  The probability of observing a fine-scale element $a \leq \eta_c$ is lower-bounded by:
%\begin{equation}
$\Pr[a\leq\eta_c] > 0.045$.
%\end{equation}
Furthermore, the variance of $\vecb$, in the directions orthogonal / parallel to $\vectheta$, conditioned on $a \leq \eta_c$, is upper-bounded by:
\begin{equation}
E(\vecb^T(I-\vectheta\vectheta^T)\vecb \:|\: a\leq\eta_c)
 \leq (n-1)\varepsilon \beta^2 ( 507 + O(\tfrac{1}{n}) ) \cdot S,
\end{equation}
%\quad 
\begin{equation}
E((\vecb\cdot\vectheta)^2 \:|\: a\leq\eta_c)
 \leq \beta^2 ( 23 + O(\tfrac{1}{n^2}) ).
\end{equation}
\end{thm}

The proof uses a similar strategy to what we showed in Section 4.  The intuition is as follows.  First write $\hat{f}(\veck) = F_0(|\veck|)$.  The important difference (compared to Section 4) is that here $F_0(r)$ decays more slowly, like $1/r^{(n-1)/2}$, for $r \lesssim 1/\delta$.  Substituting into (\ref{eqn-rf-praleqeta}), we see that a constant fraction of the probability mass lies at frequencies of order $1/\delta$.  So with constant probability, we can observe fine-scale elements $a\leq\eta_c$ where $\eta_c$ is of order $\delta/\lambda$.  Note that $\eta_c \approx \delta/\lambda \approx (const) (\delta/\beta) (n-2) \approx (const) \varepsilon (n-2)$.  So, when the shell is very thin, $\eta_c$ will be very small, and $\vecb$ will be tightly concentrated around the wavefront set.  

The rigorous proof is given in \ifthenelse{\boolean{app}}{Appendix D.2}{\cite{Liu-curvelets}}.  

}{}

%%%%%%%%%%%%%%%%%%%%%%%%%%%%%%%%%%%%%%%%%%%%%%%%%%%%%%%%%%%%%%%%%%%%%%%%%%%%%%%

\ifthenelse{\boolean{stocformat}}{
 \section{A Fast Quantum Curvelet \\ Transform}
}{
 \section{A Fast Quantum Curvelet Transform}
}

\ifthenelse{\boolean{sec6}}{

% NOTE:  see notes from 8/28/08 and following days

% In this section we present a quantum curvelet transform, acting on superposition states, and we show how such a transform can be computed efficiently.  

%\subsection{The Discrete Curvelet Transform}
\vskipline
\noindent \textbf{6.1 The Discrete Curvelet Transform:}
First, we describe the discrete curvelet transform, which has been studied in the classical setting \cite{fast-curvelet-trans}.  The discrete curvelet transform takes a function $f(\vecx)$ and returns a function $\Gamma_f(a,\vecb,\vectheta)$, where both functions are defined over finite domains.  This is constructed analogously to the continuous curvelet transform, except that one now uses the discrete Fourier transform on $(\ZZ_M)^n$, and a discrete set of scale/direction pairs $(a,\vectheta)$.  

The discrete Fourier transform is defined as follows.  We assume that $f$ is defined on a domain $Z$ consisting of a discrete grid in a finite region of $\RR^n$.  For example, let $Z = (\sigma\ZZ)^n \cap [-L,L)^n$, the intersection of a tightly-spaced square lattice and a large cube.  Also let $\hat{Z} = (\tfrac{1}{2L}\ZZ)^n \cap [-\tfrac{1}{2\sigma}, \tfrac{1}{2\sigma})^n$.  The discrete Fourier transform maps $f$ to a function $\hat{f}$ defined on $\hat{Z}$, as follows:  
%\begin{align}
$\hat{f}(\veck)
 = (\tfrac{\sigma}{2L})^{n/2} \sum_{\vecx\in Z} f(\vecx) e^{-2\pi i \veck\cdot\vecx}$, 
and 
$f(\vecx)
 = (\tfrac{\sigma}{2L})^{n/2} \sum_{\veck\in\hat{Z}} \hat{f}(\veck) e^{2\pi i \veck\cdot\vecx}$.
%\end{align}

One can argue that this is approximates the continuous Fourier transform in the following sense.  Let $f_{cont}$ be a function on $\RR^n$, and let $\hat{f}_{cont}$ be its continuous Fourier transform.  Suppose that $f_{cont}$ is supported inside the cube $[-L,L)^n$, and $\hat{f}_{cont}$ has all except an $\varepsilon$ fraction of its probability mass inside the cube $[-\tfrac{1}{2\sigma}, \tfrac{1}{2\sigma})^n$.  Then there exists a function $f_{dis}$ on $Z$, with discrete Fourier transform $\hat{f}_{dis}$, such that $f_{dis} \approx \sigma^{n/2} f_{cont} |_Z$ and $\hat{f}_{dis} = (\tfrac{1}{2L})^{n/2} \hat{f}_{cont} |_{\hat{Z}}$, up to errors whose total probability mass is roughly $\varepsilon$.  See \ifthenelse{\boolean{app}}{Appendix E.1}{\cite{Liu-curvelets}} for more details.

Now the discrete curvelet transform is given by 
\begin{equation}
\Gamma_f(a,\vecb,\vectheta)
 := (\tfrac{\sigma}{2L})^{n/2} \sum_{\veck\in\hat{Z}} \hat{f}(\veck) \chi_{a,\vectheta}(\veck) e^{2\pi i \veck\cdot\vecb}.
\end{equation}
The ``location'' variables $\vecx$ and $\vecb$ take values in $Z$, and the ``scale'' and ``direction'' variables $(a,\vectheta)$ take values in some discrete set $G$, which we will describe below.  The window functions $\chi_{a,\vectheta}$ are defined over $\hat{Z}$, and are constructed so that 
%\begin{equation}
$\sum_{a,\vectheta} \chi_{a,\vectheta}(\veck)^2 = 1$, 
%\quad 
$\forall \veck \in \hat{Z}$.
%\end{equation}
This ensures that the curvelet transform can be realized as a unitary operation on the space spanned by the states $\ket{\veck,a,\vectheta}$.  

\begin{figure}
\ifthenelse{\boolean{stocformat}}{
 \centering
 \epsfig{file=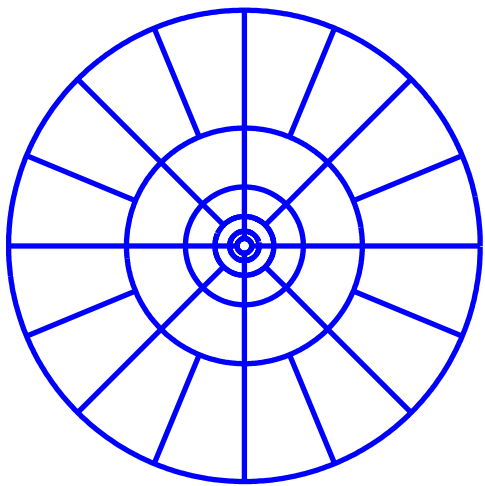, height=2in, width=2in}
}{
 \centering
 \includegraphics{tiling.eps}
}
\caption{Construction of a 2-D discrete curvelet transform --- ``tiling'' of the frequency domain into sectors $S_{a,\vectheta}$, plus special sectors $S_{low}$ and $S_{high}$ for very low and very high frequencies.  Within the annulus, $a$ takes on values $1/2^k$, for $k = 0,1,2,3,4$.  Each sector has inner radius $2^k = 1/a$, outer radius $2^{k+1} = 2/a$, and angular width $(\pi/2) / 2^{\lfloor k/2 \rfloor} \approx (\pi/2) \sqrt{a}$.}
\label{fig-tiling}
\end{figure}

Recall from Section 2 that each window function $\chi_{a,\vectheta}$ is supported on a ``sector'' $S_{a,\vectheta}$ that has angular width $\sqrt{a}$, inner radius $1/a$ and outer radius $2/a$.  To satisfy the above condition on $\chi_{a,\vectheta}$, we want to choose a discrete set of values $(a,\vectheta)$, that corresponds to a discrete collection of sectors $S_{a,\vectheta}$, that forms a ``tiling'' of the frequency domain.  Intuitively, this is done by setting $a$ equal to powers of 2, and sampling $\vectheta$ from a mesh with angular spacing $\sqrt{a}$ on the sphere $S_{n-1}$.  Then the sectors $S_{a,\vectheta}$ fit together nicely, as in Figure \ref{fig-tiling}.  (This picture is a slight oversimplification; actually, since we want the window functions $\chi_{a,\vectheta}$ to decay smoothly to zero, we should make their supports overlap slightly.)  We will describe a construction of this kind in the next section; other constructions were given in \cite{fast-curvelet-trans, fast-curvelet-trans-3d}.

This discretization affects the values of $a$ and $\vectheta$, relative to the continuous case.  Intuitively, the ``discrete'' $a$ can differ from the ``continuous'' $a$ by a constant factor, and the ``discrete'' $\vectheta$ can differ from the ``continuous'' $\vectheta$ by an additive error of size $\sqrt{a}$.

%\subsection{The Quantum Curvelet Transform}
\vskipline
\noindent \textbf{6.2 The Quantum Curvelet Transform:}
% Then, we define the quantum curvelet transform.  
% Here we encounter some new technical challenges:  implementing this transform requires preparing certain superpositions of different scales and directions $\ket{a,\vectheta}$.  This seems hard for a generic choice of the window functions $\chi_{a,\vectheta}(\veck)$.  However, we construct two specific families of window functions, for which the quantum curvelet transform can be implemented efficiently.  One family consists of indicator functions (``Haar'' curvelets), which have poor analytic properties, but are extremely simple.  The other family consists of certain ``bump'' functions, which have nicer smoothness properties, and more closely resemble the curvelets used in Sections 3, 4 and 5.  
The quantum curvelet transform is the unitary operation that maps 
\begin{equation}
\sum_{\vecx} f(\vecx) \ket{\vecx} \ket{0,\vec{0}}
 \mapsto \sum_{a,\vecb,\vectheta} \Gamma_f(a,\vecb,\vectheta) \ket{\vecb} \ket{a,\vectheta}.
\end{equation}
This can be implemented as follows:  first apply the quantum Fourier transform (QFT), then the operation $\cal{X}$ that maps 
%\begin{equation}
$\ket{\veck} \ket{0,\vec{0}}
 \mapsto \ket{\veck} \sum_{a,\vectheta} \chi_{a,\vectheta}(\veck) \ket{a,\vectheta}$,
%\end{equation}
and then the inverse QFT.  

We want to compute this in time polynomial in $n$ and $\log(M)$ (where $M = 2L/\sigma$ is the length of the discrete Fourier transform).  This is possible for the QFT.  But it is not clear how to perform the operation $\cal{X}$, for a generic choice of the window functions $\chi_{a,\vectheta}$.  (Note that we want the functions $\chi_{a,\vectheta}$ to be $C^1$-smooth, so their supports will necessarily overlap; thus the operation $\cal{X}$ must prepare a superposition containing $2^{\Theta(n)}$ terms.)  

Nonetheless, we can perform $\cal{X}$ efficiently in two cases:  (1) when the window functions are indicator functions supported on disjoint sets, and (2) when the window functions are smooth ``bump'' functions that can be expressed as products of 1-D functions using spherical coordinates.  The first case has poor analytic properties, but the second case is a reasonable approximation of the curvelets used in Sections 3-5.  Thus we get an efficient quantum curvelet transform.  See \ifthenelse{\boolean{app}}{Appendix E.2}{\cite{Liu-curvelets}} for details.

}{}

%%%%%%%%%%%%%%%%%%%%%%%%%%%%%%%%%%%%%%%%%%%%%%%%%%%%%%%%%%%%%%%%%%%%%%%%%%%%%%%

\section{Quantum Algorithms using the Curvelet Transform}

\ifthenelse{\boolean{sec7}}{

\vskipline
%\subsection{Single-shot measurement of a quantum-sample state}

\noindent \textbf{7.1 Single-shot measurement of a quantum-sample state:}
Consider the following problem.  Let $B$ be a ball of (unknown) radius $\beta$ centered at some (unknown) point $\vecc$ in $\RR^n$, for $n \geq 4$.  We are given as input:  $n$, the dimension; $\betatilde$, an estimate of the radius of the ball (we are promised that $\betatilde/2 \leq \beta \leq \betatilde$); $R$, an outer bound on the location of the center (we are promised that $|\vecc| \leq R$); $\mu$, the desired accuracy of our answer; a description of the set of grid points $G = (\sigma\ZZ)^n \cap [-L,L)^n$, such that $L \geq R+\betatilde$ and $\sigma \leq \frac{\pi e}{600} \frac{\mu^2}{\betatilde} \frac{1}{14300n+Q_1}$ (for some constant $Q_1$, to be specified later); and a quantum state $\frac{1}{\sqrt{|G \cap B|}} \sum_{\vecx \in G \cap B} \ket{\vecx}$, that is, a single quantum-sample over the ball $B$.  We are then asked to output a point $\vec{z}$ in $\RR^n$, that lies within distance $\mu$ of the center $\vecc$.
%\begin{tabbing}
%Output: \= \kill % dummy line to set tab stop
%Input: \> a natural number $n \geq 4$; \\
%\> a description of the set of grid points $G = (\sigma\ZZ)^n \cap [-L,L)^n$; \\
%\> real numbers $\beta$, $R$ and $\mu$; \\
%\> a quantum state $\frac{1}{\sqrt{|G \cap B|}} \sum_{\vecx \in G \cap B} \ket{\vecx}$, where $B$ is a ball in $\RR^n$ of radius $\beta$, \\
%\> centered at some unspecified point $\vecc$, such that $\vecc$ lies within distance $R$ of the origin. \\
%Output: \> a point $\vec{z}$ in $\RR^n$, which lies within distance $\mu$ of $\vecc$.
%\end{tabbing}

We propose the following algorithm.  Intuitively, this algorithm uses the curvelet transform to find a line that passes near the center of the ball, then guesses a random point along this line.
\begin{tabbing}
\textbf{Algorithm 1:} \\
Let $\ket{\psi}$ be the input quantum state. \\
Set $\eta = \frac{1}{6} \frac{\mu^2}{\betatilde^2} \frac{1}{14300 + (Q_1/n)}$, where $Q_1$ is some constant. \\
Apply the fast quantum curvelet transform, with \ifthenelse{\boolean{app}}{$\lambda = \frac{2\pi\betatilde e}{n}$, $s_{min} = 1$, $s_{max} = \lg\frac{1}{\eta} + 3$.}{$\lambda = \frac{2\pi\betatilde e}{n}$.} \\
Measure the scale \ifthenelse{\boolean{app}}{$a = 2^{-s}$}{$a$}, location $\vecb$ and direction $\vectheta$. \\
If $a > \eta$, then return ``no answer.'' \\
Set $Q_3 = \sqrt{3}\sqrt{242+(Q_2/n)}$, where $Q_2$ is some constant. \\
Guess some $u \in [-1,1]$ uniformly at random. \\
Return the point $\vecb' = \vecb + u Q_3 \betatilde \vectheta$.
\end{tabbing}

We are especially interested in instances where the error $\mu$ is a constant fraction of the radius $\beta$, i.e., $\mu = \nu\beta$, for some fixed $\nu < 1$.  We conjecture that, for any $\nu$, Algorithm 1 solves these instances with probability $\Omega(\nu^3)$, independent of the dimension $n$.  (In other words, the success probability has a ``heavy tail.'')  This is a sharp contrast to what happens in the classical case:  if we choose a single point uniformly at random from the ball, then the success probability is $\nu^n$, which is exponentially small in $n$.  This is because, in high dimensions, most of the volume of the ball lies near its surface.  This is bad for classical sampling, but it helps the quantum curvelet transform, which works by finding a line normal to the surface of the ball.  

We now show how our results from Section 4 support this conjecture.  We prove the following:
\begin{thm} \label{thm-ball-alg}
Consider a ``continuous'' analogue of Algorithm 1, using the continuous curvelet transform over $\RR^n$.  This algorithm succeeds with constant probability $\geq \Omega(\nu^3)$.  
\end{thm}
We will also argue, non-rigorously, that the discrete algorithm will behave like the continuous one, provided that the grid $G$ is sufficiently fine.  When the grid is chosen properly, the discrete algorithm runs in time $\poly(n, \log R, \log\betatilde, \log\frac{1}{\mu})$.  See \ifthenelse{\boolean{app}}{Appendix F.1}{\cite{Liu-curvelets}} for details.

We remark that it should be possible to achieve a better success probability, $\Omega(\nu^2)$, using the quantum curvelet transform.  Here, we showed that $\vecb$ was within distance $O(\beta)$ of the center, so in the last step of the algorithm, we simply guessed a point along the line, with success probability $\Omega(\nu)$.  But in fact, $\vecb$ should lie at distance $\approx \beta$ from the center, so we should be able to guess one of the two points $\vecb \pm \beta\vectheta$, with success probability $\Omega(1)$.  

We also remark that classical sampling becomes more powerful if one is allowed any constant number of samples, instead of just one.  By sampling $k$ random points from the ball and taking their average, one can find the center with accuracy $\pm \beta/\sqrt{k}$, with constant probability.  (However, for fixed $k$, the success probability does not have a ``heavy tail,'' i.e., one cannot expect to get better accuracy with significant probability.  This is because, in high dimensions, random sampling produces $k$ vectors that are nearly orthogonal.)

\vskipline
%\subsection{Quantum algorithm for finding the center of a radial function}

\noindent \textbf{7.2 Quantum algorithm for finding the center of a radial function:}
Let $f$ be a radial function on $\RR^n$ (where $n \geq 4$), centered at some point $\vecc$, and taking values in some arbitrary set.  Suppose that the level sets of $f$ are concentric spherical shells of thickness $\delta$ centered at $\vecc$, i.e., $f$ is constant on each shell, and $f$ takes on distinct values on different shells.  (Note, in previous versions of this paper, we made an additional assumption, that one can efficiently compute the radius of a shell, given the value of $f$ on that shell.  This assumption is no longer needed.)

Consider the following problem.  We are given as input:  $n$, the dimension; $R$, an outer bound on the location of the center (we are promised that $|\vecc| \leq R$); $\delta$, the thickness of the spherical shells; $\mu$, the desired accuracy of our answer; and an oracle that computes the radial function $f$.  We are asked to output a point $\vec{z}$ in $\RR^n$, that lies within distance $\mu$ of the center $\vecc$.
%\begin{tabbing}
%Output: \= \kill % dummy line to set tab stop
%Input: \> a natural number $n \geq 4$; \\
%\> a description of the set of grid points $G = (\sigma\ZZ)^n \cap [-L,L)^n$; \\
%\> real numbers $R$, $\delta$ and $\mu$; \\
%\> an oracle that computes a radial function $f$, where $f$ satisfies the above conditions, \\
%\> and the (unknown) center point $\vecc$ lies within distance $R$ of the origin. \\
%Output: \> a point $\vec{z}$ in $\RR^n$, which lies within distance $\mu$ of $\vecc$.
%\end{tabbing}

We propose the following algorithm.  The basic idea is to prepare a quantum superposition over a large ball around the origin, then measure the value of $f$ to get a superposition over a spherical shell centered at $\vecc$, then apply the curvelet transform, and find a line that passes near $\vecc$.  The algorithm does this twice, then returns the point on the first line that lies closest to the second line (note that, with high probability, the two lines are nearly orthogonal).
\begin{tabbing}
\textbf{Algorithm 2:} \\
Set $R' = nR$.  Let $B$ be a ball of radius $R'$ around $\vec{0}$. \\
Choose the grid $G = (\sigma\ZZ)^n \cap [-L,L)^n$, \ifthenelse{\boolean{stocformat}}{\\}{}
where $L = 2R'$ and $\sigma = \delta/400$. \\
For \= $i \in \set{1,2}$, do the following: \\
\> Prepare the state $\frac{1}{\sqrt{|G \cap B|}} \sum_{\vecx \in G \cap B} \ket{\vecx}$, \ifthenelse{\boolean{stocformat}}{\\ \>}{} using the methods of \cite{aharonov-tashma} or \cite{grover-rudolph}. \\
\> Compute the value of $f$ in an auxiliary register, \ifthenelse{\boolean{stocformat}}{\\ \>}{} and measure it; call this $y^{(i)}$. \\
\> Set $\betatilde^{(i)} = R'-R$. \\
\> Set $\eta^{(i)} = (\delta/\betatilde^{(i)}) (n-2)/e$. \\
\> Apply the fast quantum curvelet transform, \\
\> with \ifthenelse{\boolean{app}}{$\lambda^{(i)} = \frac{2\pi\betatilde^{(i)} e}{n-2}$, $s_{min}^{(i)} = 1$, $s_{max}^{(i)} = \lg\frac{1}{\eta^{(i)}} + 3$.}{$\lambda^{(i)} = \frac{2\pi\betatilde^{(i)} e}{n-2}$.} \\
\> Measure the scale \ifthenelse{\boolean{app}}{$a^{(i)} = 2^{-s^{(i)}}$}{$a^{(i)}$}, location $\vecb^{(i)}$ and direction $\vectheta^{(i)}$. \\
\> If $a^{(i)} > \eta^{(i)}$, then return ``no answer.'' \\
End for. \\
If $|\vectheta^{(1)} \cdot \vectheta^{(2)}| > 3/4$, then return ``no answer.'' \\
Set $r = \vectheta^{(1)} \cdot \vectheta^{(2)}$, $s = \vectheta^{(1)} \cdot (\vecb^{(1)}-\vecb^{(2)})$, $t = \vectheta^{(2)} \cdot (\vecb^{(1)}-\vecb^{(2)})$. \\
Return the point $\frac{-s+rt}{1-r^2} \vectheta^{(1)} + \vecb^{(1)}$.
\end{tabbing}

We conjecture that Algorithm 2 finds the center with arbitrary precision $\mu$, provided that $\delta$ is sufficiently small, i.e., the radial function $f$ computed by the oracle is sufficiently ``precise.''  Let us assume that 
%\begin{equation}
$\delta \leq \frac{1}{192} \cdot \frac{\mu^2}{(n-1)^2R} \cdot \frac{1}{761+(Q_1/n)}$,
%\label{eqn-alg2-delta}
%\end{equation}
for some constant $Q_1$ to be defined later.  We conjecture that Algorithm 2 then finds a solution with constant probability, independent of the dimension $n$.  Thus, only $O(1)$ oracle queries are needed.  This is an improvement over the classical case, where $\tilde{\Omega}(n \log \frac{R}{\mu})$ queries are required.  (The $\tilde{\Omega}$ indicates that we are omitting some log factors.)  % Intuitively, this is because the curvelet transform uses constructive interference to find a direction in $\RR^n$ from just one quantum query.  

We now show how our results from Section 5 support this conjecture.  We prove the following:  
\begin{thm} \label{thm-crfn-alg}
Consider a ``continuous'' analogue of Algorithm 2, using the continuous curvelet transform over $\RR^n$.  This algorithm succeeds with constant probability.  % (This holds provided that $\delta$ is small, so the Gaussian approximation in Section 5 is valid.)  
% This holds provided that the spherical shell with square cross-section can be approximated by one with Gaussian cross-section, as in Section 5.  This approximation is accurate when $\delta$ is small.  
\end{thm}
We will also argue, non-rigorously, that the discrete algorithm will behave like the continuous one, provided that the grid $G$ is sufficiently fine.  When the grid is chosen properly, the algorithm runs in time $\poly(n, \log R, \log\frac{1}{\mu})$.  See \ifthenelse{\boolean{app}}{Appendix F.2}{\cite{Liu-curvelets}} for details.

\vskipline
%\subsection{Classical lower bound}

\noindent \textbf{7.3 Classical lower bound:}
We claim that any classical algorithm for finding the center of a radial function must use at least $\tilde{\Omega}(n \log \frac{R}{\mu})$ queries.  (The $\tilde{\Omega}$ indicates that we are omitting some log factors.)  

Our intuition is as follows.  Any algorithm can be described as a decision tree, where each node represents a query to the oracle $f$, and the algorithm chooses which branch to follow depending on the oracle's answer.  However, the values of the function $f$ are meaningless by themselves, so when the algorithm receives an answer from the oracle, the algorithm cannot do anything besides comparing this answer with the answers returned previously.  Thus, after its $k$'th query, the algorithm can choose one of at most $k$ distinct branches.  

It follows that, if the algorithm makes $\ell$ queries, the number of possible outputs (i.e., the number of leaves in the tree) is at most $\ell!$.  In order to solve this problem, however, the algorithm must be able to output at least $(R/\mu)^n$ different points.  So we have $(R/\mu)^n \leq \ell!$, which implies $\ell \geq \tilde{\Omega}(n \log \frac{R}{\mu})$.  

A formal statement and proof of this result is given in \ifthenelse{\boolean{app}}{Appendix F.3}{\cite{Liu-curvelets}}.

\vskipline
%\subsection{Finding the center through multiple iterations}

\noindent \textbf{7.4 Finding the center through multiple iterations:}
We now describe a variant of Algorithm 2, for finding the center of a radial function.  This new algorithm will use multiple iterations, and a larger number of queries, but it has a less demanding requirement on the thickness of the shells that form the level sets of the radial function $f$.  

First, we describe a single iteration of the new algorithm.  We call this procedure OneRound().  This is similar to Algorithm 2, but it starts out with a promise that the center lies within distance $R$ of some point $\vec{p}$, and it returns a point $\vec{q}$ that lies within distance $\sqrt{R} \sqrt{\mu/2}$ of the center.  OneRound() also takes a parameter $S \geq 1$ that controls the accuracy and success probability:  OneRound() returns a point $\vec{q}$ (instead of ``no answer'') with constant probability, and when this happens, the point $\vec{q}$ is accurate with probability $\geq 1-O(\frac{1}{S})$.
\begin{tabbing}
\textbf{Procedure OneRound}($R$, $\vec{p}$, $S$): \\
Set $R' = nSR$.  Let $B$ be a ball of radius $R'$ around $\vec{0}$. \\
Choose the grid $G = (\sigma\ZZ)^n \cap [-L,L)^n$, \ifthenelse{\boolean{stocformat}}{\\}{} where $L = 2R'$ and $\sigma = \delta/400$. \\
For \= $i \in \set{1,2}$, do the following: \\
\> Prepare the state $\frac{1}{\sqrt{|G \cap B|}} \sum_{\vecx \in G \cap B} \ket{\vecx}$, \ifthenelse{\boolean{stocformat}}{\\ \>}{} using the methods of \cite{aharonov-tashma} or \cite{grover-rudolph}. \\
\> Define the function $g(\vecx) = f(\vecx+\vec{p})$. \\
\> Compute the value of $g$ in an auxiliary register, \ifthenelse{\boolean{stocformat}}{\\ \>}{} and measure it; call this $y^{(i)}$. \\
\> Set $\betatilde^{(i)} = R'-R$. \\
\> Set $\eta^{(i)} = (\delta/\betatilde^{(i)}) (n-2)/e$. \\
\> Apply the fast quantum curvelet transform, \\
\> with \ifthenelse{\boolean{app}}{$\lambda^{(i)} = \frac{2\pi\betatilde^{(i)} e}{n-2}$, $s_{min}^{(i)} = 1$, $s_{max}^{(i)} = \lg\frac{1}{\eta^{(i)}} + 3$.}{$\lambda^{(i)} = \frac{2\pi\betatilde^{(i)} e}{n-2}$.} \\
\> Measure the scale \ifthenelse{\boolean{app}}{$a^{(i)} = 2^{-s^{(i)}}$}{$a^{(i)}$}, location $\vecb^{(i)}$ and direction $\vectheta^{(i)}$. \\
\> If $a^{(i)} > \eta^{(i)}$, then return ``no answer.'' \\
End for. \\
If $|\vectheta^{(1)} \cdot \vectheta^{(2)}| > 3/4$, then return ``no answer.'' \\
Set $r = \vectheta^{(1)} \cdot \vectheta^{(2)}$, $s = \vectheta^{(1)} \cdot (\vecb^{(1)}-\vecb^{(2)})$, $t = \vectheta^{(2)} \cdot (\vecb^{(1)}-\vecb^{(2)})$. \\
Return the point $\vec{q} = \vec{p} + \frac{-s+rt}{1-r^2} \vectheta^{(1)} + \vecb^{(1)}$.
\end{tabbing}

Now we describe the full algorithm, with multiple iterations.  This algorithm begins with a point at distance $R$ from the center, then uses OneRound() to find a point at distance $\sqrt{R} \sqrt{\mu/2}$ from the center, and by repeating the procedure, shrinks the distance to $R^{1/4} (\mu/2)^{3/4}$, $R^{1/8} (\mu/2)^{7/8}$ and so on.  It may seem surprising that the distance decreases by more than a constant factor during each iteration.  Intuitively, this is because the spherical shells used by the algorithm are not exact dilations of each other.  Recall that the shells have different radii $\beta$, but they all have the same thickness $\delta$.  The larger the radius $\beta$, the smaller the ratio $\varepsilon = \delta/\beta$; so a larger shell allows a significantly more precise determination of its center.  In a sense, the algorithm makes more progress during the early iterations, when the spherical shells are larger.
\begin{tabbing}
\textbf{Algorithm 3:} \\
Set $R_{cur} = R$ and $\vec{p}_{cur} = \vec{0}$. \\
Set $n_{iter} = \lceil \lg\lg \frac{2R}{\mu} \rceil$, $S = (9.4) n_{iter}$ and $n_{tries} = 910\log S$. \\
Whil\=e $R_{cur} \geq \mu$ do: \\
\> Try running $\text{OneRound}(R_{cur}, \vec{p}_{cur}, S)$ up to $n_{tries}$ times. \\
\> If OneRound() returns ``no answer'' on every attempt, \\
\> then return ``no answer.'' \\
\> Let $\vec{q}$ be the point returned by OneRound() \ifthenelse{\boolean{stocformat}}{\\ \>}{} on one of the successful attempts. \\
\> Set $R_{cur} = \sqrt{R_{cur}} \sqrt{\mu/2}$ and $\vec{p}_{cur} = \vec{q}$. \\
End while. \\
Return $\vec{p}_{cur}$.
\end{tabbing}

We conjecture that Algorithm 3 will succeed when 
\begin{equation}
\delta < \frac{\mu}{128 (10\lg\lg \frac{2R}{\mu})^3 n^2 (507+\frac{Q_1}{n})}, 
\end{equation}
which is a weaker requirement than that of Algorithm 2, where $\delta$ had to scale like $1/R$.  We conjecture that this algorithm then finds a solution with constant probability.  Note that this algorithm uses $O(\lg\lg \frac{2R}{\mu} \lg\lg\lg \frac{2R}{\mu})$ queries, which still beats the classical lower bound of $\tilde{\Omega}(n \log \frac{R}{\mu})$ queries.  

% We conjecture that Algorithm 3 finds the center with arbitrary precision $\mu$, provided that $\delta$ is sufficiently small, as described above.  We conjecture that this algorithm then finds a solution with constant probability, independent of the dimension $n$.  Thus, only $O(1)$ oracle queries are needed.  This is an improvement over the classical case, where $\Omega(n)$ queries seem to be required.  % Intuitively, this is because the curvelet transform uses constructive interference to find a direction in $\RR^n$ from just one quantum query.  

We now show how our results from Section 5 support this conjecture.  We prove the following:  
\begin{thm} \label{thm-crfn-alg2}
Consider a ``continuous'' analogue of Algorithm 3, using the continuous curvelet transform over $\RR^n$.  This algorithm succeeds with constant probability.  % (This holds provided that $\delta$ is small, so the Gaussian approximation in Section 5 is valid.)  
% This holds provided that the spherical shell with square cross-section can be approximated by one with Gaussian cross-section, as in Section 5.  This approximation is accurate when $\delta$ is small.  
\end{thm}
We will also argue, non-rigorously, that the discrete algorithm will behave like the continuous one, provided that the grid $G$ is sufficiently fine.  When the grid is chosen properly, the algorithm runs in time $\poly(n, \log R, \log\frac{1}{\mu})$.  See \ifthenelse{\boolean{app}}{Appendix F.4}{\cite{Liu-curvelets}} for details.

}{}

%%%%%%%%%%%%%%%%%%%%%%%%%%%%%%%%%%%%%%%%%%%%%%%%%%%%%%%%%%%%%%%%%%%%%%%%%%%%%%%

\section{Conclusions}

We introduced the curvelet transform as a tool for quantum algorithms, and demonstrated how it can solve problems involving geometric objects in $\RR^n$.  We showed that:  (1) for functions with radial symmetry, the continuous curvelet transform concentrates probability mass near the wavefront set; (2) a quantum curvelet transform (which is a discrete approximation of the continuous curvelet transform) can be implemented efficiently; (3) this leads to quantum algorithms for approximately finding the center of a ball in $\RR^n$, given a single quantum-sample state, and for exactly finding the center of a radial function in $\RR^n$, using $O(1)$ oracle queries.

There are several ways in which these results might be extended.  Perhaps one can adapt these quantum algorithms to solve more general problems, like finding the center of an ellipsoid.  Perhaps the quantum speed-up can be amplified using a recursive construction, as in \cite{bernstein-vazirani, hallgren-harrow}.  % Also, our analysis of the ball / spherical shell can probably be improved; in particular, we showed that the variance of $\vecb\cdot\vectheta$ is at most $O(\beta^2)$, but in fact, $\vecb\cdot\vectheta$ should be concentrated around $\beta$ and $-\beta$ with high probability.  

A general open problem is to understand the behavior of the curvelet transform on more complicated shapes.  Can one prove that the probability mass of the curvelet transform is concentrated near the wavefront set, for arbitrary functions on $\RR^n$?  That would generalize the results of this paper, \cite{curvelets-2002} and \cite{cont-curvelet-trans-1}.  Also, can one rigorously bound the approximation of the continuous curvelet transform by a discrete one?

Another problem is to find new quantum algorithms based on the curvelet transform.  For example, can one construct a curvelet transform over $\mathbb{F}_q^n$, that could help to solve hidden polynomial problems \cite{childs-nonlin}?  Are there quantum states with ``wavefront'' features, from which the curvelet transform could extract useful information?  Some candidates are quantum-sample states over convex polytopes \cite{aharonov-tashma, grover-rudolph}, and states produced by the evolution of a quantum walk.

One might also try to use the output of the curvelet transform in a more sophisticated way.  In this paper, we simply measured the output state, and we made very little use of the scale variable $a$, which measures the ``sharpness'' of the wavefront discontinuity.

\vskipline

\noindent
\textit{Acknowledgements}:  The author is grateful to R. Koenig, J. Preskill, L. Schulman, A. Childs, D. Meyer, N. Wallach, S. Jordan, E. Cand\`es, U. Vazirani (who suggested the iterative algorithm in section 7.4), Z. Landau, D. Aharonov (who pointed out reference \cite{bernstein-vazirani}), E. Eban, T. Vidick, and the anonymous referees, for helpful discussions and comments.  Supported by an NSF Mathematical Sciences Postdoctoral Fellowship.

%%%%%%%%%%%%%%%%%%%%%%%%%%%%%%%%%%%%%%%%%%%%%%%%%%%%%%%%%%%%%%%%%%%%%%%%%%%%%%%

\ifthenelse{\boolean{stocformat}}{ % STOC format

% BEGIN PASTE FROM .bbl FILE

% END PASTE FROM .bbl FILE

}{ % Regular (non-STOC) format

\bibliographystyle{plain}
\bibliography{curvelets}

}

%\begin{thebibliography}{99}
%\end{thebibliography}

%%%%%%%%%%%%%%%%%%%%%%%%%%%%%%%%%%%%%%%%%%%%%%%%%%%%%%%%%%%%%%%%%%%%%%%%%%%%%%%

\ifthenelse{\boolean{app}}{

\appendix

%%%%%%%%%%%%%%%%%%%%%%%%%%%%%%%%%%%%%%%%%%%%%%%%%%%%%%%%%%%%%%%%%%%%%%%%%%%%%%%

\section{The Continuous Curvelet Transform}

\ifthenelse{\boolean{sec2}}{

%Figure \ref{fig-tiling} shows the ``tiling'' of the frequency domain into sectors, which are the supports of the window functions $\chi_{a,\vectheta}$.  
%\begin{figure}
%\includegraphics{tiling.eps}
%\caption{``Tiling'' of the frequency domain into sectors with depth $\approx 1/a$ and width $\approx 1/\sqrt{a}$ (note that this corresponds to angular width $\approx \sqrt{a}$), where $a = 1, \frac{1}{2}, \frac{1}{4}, \frac{1}{8}, \ldots$.}
%\label{fig-tiling}
%\end{figure}

%\vskipline

We now sketch the proofs of Theorems \ref{thm-cct-1} and \ref{thm-cct-2}.

% NOTE:  see our notes from 6/3/08, 5/5/08

First, for any $a$ and $\vectheta$, let us define 
\begin{equation}
g_{a,\vectheta}(\vecx)
 := \int_{\RR^n} \langle \gamma_{a,\vecb,\vectheta}, f \rangle \gamma_{a,\vecb,\vectheta}(\vecx) d\vecb.
\end{equation}
We claim that 
\begin{equation}
\hat{g}_{a,\vectheta}(\veck)
 = \bigl| \chi_{a,\vectheta}(\veck) \bigr|^2 \hat{f}(\veck).
\label{eqn-cct-lemma1}
\end{equation}

To see this, write 
\begin{equation*}
\langle \gamma_{a,\vecb,\vectheta}, f \rangle
 = \int_{\RR^n} \gamma_{a,\vec{0},\vectheta}^*(\vec{y}-\vecb) f(\vec{y}) d\vec{y}
 = (\tilde{\gamma}_{a,\vec{0},\vectheta}^* * f)(\vecb),
\end{equation*}
where we define 
\begin{equation*}
\tilde{\gamma}_{a,\vec{0},\vectheta}(\vecx) = \gamma_{a,\vec{0},\vectheta}(-\vecx).
\end{equation*}
Thus we can write $g_{a,\vectheta}(\vecx)$ as
\begin{equation*}
g_{a,\vectheta}(\vecx)
 = \int_{\RR^n} \gamma_{a,\vec{0},\vectheta}(\vecx-\vecb) (\tilde{\gamma}_{a,\vec{0},\vectheta}^* * f)(\vecb) d\vecb
 = (\gamma_{a,\vec{0},\vectheta} * \tilde{\gamma}_{a,\vec{0},\vectheta}^* * f)(\vecb).
\end{equation*}

Taking the Fourier transform, we get
\begin{equation*}
\hat{g}_{a,\vectheta}(\veck)
 = \hat{\gamma}_{a,\vec{0},\vectheta}(\veck) (\tilde{\gamma}_{a,\vec{0},\vectheta}^*)\hat{\;}(\veck) \hat{f}(\veck)
 = \bigl| \hat{\gamma}_{a,\vec{0},\vectheta}(\veck) \bigr|^2 \hat{f}(\veck),
\end{equation*}
using the fact that 
\begin{equation*}
(\tilde{\gamma}_{a,\vec{0},\vectheta}^*)\hat{\;}(\veck)
 = \int_{\RR^n} \gamma_{a,\vec{0},\vectheta}^*(-\vecx) e^{-2\pi i \veck\cdot\vecx} d\vecx
 = \int_{\RR^n} \gamma_{a,\vec{0},\vectheta}^*(\vecx) e^{2\pi i \veck\cdot\vecx} d\vecx
 = \bigl( \hat{\gamma}_{a,\vec{0},\vectheta}(\veck) \bigr)^*.
\end{equation*}
This proves equation (\ref{eqn-cct-lemma1}).

Next, we claim that, for all $|\veck| \geq 1/\lambda$, 
\begin{equation}
\int_0^1 \int_{S^{n-1}} |\chi_{a,\vectheta}(\veck)|^2 d\sigma(\vectheta) \frac{da}{a^{n+1}} = 1.
\label{eqn-cct-lemma2}
\end{equation}

To see this, proceed as follows.  Fix $\veck$, and write $\vectheta$ in spherical coordinates centered around $\veck$, such that $\theta_1$ is the angle between $\vectheta$ and $\veck$.  Then we have 
\begin{equation*}
\chi_{a,\vectheta}(\veck) = W(\lambda a |\veck|) V(\theta_1/\sqrt{a}) a^{(n+1)/4} \Bigl( \frac{\sin(\theta_1/\sqrt{a}) \sqrt{a}}{\sin(\theta_1)} \Bigr)^{(n-2)/2}.
\end{equation*}
Then substitute into the integral:
\begin{equation*}
\begin{split}
\int_0^1 \int_{S^{n-1}} & |\chi_{a,\vectheta}(\veck)|^2 d\sigma(\vectheta) \frac{da}{a^{n+1}} \\
 &= \int_0^1 \int_{S^{n-1}} W(\lambda a |\veck|)^2 V(\theta_1/\sqrt{a})^2 a^{(n+1)/2} \Bigl(\frac{\sin(\theta_1/\sqrt{a})}{\sin(\theta_1)}\Bigr)^{n-2} a^{(n-2)/2} d\sigma(\vectheta) \frac{da}{a^{n+1}} \\
 &= \int_0^1 \int_{S^{n-1}} W(\lambda a |\veck|)^2 V(\theta_1/\sqrt{a})^2 \sin^{n-2}(\theta_1/\sqrt{a}) d\theta_1 d\sigma(\theta_2,\ldots,\theta_{n-1}) \frac{da}{a^{3/2}}.
\end{split}
\end{equation*}
Note that $V(\theta_1/\sqrt{a})$ is nonzero only when $\theta_1 \in [0,\pi\sqrt{a}]$, so we can restrict the integral to this range.  Then change variables, $\theta_1' = \theta_1/\sqrt{a}$, to get:
\begin{equation*}
\begin{split}
\int_0^1 \int_{S^{n-1}} & |\chi_{a,\vectheta}(\veck)|^2 d\sigma(\vectheta) \frac{da}{a^{n+1}} \\
 &= \int_0^1 \int_{S^{n-1}} W(\lambda a |\veck|)^2 V(\theta_1')^2 \sin^{n-2}(\theta_1') d\theta_1' \sqrt{a} \; d\sigma(\theta_2,\ldots,\theta_{n-1}) \frac{da}{a^{3/2}} \\
 &= \int_0^1 \int_{S^{n-1}} W(\lambda a |\veck|)^2 V(\theta_1)^2 d\sigma(\vectheta) \frac{da}{a} \\
 &= \int_0^{\lambda |\veck|} W(a)^2 \frac{da}{a} \int_{S^{n-1}} V(\theta_1)^2 d\sigma(\vectheta) \\
 &= 1,
\end{split}
\end{equation*}
using the admissibility conditions.  This proves equation (\ref{eqn-cct-lemma2}).

We now prove Theorem \ref{thm-cct-1}.  We write 
\begin{equation*}
\begin{split}
\int_{a<1} & \Gamma_f(a,\vecb,\vectheta) \gamma_{a,\vecb,\vectheta}(\vecx) d\mu(a,\vecb,\vectheta) \\
 &= \int_0^1 \int_{S^{n-1}} \int_{\RR^n} \langle \gamma_{a,\vecb,\vectheta}, f \rangle \gamma_{a,\vecb,\vectheta}(\vecx) d\vecb d\sigma(\vectheta) \frac{da}{a^{n+1}} \\
 &= \int_0^1 \int_{S^{n-1}} g_{a,\vectheta}(\vecx) d\sigma(\vectheta) \frac{da}{a^{n+1}},
\end{split}
\end{equation*}
and we claim that this equals $f(\vecx)$.  Taking the Fourier transform, we get 
\begin{equation*}
\int_0^1 \int_{S^{n-1}} \hat{g}_{a,\vectheta}(\veck) d\sigma(\vectheta) \frac{da}{a^{n+1}},
\end{equation*}
and we claim that this equals $\hat{f}(\veck)$.  Using equations (\ref{eqn-cct-lemma1}) and (\ref{eqn-cct-lemma2}), we rewrite this integral as:
\begin{equation*}
\int_0^1 \int_{S^{n-1}} |\chi_{a,\vectheta}(\veck)|^2 \hat{f}(\veck) d\sigma(\vectheta) \frac{da}{a^{n+1}}
 = \hat{f}(\veck).
\end{equation*}
(The last equality holds because of (\ref{eqn-cct-lemma2}) when $|\veck| \geq 1/\lambda$, and because $\hat{f}(\veck) = 0$ when $|\veck| < 1/\lambda$.)  This proves Theorem \ref{thm-cct-1}.

Finally, we prove Theorem \ref{thm-cct-2}.  We write 
\begin{equation*}
\int_{a<1} |\Gamma_f(a,\vecb,\vectheta)|^2 d\mu(a,\vecb,\vectheta)
 = \int_0^1 \int_{S^{n-1}} \int_{\RR^n} | \langle \gamma_{a,\vecb,\vectheta}, f \rangle |^2 d\vecb d\sigma(\vectheta) \frac{da}{a^{n+1}}.
\end{equation*}
We rewrite the innermost integral, applying some of the identities used to prove (\ref{eqn-cct-lemma1}):
\begin{equation*}
\begin{split}
\int_{\RR^n} \bigl| \langle \gamma_{a,\vecb,\vectheta}, f \rangle \bigr|^2 d\vecb
 &= \int_{\RR^n} \bigl| (\tilde{\gamma}_{a,\vec{0},\vectheta}^* * f)(\vecb) \bigr|^2 d\vecb \\
 &= \int_{\RR^n} \bigl| (\tilde{\gamma}_{a,\vec{0},\vectheta}^*)\hat{\;}(\veck) \hat{f}(\veck) \bigr|^2 d\veck \\
 &= \int_{\RR^n} \bigl| (\hat{\gamma}_{a,\vec{0},\vectheta}(\veck))^* \hat{f}(\veck) \bigr|^2 d\veck \\
 &= \int_{\RR^n} |\chi_{a,\vectheta}(\veck)|^2 |\hat{f}(\veck)|^2 d\veck.
\end{split}
\end{equation*}
Substituting in, and using (\ref{eqn-cct-lemma2}), we get:
\begin{equation*}
\begin{split}
\int_{a<1} & |\Gamma_f(a,\vecb,\vectheta)|^2 d\mu(a,\vecb,\vectheta) \\
 &= \int_0^1 \int_{S^{n-1}} \int_{\RR^n} |\chi_{a,\vectheta}(\veck)|^2 |\hat{f}(\veck)|^2 d\veck d\sigma(\vectheta) \frac{da}{a^{n+1}} \\
 &= \int_{\RR^n} \int_0^1 \int_{S^{n-1}} |\chi_{a,\vectheta}(\veck)|^2 d\sigma(\vectheta) \frac{da}{a^{n+1}} |\hat{f}(\veck)|^2 d\veck \\
 &= \int_{\RR^n} |\hat{f}(\veck)|^2 d\veck = \int_{\RR^n} |f(\vecx)|^2 d\vecx.
\end{split}
\end{equation*}
This proves Theorem \ref{thm-cct-2}.

}{}

%%%%%%%%%%%%%%%%%%%%%%%%%%%%%%%%%%%%%%%%%%%%%%%%%%%%%%%%%%%%%%%%%%%%%%%%%%%%%%%

\section{The Curvelet Transform of a Radial Function}

\ifthenelse{\boolean{sec3}}{

\subsection{The variance of $\vecb$ perpendicular to $\vectheta$}

This is a continuation of Section 3.3.  Recall that we have 
%\begin{equation}
$L(r) = F_0(r) W(\lambda ar)$, 
and %\quad
$M(\phi_1) = V(\phi_1/\sqrt{a}) \Lambda_a(\phi_1)$.  
%\end{equation}
Then we write:
\begin{equation}
\begin{split}
\int_{\RR^n} & |\tfrac{\partial}{\partial k_2} \hat{\Phi}_{a,\vec{u}}(\veck) |^2 d\veck \\
 &= \int_{S^{n-1}} \int_0^\infty \Bigl( L'(r) M(\phi_1) \sin\phi_1 \cos\phi_2 + \\
 &\qquad L(r) M'(\phi_1) r^{-1} \cos\phi_1 \cos\phi_2 \Bigr)^2 r^{n-1} dr d\sigma(\vec{\phi}) \\
 &= I_{Ar}I_{A1}I_2 + 2I_{Br}I_{B1}I_2 + I_{Cr}I_{C1}I_2, 
\end{split}
\end{equation}
where we define 
\begin{equation}
I_{Ar} = \int_0^\infty L'(r)^2 r^{n-1} dr, 
%\end{equation}
\qquad
%\begin{equation}
I_{A1} = \int_0^\pi M(\phi_1)^2 \sin^n\phi_1 d\phi_1, 
\end{equation}
\begin{equation}
I_{Br} = \int_0^\infty L'(r)L(r) r^{n-2} dr, 
%\end{equation}
\qquad
%\begin{equation}
I_{B1} = \int_0^\pi M'(\phi_1)M(\phi_1) \cos\phi_1 \sin^{n-1}\phi_1 d\phi_1, 
\end{equation}
\begin{equation}
I_{Cr} = \int_0^\infty L(r)^2 r^{n-3} dr, 
%\end{equation}
\qquad
%\begin{equation}
I_{C1} = \int_0^\pi M'(\phi_1)^2 \cos^2\phi_1 \sin^{n-2}\phi_1 d\phi_1, 
\end{equation}
\begin{equation}
I_2 = \int_{S^{n-2}} \cos^2\phi_2 d\sigma(\phi_2,\ldots,\phi_{n-1}).
\end{equation}

This shows that the variance of $\vecb$ perpendicular to $\vectheta$ is:
\begin{equation}
\text{E}(\vecb^T (I-\vectheta\vectheta^T) \vecb) = 
 \frac{n-1}{(2\pi)^2} S_0 \int_0^1 \Bigl( I_{Ar}I_{A1}I_2 + 2I_{Br}I_{B1}I_2 + I_{Cr}I_{C1}I_2 \Bigr) \frac{da}{a^{n+1}}.
\end{equation}
A similar formula gives the variance conditioned on observing $a \leq \eta$:
\begin{equation}
\text{E}(\vecb^T (I-\vectheta\vectheta^T) \vecb \:|\: a \leq \eta) = \frac{1}{\Pr[a \leq \eta]} 
 \frac{n-1}{(2\pi)^2} S_0 \int_0^\eta \Bigl( I_{Ar}I_{A1}I_2 + 2I_{Br}I_{B1}I_2 + I_{Cr}I_{C1}I_2 \Bigr) \frac{da}{a^{n+1}}.
\label{eqn-varbperp-conda}
\end{equation}
We will now upper-bound the various integrals appearing on the right hand side of equation (\ref{eqn-varbperp-conda}).  

\subsubsection{}

We begin with the integral $\int_0^\eta I_{Br}I_{B1}I_2 da/a^{n+1}$.  A straightforward calculation shows that 
\begin{equation}
I_2 = \frac{1}{n-1} \int_{S^{n-2}} d\sigma(\phi_2,\ldots,\phi_{n-1}).
\end{equation}

We can simplify $I_{B1}$, by integrating by parts:
%\begin{equation*}
%\begin{split}
$I_{B1}
% &= \tfrac{1}{2}M(\phi_1)^2 \cos\phi_1 \sin^{n-1}\phi_1 \Big|_0^\pi - \int_0^\pi \tfrac{1}{2}M(\phi_1)^2 (-\sin^n\phi_1 + (n-1)\cos^2\phi_1 \sin^{n-2}\phi_1) d\phi_1 \\
 = -\int_0^\pi \tfrac{1}{2}M(\phi_1)^2 (-1 + n\cos^2\phi_1) \sin^{n-2}\phi_1 d\phi_1$.
%\end{split}
%\end{equation*}
Substituting in the definition of $M(\phi_1)$, changing variables, and using the fact that $V$ is supported on the interval $[0,\pi/2]$, we get:
\begin{equation*}
\begin{split}
I_{B1}
 &= -\int_0^\pi \tfrac{1}{2} V(\phi_1/\sqrt{a})^2 a^{(2n-1)/2} \Bigl(\frac{\sin(\phi_1/\sqrt{a})}{\sin\phi_1}\Bigr)^{n-2} (-1 + n\cos^2\phi_1) \sin^{n-2}\phi_1 d\phi_1 \\
 &= -\int_0^\pi \tfrac{1}{2} V(\phi_1/\sqrt{a})^2 a^{(2n-1)/2} \sin(\phi_1/\sqrt{a})^{n-2} (-1 + n\cos^2\phi_1) d\phi_1 \\
 &= -\int_0^{\pi/\sqrt{a}} \tfrac{1}{2} V(\omega_1)^2 a^n \sin^{n-2}\omega_1 (-1 + n\cos^2(\omega_1\sqrt{a})) d\omega_1 \\
 &= \int_0^{\pi/2} \tfrac{1}{2} V(\omega_1)^2 a^n \sin^{n-2}\omega_1 (1 - n\cos^2(\omega_1\sqrt{a})) d\omega_1, 
\end{split}
\end{equation*}
hence 
\begin{equation*}
|I_{B1}|
 \leq \tfrac{1}{2} a^n (n-1) \int_0^{\pi/2} V(\omega_1)^2 \sin^{n-2}\omega_1 d\omega_1.
\end{equation*}
Combining this with $I_2$, we get:
\begin{equation}
|I_{B1} I_2| \leq \tfrac{1}{2} a^n \int_{S^{n-1}} V(\omega_1)^2 d\sigma(\omega_1,\phi_2,\ldots,\phi_{n-1}) = \tfrac{1}{2} a^n.
\end{equation}

Next, we can simplify $I_{Br}$, by integrating by parts:
%\begin{equation*}
%\begin{split}
$I_{Br}
% &= \tfrac{1}{2}L(r)^2 r^{n-2} \Big|_0^\infty - \int_0^\infty \tfrac{1}{2}L(r)^2 (n-2)r^{n-3} dr \\
 = -\int_0^\infty \tfrac{1}{2}L(r)^2 (n-2)r^{n-3} dr$.  
%\end{split}
%\end{equation*}
Combining this with $I_{B1}$ and $I_2$, substituting in the definition of $L(r)$, and exchanging the integrals, we get:
\begin{equation*}
\begin{split}
\Bigl| \int_0^\eta I_{Br}I_{B1}I_2 \frac{da}{a^{n+1}} \Bigr|
 &\leq \int_0^\eta |I_{Br}| \tfrac{1}{2} \frac{da}{a}
 = \int_0^\eta \int_0^\infty \tfrac{1}{2}L(r)^2 (n-2)r^{n-3} dr \tfrac{1}{2} \frac{da}{a} \\
 &= \int_0^\eta \int_0^\infty \tfrac{1}{2} F_0(r)^2 W(\lambda ar)^2 (n-2)r^{n-3} dr \tfrac{1}{2} \frac{da}{a} \\
 &\leq \int_0^\infty \tfrac{1}{2} F_0(r)^2 \int_0^\eta W(\lambda ar)^2 \tfrac{1}{2} \frac{da}{a} (n-2)r^{n-3} dr.
\end{split}
\end{equation*}
By the definition of $W$, 
\begin{equation}
\int_0^\eta W(\lambda ar)^2 \tfrac{1}{2} \frac{da}{a}
 = \int_0^{\lambda\eta r} W(\alpha)^2 \tfrac{1}{2} \frac{d\alpha}{\alpha}
 \leq \tfrac{1}{2},
\label{eqn-W2-dalphaoveralpha}
\end{equation}
and vanishes when $r \leq 1/(\lambda\eta e)$.  Thus we have:
\begin{equation}
\boxed{
\Bigl| \int_0^\eta I_{Br}I_{B1}I_2 \frac{da}{a^{n+1}} \Bigr|
 \leq \tfrac{1}{4} (n-2) \int_{1/(\lambda\eta e)}^\infty F_0(r)^2 r^{n-3} dr.
}
\label{eqn-I-B}
\end{equation}

\subsubsection{}

Next, consider the integral $\int_0^\eta I_{Ar}I_{A1}I_2 da/a^{n+1}$.  We already have a bound for $I_2$.  For $I_{A1}$, we write:
\begin{equation*}
\begin{split}
I_{A1}
 &= \int_0^\pi V(\phi_1/\sqrt{a})^2 a^{(2n-1)/2} \Bigl(\frac{\sin(\phi_1/\sqrt{a})}{\sin\phi_1}\Bigr)^{n-2} \sin^n\phi_1 d\phi_1 \\
 &= \int_0^{\pi/\sqrt{a}} V(\omega_1)^2 a^n \sin^{n-2}\omega_1 \sin^2(\omega_1\sqrt{a}) d\omega_1.
\end{split}
\end{equation*}
Using the fact that $V$ is supported on $[0,\pi/2]$, and the simple bound $\sin^2(\omega_1\sqrt{a}) \leq \omega_1^2 a \leq \tfrac{\pi^2}{4} a$, we get:
\begin{equation*}
0 \leq I_{A1}
 \leq \int_0^{\pi/2} V(\omega_1)^2 a^n \sin^{n-2}\omega_1 d\omega_1 \tfrac{\pi^2}{4} a.
\end{equation*}
Combining with $I_2$, we get:
\begin{equation}
0 \leq I_{A1} I_2
 \leq \frac{\pi^2}{4} \frac{a^{n+1}}{n-1} \int_{S^{n-1}} V(\omega_1)^2 d\sigma(\omega_1,\phi_2,\ldots,\phi_{n-1})
  = \frac{\pi^2}{4} \frac{a^{n+1}}{n-1}.
\end{equation}

We now turn to $I_{Ar}$.  First, combining with $I_{A1}$ and $I_2$, we have 
\begin{equation}
0 \leq \int_0^\eta I_{Ar}I_{A1}I_2 \frac{da}{a^{n+1}}
 \leq \int_0^\eta I_{Ar} da \cdot \frac{\pi^2}{4(n-1)}.
\end{equation}
We can upper-bound $I_{Ar}$ as follows.  Note that, for any two $L^2$ functions, the Cauchy-Schwarz inequality implies that 
\begin{equation}
\norm{f+g}^2
 = \norm{f}^2 + 2 \langle f,g \rangle + \norm{g}^2
 \leq \norm{f}^2 + 2 \norm{f} \norm{g} + \norm{g}^2
 \leq 2\norm{f}^2 + 2\norm{g}^2;
\end{equation}
in the last step we used the arithmetic-geometric mean inequality, $\sqrt{ab} \leq \tfrac{a+b}{2}$ for $a,b \geq 0$, with $a = \norm{f}^2$ and $b = \norm{g}^2$.  Thus we can write
\begin{equation}
\begin{split}
0 \leq I_{Ar}
 &= \int_0^\infty \Bigl( F'_0(r) W(\lambda a r) + F_0(r) W'(\lambda a r) \lambda a \Bigr)^2 r^{n-1} dr \\
 &= \norm{G_1+G_2}^2 \leq 2\norm{G_1}^2 + 2\norm{G_2}^2,
\end{split}
\end{equation}
where we define 
\begin{align}
G_1(r) &= F'_0(r) W(\lambda a r) r^{(n-1)/2} \\
G_2(r) &= F_0(r) W'(\lambda a r) \lambda a r^{(n-1)/2}.
\end{align}
Thus we have 
\begin{equation}
\boxed{
0 \leq \int_0^\eta I_{Ar}I_{A1}I_2 \frac{da}{a^{n+1}}
 \leq \Bigl( \int_0^\eta \norm{G_1}^2 da + \int_0^\eta \norm{G_2}^2 da \Bigr) \cdot \frac{\pi^2}{2(n-1)}.
}
\label{eqn-I-A}
\end{equation}
We then want to upper-bound the integrals $\int_0^\eta \norm{G_1}^2 da$ and $\int_0^\eta \norm{G_2}^2 da$.  

For the first one, we have:
\begin{equation*}
\begin{split}
\int_0^\eta \norm{G_1}^2 da
 &= \int_0^\eta \int_0^\infty F'_0(r)^2 W(\lambda a r)^2 r^{n-1} dr da \\
 &= \int_0^\infty F'_0(r)^2 \int_0^\eta W(\lambda a r)^2 da \: r^{n-1} dr.
\end{split}
\end{equation*}
Using the fact that $W$ is supported on $[1/e,1]$, we can write 
\begin{equation}
\int_0^\eta W(\lambda a r)^2 da
 = \int_0^{\lambda\eta r} W(\alpha)^2 d\alpha \frac{1}{\lambda r}
 \leq \int_0^{\lambda\eta r} W(\alpha)^2 \frac{d\alpha}{\alpha} \frac{1}{\lambda r}
 \leq \frac{1}{\lambda r},
\label{eqn-W2-dalpha}
\end{equation}
and vanishes when $r \leq 1/(\lambda \eta e)$.  Hence,
\begin{equation}
\boxed{
\int_0^\eta \norm{G_1}^2 da
 \leq \frac{1}{\lambda} \int_{1/(\lambda \eta e)}^\infty F'_0(r)^2 r^{n-2} dr.
}
\label{eqn-I-A-G1}
\end{equation}

For the second integral, we have:
\begin{equation*}
\begin{split}
\int_0^\eta \norm{G_2}^2 da
 &= \int_0^\eta \int_0^\infty F_0(r)^2 W'(\lambda a r)^2 \lambda^2 a^2 r^{n-1} dr da \\
 &= \int_0^\infty F_0(r)^2 \int_0^\eta W'(\lambda a r)^2 \lambda^2 a^2 da \: r^{n-1} dr.
\end{split}
\end{equation*}
Note that the derivative of $W$ is given by 
\begin{equation}
W'(r) = \begin{cases}
  C_w \sin(2\pi\log r) \pi/r, & 1/e \leq r \leq 1, \\
  0, & \text{otherwise},
\end{cases}
\label{eqn-Wprime}
\end{equation}
where $C_w = \sqrt{8/3}$.  So we can write 
\begin{equation}
\int_0^\eta W'(\lambda a r)^2 \lambda^2 a^2 da
 = \int_0^{\lambda\eta r} W'(\alpha)^2 \alpha^2 d\alpha \frac{1}{\lambda r^3}
 \leq \int_{1/e}^1 C_w^2 (\pi/\alpha)^2 \alpha^2 d\alpha \frac{1}{\lambda r^3}
 = \tfrac{8}{3} \pi^2 (1-\tfrac{1}{e}) \frac{1}{\lambda r^3},
\label{eqn-Wprime2-alpha2dalpha}
\end{equation}
and vanishes when $r \leq 1/(\lambda \eta e)$.  Hence,
\begin{equation}
\boxed{
\int_0^\eta \norm{G_2}^2 da
 \leq \tfrac{8}{3} \pi^2 (1-\tfrac{1}{e}) \frac{1}{\lambda} \int_{1/(\lambda \eta e)}^\infty F_0(r)^2 r^{n-4} dr.
}
\label{eqn-I-A-G2}
\end{equation}

\subsubsection{}

Finally, we consider the integral $\int_0^\eta I_{Cr}I_{C1}I_2 da/a^{n+1}$.  We already have a bound for $I_2$.  For $I_{C1}$ we can write:
\begin{equation}
\begin{split}
0 \leq I_{C1}
 &= \int_0^\pi \Bigl( V'(\phi_1/\sqrt{a})(1/\sqrt{a}) \Lambda_a(\phi_1) + V(\phi_1/\sqrt{a}) \Lambda'_a(\phi_1) \Bigr)^2 \cos^2 \phi_1 \sin^{n-2} \phi_1 d\phi_1 \\
 &= \norm{U_1+U_2}^2 \leq 2\norm{U_1}^2 + 2\norm{U_2}^2,
\end{split}
\end{equation}
where we define 
\begin{align}
U_1(\phi_1) &= V'(\phi_1/\sqrt{a})(1/\sqrt{a}) \Lambda_a(\phi_1) \cos\phi_1 \sin^{(n-2)/2}\phi_1 \\
U_2(\phi_1) &= V(\phi_1/\sqrt{a}) \Lambda'_a(\phi_1) \cos\phi_1 \sin^{(n-2)/2}\phi_1.
\end{align}
Then 
\begin{equation}
0 \leq I_{C1} I_2 \leq 2\norm{U_1}^2 I_2 + 2\norm{U_2}^2 I_2.
\label{eqn-IC1-IB2}
\end{equation}
We now evaluate $\norm{U_1}^2$ and $\norm{U_2}^2$.

For $\norm{U_1}^2$, we can write 
\begin{equation*}
\begin{split}
\norm{U_1}^2
 &= \int_0^\pi V'(\phi_1/\sqrt{a})^2 a^{-1} a^{(2n-1)/2} \Bigl(\frac{\sin(\phi_1/\sqrt{a})}{\sin(\phi_1)}\Bigr)^{n-2} \cos^2\phi_1 \sin^{n-2}\phi_1 d\phi_1 \\
 &= \int_0^\pi V'(\phi_1/\sqrt{a})^2 a^{(2n-3)/2} \sin^{n-2}(\phi_1/\sqrt{a}) \cos^2\phi_1 d\phi_1 \\
 &= \int_0^{\pi/\sqrt{a}} V'(\omega_1)^2 a^{n-1} \sin^{n-2}\omega_1 \cos^2(\omega_1\sqrt{a}) d\omega_1 \\
 &\leq a^{n-1} \int_0^{\pi/2} V'(\omega_1)^2 \sin^{n-2}\omega_1 d\omega_1.
\end{split}
\end{equation*}

The derivative of $V$ is given by 
\begin{equation}
V'(t) = \begin{cases}
  -2C_v\cos(t)\sin(t), & 0 \leq t \leq \pi/2, \\
  0, & \text{otherwise},
\end{cases}
\end{equation}
where $C_v = \sqrt{\frac{2(n+2)n}{3S_0}}$.  
% NOTE:  see notes 8/20/08.  
Using these formulas and (\cite{abram-stegun}, eqn. 4.3.127), a straightforward calculation shows that 
\begin{equation}
\int_0^{\pi/2} V'(\omega_1)^2 \sin^{n-2}\omega_1 d\omega_1
 = \frac{4n}{3S_0} \Bigl(1-\frac{1}{n}\Bigr) \int_0^\pi \sin^{n-2}\omega_1 d\omega_1.
\label{eqn-Vprime2-sinnminus2-domega}
\end{equation}

So we have 
\begin{equation*}
\norm{U_1}^2
 \leq a^{n-1} \cdot \frac{4n}{3S_0} \Bigl(1-\frac{1}{n}\Bigr) \int_0^\pi \sin^{n-2}\omega_1 d\omega_1.  
\end{equation*}
Combining with $I_2$, we get
\begin{equation}
0 \leq \norm{U_1}^2 I_2
 \leq a^{n-1} \cdot \frac{4n}{3S_0} \Bigl(1-\frac{1}{n}\Bigr) \cdot \frac{1}{n-1} \int_{S^{n-1}} d\sigma(\omega_1, \phi_2, \ldots, \phi_{n-1})
 = \tfrac{4}{3} a^{n-1}.  
\end{equation}

Next we evaluate $\norm{U_2}^2$.  The derivative of $\Lambda_a$ is given by:
\begin{equation}
\Lambda'_a(\phi_1) = a^{(2n-1)/4} (\tfrac{n-2}{2}) \Bigl(\frac{\sin(\phi_1/\sqrt{a})}{\sin(\phi_1)}\Bigr)^{(n-4)/2} \Bigl( \frac{\cos(\phi_1/\sqrt{a})}{\sqrt{a}\sin(\phi_1)} - \frac{\sin(\phi_1/\sqrt{a})\cos(\phi_1)}{\sin^2(\phi_1)} \Bigr),
\label{eqn-Lambdaprime}
\end{equation}
hence 
\begin{equation*}
U_2(\phi_1) = V(\phi_1/\sqrt{a}) a^{(2n-1)/4} (\tfrac{n-2}{2}) \sin^{(n-4)/2}(\phi_1/\sqrt{a}) \Bigl( \frac{\cos(\phi_1/\sqrt{a})}{\sqrt{a}} - \frac{\sin(\phi_1/\sqrt{a})\cos(\phi_1)}{\sin(\phi_1)} \Bigr) \cos\phi_1,
\end{equation*}
and
\begin{equation*}
\begin{split}
\norm{U_2}^2
 &= \int_0^\pi V(\phi_1/\sqrt{a})^2 a^{(2n-1)/2} (\tfrac{n-2}{2})^2 \sin^{n-4}(\phi_1/\sqrt{a}) \Bigl( \frac{\cos(\phi_1/\sqrt{a})}{\sqrt{a}} - \frac{\sin(\phi_1/\sqrt{a})\cos(\phi_1)}{\sin(\phi_1)} \Bigr)^2 \cos^2\phi_1 d\phi_1 \\
 &= \int_0^{\pi/\sqrt{a}} V(\omega_1)^2 a^n (\tfrac{n-2}{2})^2 \sin^{n-4}(\omega_1) \Bigl( \frac{\cos(\omega_1)}{\sqrt{a}} - \frac{\sin(\omega_1)\cos(\omega_1\sqrt{a})}{\sin(\omega_1\sqrt{a})} \Bigr)^2 \cos^2(\omega_1\sqrt{a}) d\omega_1.
\end{split}
\end{equation*}
Recall that $V$ is supported on $[0,\pi/2]$.  For $\omega_1$ in this range, we have the following crude bound:  (using \cite{abram-stegun}, eqn. 4.3.81)
\begin{equation*}
\Bigl| \frac{\cos(\omega_1)}{\sqrt{a}} - \frac{\sin(\omega_1)\cos(\omega_1\sqrt{a})}{\sin(\omega_1\sqrt{a})} \Bigr|
 \leq \frac{1}{\sqrt{a}} + \frac{\sin(\omega_1)}{\sin(\omega_1\sqrt{a})} \cdot \frac{\sin(\omega_1\sqrt{a})}{\omega_1\sqrt{a}}
 \leq \frac{1}{\sqrt{a}} + \frac{1}{\sqrt{a}}.
\end{equation*}
Also, we have $\cos^2(\omega_1\sqrt{a}) \leq 1$.  Hence,
\begin{equation*}
\begin{split}
\norm{U_2}^2
 &\leq \int_0^{\pi/2} V(\omega_1)^2 a^n (\tfrac{n-2}{2})^2 \sin^{n-4}(\omega_1) (\tfrac{2}{\sqrt{a}})^2 d\omega_1 \\
 &= 4 a^{n-1} (\tfrac{n-2}{2})^2 \int_0^{\pi/2} V(\omega_1)^2 \sin^{n-4}\omega_1 d\omega_1.
\end{split}
\end{equation*}

Note that 
\begin{equation}
\int_0^{\pi/2} V(\omega_1)^2 \sin^{n-4}\omega_1 d\omega_1
 = \Bigl(1+\frac{5}{n-3}\Bigr) \int_0^{\pi/2} V(\omega_1)^2 \sin^{n-2}\omega_1 d\omega_1, 
\label{eqn-V2-sinnminus4-domega}
\end{equation}
using the definition of $V$, and integration by parts.  
% NOTE:  see notes, 8/21/08.

So we have:
\begin{equation*}
\norm{U_2}^2
 \leq 4 a^{n-1} (\tfrac{n-2}{2})^2 (1+\tfrac{5}{n-3}) \int_0^{\pi/2} V(\omega_1)^2 \sin^{n-2}\omega_1 d\omega_1.
\end{equation*}
Combining with $I_2$, we get:
\begin{equation}
\begin{split}
0 \leq \norm{U_2}^2 I_2
 &\leq 4 a^{n-1} (\tfrac{n-2}{2})^2 (1+\tfrac{5}{n-3}) \tfrac{1}{n-1} \int_{S^{n-1}} V(\omega_1)^2 d\sigma(\omega_1,\phi_2,\ldots,\phi_{n-1}) \\
 &\leq a^{n-1} (n-2) (1+\tfrac{5}{n-3}).
\end{split}
\end{equation}

So, by substituting into (\ref{eqn-IC1-IB2}), we have 
\begin{equation}
0 \leq I_{C1} I_2 \leq 2 \cdot \tfrac{4}{3}a^{n-1} + 2 \cdot a^{n-1} (n-2) (1+\tfrac{5}{n-3}).
\end{equation}

Finally, we turn to $I_{Cr}$.  Combining it with $I_{C1}$ and $I_2$, we have
\begin{equation*}
0 \leq \int_0^\eta I_{Cr}I_{C1}I_2 \frac{da}{a^{n+1}}
 \leq \Bigl( \tfrac{8}{3} + 2 (n-2) (1+\tfrac{5}{n-3}) \Bigr) \int_0^\eta I_{Cr} \frac{da}{a^2}.
\end{equation*}
We can bound the integral on the right hand side as follows.
\begin{equation*}
\begin{split}
\int_0^\eta I_{Cr} \frac{da}{a^2}
 &= \int_0^\eta \int_0^\infty F_0(r)^2 W(\lambda a r)^2 r^{n-3} dr \frac{da}{a^2} \\
 &= \int_0^\infty F_0(r)^2 \int_0^\eta W(\lambda a r)^2 \frac{da}{a^2} r^{n-3} dr.
\end{split}
\end{equation*}
Using the fact that $W$ is supported on $[1/e,1]$, 
\begin{equation}
\int_0^\eta W(\lambda a r)^2 \frac{da}{a^2}
 = \int_0^{\lambda\eta r} W(\alpha)^2 \frac{d\alpha}{\alpha^2} \lambda r
 \leq \int_0^{\lambda\eta r} W(\alpha)^2 \frac{d\alpha}{\alpha} e\lambda r
 \leq e\lambda r,
\label{eqn-W2-dalphaoveralpha2}
\end{equation}
and vanishes when $r \leq 1/(\lambda\eta e)$.  Hence 
\begin{equation*}
\int_0^\eta I_{Cr} \frac{da}{a^2}
 \leq e\lambda \int_{1/(\lambda\eta e)}^\infty F_0(r)^2 r^{n-2} dr.
\end{equation*}
\begin{equation}
\boxed{
0 \leq \int_0^\eta I_{Cr}I_{C1}I_2 \frac{da}{a^{n+1}}
 \leq \Bigl( \tfrac{8}{3} + 2 (n-2) (1+\tfrac{5}{n-3}) \Bigr) e\lambda \int_{1/(\lambda\eta e)}^\infty F_0(r)^2 r^{n-2} dr.
}
\label{eqn-I-C}
\end{equation}

%%%%%%%%%%%%%%%%%%%%%%%%%%%%%%%%%%%%%%%%%%%%%%%%%%%%%%%%%%%%%%%%%%%%%%%

\subsection{The variance of $\vecb$ parallel to $\vectheta$}

% NOTE:  see our notes from 8/20/08, 8/21/08

Finally, we seek to bound the variance of $\vecb$, in the direction parallel to $\vectheta$.  The analysis is similar to the previous case (i.e., the variance of $\vecb$ orthogonal to $\vectheta$).

The variance of $\vecb$ parallel to $\vectheta$ is:
\begin{equation}
\text{E}((\vecb\cdot\vectheta)^2)
 = \int_{S^{n-1}} \int_0^1 \int_{\RR^n} (\vecb\cdot\vectheta)^2 |\Gamma_f(a,\vecb,\vectheta)|^2 d\vecb \frac{da}{a^{n+1}} d\sigma(\vectheta).
\end{equation}
We can simplify this by taking advantage of rotational symmetry.  Fix a vector $\vec{u} = (1,0,\ldots,0)$, and for each $\vectheta$, let $R$ be a rotation that maps $\vectheta$ to $\vec{u}$.  Then 
\begin{equation}
\text{E}((\vecb\cdot\vectheta)^2)
 = \int_{S^{n-1}} \int_0^1 \int_{\RR^n} (R(\vecb)\cdot\vec{u})^2 |\Gamma_f(a,R(\vecb),\vec{u})|^2 d\vecb \frac{da}{a^{n+1}} d\sigma(\vectheta).
\end{equation}
Then change variables $\vecb \mapsto R^{-1}(\vecb)$.  The integrand is now independent of $\vectheta$, so we can do the $\vectheta$ integral.  We get:
\begin{equation}
\text{E}((\vecb\cdot\vectheta)^2)
 = S_0 \int_0^1 \int_{\RR^n} b_1^2 |\Gamma_f(a,\vecb,\vec{u})|^2 d\vecb \frac{da}{a^{n+1}}.
\end{equation}

We now introduce some new notation, 
\begin{equation}
\Phi_{a,\vectheta}(\vecb) := \Gamma_f(a,\vecb,\vectheta), 
\end{equation}
to emphasize that we view this as a function of $\vecb$.  By equation (\ref{eqn-cct-def1}), the Fourier transform of $\Phi_{a,\vectheta}$ is given by 
\begin{equation}
\hat{\Phi}_{a,\vectheta}(\veck) = \hat{f}(\veck) \chi_{a,\vectheta}(\veck).  
\end{equation}
And we have, by Plancherel's theorem:
\begin{equation}
\begin{split}
\text{E}((\vecb\cdot\vectheta)^2)
 &= S_0 \int_0^1 \int_{\RR^n} b_1^2 |\Phi_{a,\vec{u}}(\vecb)|^2 d\vecb \frac{da}{a^{n+1}} \\
 &= S_0 \int_0^1 \int_{\RR^n} \Bigl| \frac{1}{2\pi i} \frac{\partial}{\partial k_1} \hat{\Phi}_{a,\vec{u}}(\veck) \Bigr|^2 d\veck \frac{da}{a^{n+1}}.  
\end{split}
\end{equation}

Now, using spherical coordinates $\veck = (r,\phi_1,\ldots,\phi_{n-1})$, we write $\hat{\Phi}_{a,\vec{u}}(\veck)$ as a product of a radial part and an angular part:
\begin{equation}
\hat{\Phi}_{a,\vec{u}}(\veck) = L(r) M(\phi_1), 
\end{equation}
where 
\begin{equation}
L(r) = F_0(r) W(\lambda ar), \quad
M(\phi_1) = V(\phi_1/\sqrt{a}) \Lambda_a(\phi_1).  
\end{equation}
Then we have 
\begin{equation}
\frac{\partial}{\partial k_1} \hat{\Phi}_{a,\vec{u}}(\veck)
 = L'(r) M(\phi_1) \frac{\partial r}{\partial k_1} + L(r) M'(\phi_1) \frac{\partial \phi_1}{\partial k_1}, 
\end{equation}
where 
\begin{equation}
\frac{\partial r}{\partial k_1} = \cos\phi_1, \quad
\frac{\partial \phi_1}{\partial k_1} = -\frac{\sin\phi_1}{r}.
\end{equation}
Now we can expand out the following integral:  (note that $\hat{\Phi}_{a,\vec{u}}(\veck)$ is real)
\begin{equation}
\begin{split}
\int_{\RR^n} \Bigl|\frac{\partial}{\partial k_1} \hat{\Phi}_{a,\vec{u}}(\veck) \Bigr|^2 d\veck
 &= \int_{S^{n-1}} \int_0^\infty \Bigl( L'(r) M(\phi_1) \cos\phi_1 \\
 &\qquad - L(r) M'(\phi_1) r^{-1} \sin\phi_1 \Bigr)^2 r^{n-1} dr d\sigma(\vec{\phi}) \\
 &= K_{Ar}K_{A1}K_2 - 2K_{Br}K_{B1}K_2 + K_{Cr}K_{C1}K_2, 
\end{split}
\end{equation}
where we define 
\begin{align}
K_{Ar} &= \int_0^\infty L'(r)^2 r^{n-1} dr \\
K_{A1} &= \int_0^\pi M(\phi_1)^2 \cos^2\phi_1 \sin^{n-2}\phi_1 d\phi_1 \\
K_2 &= \int_{S^{n-2}} d\sigma(\phi_2,\ldots,\phi_{n-1}) \\
K_{Br} &= \int_0^\infty L'(r)L(r) r^{n-2} dr \\
K_{B1} &= \int_0^\pi M'(\phi_1)M(\phi_1) \cos\phi_1 \sin^{n-1}\phi_1 d\phi_1 \\
K_{Cr} &= \int_0^\infty L(r)^2 r^{n-3} dr \\
K_{C1} &= \int_0^\pi M'(\phi_1)^2 \sin^n\phi_1 d\phi_1.
\end{align}

Thus we can write the variance of $\vecb$ parallel to $\vectheta$ as:
\begin{equation}
\text{E}((\vecb\cdot\vectheta)^2)
 = \frac{S_0}{(2\pi)^2} \int_0^1 \Bigl( K_{Ar}K_{A1}K_2 - 2K_{Br}K_{B1}K_2 + K_{Cr}K_{C1}K_2 \Bigr) \frac{da}{a^{n+1}}.
\end{equation}
A similar formula gives the variance conditioned on observing $a \leq \eta$:
\begin{multline}
\text{E}((\vecb\cdot\vectheta)^2 \:|\: a \leq \eta) \\
 = \frac{1}{\Pr[a \leq \eta]} \frac{S_0}{(2\pi)^2} \int_0^\eta \Bigl( K_{Ar}K_{A1}K_2 - 2K_{Br}K_{B1}K_2 + K_{Cr}K_{C1}K_2 \Bigr) \frac{da}{a^{n+1}}.
\label{eqn-varbpara-conda}
\end{multline}
We would then like to bound the various integrals appearing on the right hand side.  

\subsubsection{}

We begin with the integral $\int_0^\eta K_{Br}K_{B1}K_2 da/a^{n+1}$.  

Note that $K_2 = (n-1) I_2$, while $K_{B1} = I_{B1}$ and $K_{Br} = I_{Br}$.  Using the argument from the previous section, we have:
\begin{equation}
\boxed{
\Bigl| \int_0^\eta K_{Br}K_{B1}K_2 \frac{da}{a^{n+1}} \Bigr|
 \leq \tfrac{1}{4} (n-1) (n-2) \int_{1/(\lambda\eta e)}^\infty F_0(r)^2 r^{n-3} dr.
}
\label{eqn-K-B}
\end{equation}

\subsubsection{}

Next, consider the integral $\int_0^\eta K_{Ar}K_{A1}K_2 da/a^{n+1}$.  We already have a bound for $K_2$.  For $K_{A1}$, we write:
\begin{equation*}
\begin{split}
K_{A1}
 &= \int_0^\pi V(\phi_1/\sqrt{a})^2 \Lambda_a(\phi_1)^2 \cos^2\phi_1 \sin^{n-2}\phi_1 d\phi_1 \\
 &= \int_0^\pi V(\phi_1/\sqrt{a})^2 a^{(2n-1)/2} \Bigl(\frac{\sin(\phi_1/\sqrt{a})}{\sin\phi_1}\Bigr)^{n-2} \cos^2\phi_1 \sin^{n-2}\phi_1 d\phi_1 \\
 &= \int_0^{\pi/\sqrt{a}} V(\omega_1)^2 a^n \sin^{n-2}\omega_1 \cos^2(\omega_1\sqrt{a}) d\omega_1.
\end{split}
\end{equation*}
Using the fact that $V$ is supported on $[0,\pi/2]$, and the simple bound $\cos^2(\omega_1\sqrt{a}) \leq 1$, we get:
\begin{equation*}
0 \leq K_{A1}
 \leq \int_0^{\pi/2} V(\omega_1)^2 a^n \sin^{n-2}\omega_1 d\omega_1.
\end{equation*}
Combining with $K_2$, we get:
\begin{equation}
0 \leq K_{A1} K_2
 \leq a^n \int_{S^{n-1}} V(\omega_1)^2 d\sigma(\omega_1,\phi_2,\ldots,\phi_{n-1})
  = a^n.
\end{equation}

We now turn to $K_{Ar}$.  First, combining with $K_{A1}$ and $K_2$, we have 
\begin{equation}
0 \leq \int_0^\eta K_{Ar}K_{A1}K_2 \frac{da}{a^{n+1}}
 \leq \int_0^\eta K_{Ar} \frac{da}{a}.
\end{equation}
Note that $K_{Ar} = I_{Ar}$, so we can upper-bound $K_{Ar}$ as in the previous section:
\begin{equation}
0 \leq K_{Ar} \leq 2\norm{G_1}^2 + 2\norm{G_2}^2,
\end{equation}
where we define 
\begin{align}
G_1(r) &= F'_0(r) W(\lambda a r) r^{(n-1)/2} \\
G_2(r) &= F_0(r) W'(\lambda a r) \lambda a r^{(n-1)/2}.
\end{align}
Thus we have 
\begin{equation}
\boxed{
0 \leq \int_0^\eta K_{Ar}K_{A1}K_2 \frac{da}{a^{n+1}}
 \leq 2\Bigl( \int_0^\eta \norm{G_1}^2 \frac{da}{a} + \int_0^\eta \norm{G_2}^2 \frac{da}{a} \Bigr).
}
\label{eqn-K-A}
\end{equation}
We then want to upper-bound the integrals $\int_0^\eta \norm{G_1}^2 da/a$ and $\int_0^\eta \norm{G_2}^2 da/a$.  

For the first one, we have:
\begin{equation*}
\begin{split}
\int_0^\eta \norm{G_1}^2 \frac{da}{a}
 &= \int_0^\eta \int_0^\infty F'_0(r)^2 W(\lambda a r)^2 r^{n-1} dr \frac{da}{a} \\
 &= \int_0^\infty F'_0(r)^2 \int_0^\eta W(\lambda a r)^2 \frac{da}{a} r^{n-1} dr.
\end{split}
\end{equation*}
Using equation (\ref{eqn-W2-dalphaoveralpha}), we get:
\begin{equation}
\boxed{
\int_0^\eta \norm{G_1}^2 \frac{da}{a}
 \leq \int_{1/(\lambda \eta e)}^\infty F'_0(r)^2 r^{n-1} dr.
}
\label{eqn-K-A-G1}
\end{equation}

For the second integral, we have:
\begin{equation*}
\begin{split}
\int_0^\eta \norm{G_2}^2 \frac{da}{a}
 &= \int_0^\eta \int_0^\infty F_0(r)^2 W'(\lambda a r)^2 \lambda^2 a^2 r^{n-1} dr \frac{da}{a} \\
 &= \int_0^\infty F_0(r)^2 \int_0^\eta W'(\lambda a r)^2 \lambda^2 a da \: r^{n-1} dr.
\end{split}
\end{equation*}
Recall that the derivative of $W$ is given by equation (\ref{eqn-Wprime}).  So we can write 
\begin{equation}
\int_0^\eta W'(\lambda a r)^2 \lambda^2 a da
 = \int_0^{\lambda\eta r} W'(\alpha)^2 \alpha d\alpha \frac{1}{r^2}
 \leq \int_{1/e}^1 C_w^2 (\pi/\alpha)^2 \alpha d\alpha \frac{1}{r^2}
 = \tfrac{8}{3} \pi^2 \frac{1}{r^2},
\label{eqn-Wprime2-alphadalpha}
\end{equation}
and vanishes when $r \leq 1/(\lambda \eta e)$.  Hence,
\begin{equation}
\boxed{
\int_0^\eta \norm{G_2}^2 \frac{da}{a}
 \leq \tfrac{8}{3} \pi^2 \int_{1/(\lambda \eta e)}^\infty F_0(r)^2 r^{n-3} dr.
}
\label{eqn-K-A-G2}
\end{equation}

\subsubsection{}

Finally, we consider the integral $\int_0^\eta K_{Cr}K_{C1}K_2 da/a^{n+1}$.  We already have a bound for $K_2$.  For $K_{C1}$ we can write:
\begin{equation}
\begin{split}
0 \leq K_{C1}
 &= \int_0^\pi \Bigl( V'(\phi_1/\sqrt{a})(1/\sqrt{a}) \Lambda_a(\phi_1) + V(\phi_1/\sqrt{a}) \Lambda'_a(\phi_1) \Bigr)^2 \sin^n\phi_1 d\phi_1 \\
 &= \norm{\tilde{U}_1+\tilde{U}_2}^2 \leq 2\norm{\tilde{U}_1}^2 + 2\norm{\tilde{U}_2}^2,
\end{split}
\end{equation}
where we define 
\begin{align}
\tilde{U}_1(\phi_1) &= V'(\phi_1/\sqrt{a})(1/\sqrt{a}) \Lambda_a(\phi_1) \sin^{n/2}\phi_1 \\
\tilde{U}_2(\phi_1) &= V(\phi_1/\sqrt{a}) \Lambda'_a(\phi_1) \sin^{n/2}\phi_1.
\end{align}
Then 
\begin{equation}
0 \leq K_{C1} K_2 \leq 2\norm{\tilde{U}_1}^2 K_2 + 2\norm{\tilde{U}_2}^2 K_2.
\label{eqn-KC1-K2}
\end{equation}
We now evaluate $\norm{\tilde{U}_1}^2$ and $\norm{\tilde{U}_2}^2$.

For $\norm{\tilde{U}_1}^2$, we can write 
\begin{equation*}
\begin{split}
\norm{\tilde{U}_1}^2
 &= \int_0^\pi V'(\phi_1/\sqrt{a})^2 a^{-1} a^{(2n-1)/2} \Bigl(\frac{\sin(\phi_1/\sqrt{a})}{\sin(\phi_1)}\Bigr)^{n-2} \sin^n\phi_1 d\phi_1 \\
 &= \int_0^\pi V'(\phi_1/\sqrt{a})^2 a^{(2n-3)/2} \sin^{n-2}(\phi_1/\sqrt{a}) \sin^2\phi_1 d\phi_1 \\
 &= \int_0^{\pi/\sqrt{a}} V'(\omega_1)^2 a^{n-1} \sin^{n-2}\omega_1 \sin^2(\omega_1\sqrt{a}) d\omega_1 \\
 &\leq \frac{\pi^2}{4} a^n \int_0^{\pi/2} V'(\omega_1)^2 \sin^{n-2}\omega_1 d\omega_1.
\end{split}
\end{equation*}
(In the last step we used the bound $\sin^2(\omega_1\sqrt{a}) \leq \omega_1^2 a \leq \frac{\pi^2}{4} a$.)  

Then, using equation (\ref{eqn-Vprime2-sinnminus2-domega}), we have 
\begin{equation*}
\norm{\tilde{U}_1}^2
 \leq \frac{\pi^2}{4} a^n \cdot \frac{4n}{3S_0} \Bigl(1-\frac{1}{n}\Bigr) \int_0^\pi \sin^{n-2}\omega_1 d\omega_1.  
\end{equation*}
Combining with $K_2$, we get
\begin{equation}
0 \leq \norm{\tilde{U}_1}^2 K_2
 \leq \frac{\pi^2}{4} a^n \cdot \frac{4n}{3S_0} \Bigl(1-\frac{1}{n}\Bigr) \cdot \int_{S^{n-1}} d\sigma(\omega_1, \phi_2, \ldots, \phi_{n-1})
 = \tfrac{\pi^2}{3} a^n (n-1).  
\end{equation}

Next we evaluate $\norm{\tilde{U}_2}^2$.  The derivative of $\Lambda_a$ is given by equation (\ref{eqn-Lambdaprime}), hence 
\begin{equation*}
\tilde{U}_2(\phi_1)
 = V(\phi_1/\sqrt{a}) a^{(2n-1)/4} (\tfrac{n-2}{2}) \sin^{(n-4)/2}(\phi_1/\sqrt{a}) \Bigl( \frac{\cos(\phi_1/\sqrt{a})}{\sqrt{a}} - \frac{\sin(\phi_1/\sqrt{a})\cos(\phi_1)}{\sin(\phi_1)} \Bigr) \sin\phi_1,
\end{equation*}
and
\begin{equation*}
\begin{split}
\norm{\tilde{U}_2}^2
 &= \int_0^\pi V(\phi_1/\sqrt{a})^2 a^{(2n-1)/2} (\tfrac{n-2}{2})^2 \sin^{n-4}(\phi_1/\sqrt{a}) \Bigl( \frac{\cos(\phi_1/\sqrt{a})}{\sqrt{a}} - \frac{\sin(\phi_1/\sqrt{a})\cos(\phi_1)}{\sin(\phi_1)} \Bigr)^2 \sin^2\phi_1 d\phi_1 \\
 &= \int_0^{\pi/\sqrt{a}} V(\omega_1)^2 a^n (\tfrac{n-2}{2})^2 \sin^{n-4}(\omega_1) \Bigl( \frac{\cos(\omega_1)}{\sqrt{a}} - \frac{\sin(\omega_1)\cos(\omega_1\sqrt{a})}{\sin(\omega_1\sqrt{a})} \Bigr)^2 \sin^2(\omega_1\sqrt{a}) d\omega_1.
\end{split}
\end{equation*}
Recall that $V$ is supported on $[0,\pi/2]$.  For $\omega_1$ in this range, we have the following crude bound:  (using \cite{abram-stegun}, eqn. 4.3.81)
\begin{equation*}
\Bigl| \frac{\cos(\omega_1)}{\sqrt{a}} - \frac{\sin(\omega_1)\cos(\omega_1\sqrt{a})}{\sin(\omega_1\sqrt{a})} \Bigr|
 \leq \frac{1}{\sqrt{a}} + \frac{\sin(\omega_1)}{\sin(\omega_1\sqrt{a})} \cdot \frac{\sin(\omega_1\sqrt{a})}{\omega_1\sqrt{a}}
 \leq \frac{1}{\sqrt{a}} + \frac{1}{\sqrt{a}}.
\end{equation*}
Also, we have $\sin^2(\omega_1\sqrt{a}) \leq \omega_1^2 a \leq \frac{\pi^2}{4} a$.  Hence,
\begin{equation*}
\begin{split}
\norm{\tilde{U}_2}^2
 &\leq \int_0^{\pi/2} V(\omega_1)^2 a^n (\tfrac{n-2}{2})^2 \sin^{n-4}(\omega_1) (\tfrac{2}{\sqrt{a}})^2 \frac{\pi^2}{4} a d\omega_1 \\
 &= \pi^2 a^n (\tfrac{n-2}{2})^2 \int_0^{\pi/2} V(\omega_1)^2 \sin^{n-4}\omega_1 d\omega_1.
\end{split}
\end{equation*}

By equation (\ref{eqn-V2-sinnminus4-domega}), we have:
\begin{equation*}
\norm{\tilde{U}_2}^2
 \leq \pi^2 a^n (\tfrac{n-2}{2})^2 (1+\tfrac{5}{n-3}) \int_0^{\pi/2} V(\omega_1)^2 \sin^{n-2}\omega_1 d\omega_1.
\end{equation*}
Combining with $K_2$, we get:
\begin{equation}
\begin{split}
0 \leq \norm{\tilde{U}_2}^2 K_2
 &\leq \pi^2 a^n (\tfrac{n-2}{2})^2 (1+\tfrac{5}{n-3}) \int_{S^{n-1}} V(\omega_1)^2 d\sigma(\omega_1,\phi_2,\ldots,\phi_{n-1}) \\
 &\leq \tfrac{\pi^2}{4} a^n (n-2)^2 (1+\tfrac{5}{n-3}).
\end{split}
\end{equation}

So, by substituting into (\ref{eqn-KC1-K2}), we have 
\begin{equation}
0 \leq K_{C1} K_2 \leq 2 \cdot \tfrac{\pi^2}{3}a^n (n-1) + 2 \cdot \tfrac{\pi^2}{4} a^n (n-2)^2 (1+\tfrac{5}{n-3}).
\end{equation}

Finally, we turn to $K_{Cr}$.  Combining it with $K_{C1}$ and $K_2$, we have
\begin{equation*}
0 \leq \int_0^\eta K_{Cr}K_{C1}K_2 \frac{da}{a^{n+1}}
 \leq \Bigl( \tfrac{2\pi^2}{3} (n-1) + \tfrac{\pi^2}{2} (n-2)^2 (1+\tfrac{5}{n-3}) \Bigr) \int_0^\eta K_{Cr} \frac{da}{a}.
\end{equation*}
We can bound the integral on the right hand side as follows.
\begin{equation*}
\begin{split}
\int_0^\eta K_{Cr} \frac{da}{a}
 &= \int_0^\eta \int_0^\infty F_0(r)^2 W(\lambda a r)^2 r^{n-3} dr \frac{da}{a} \\
 &= \int_0^\infty F_0(r)^2 \int_0^\eta W(\lambda a r)^2 \frac{da}{a} r^{n-3} dr.
\end{split}
\end{equation*}
Using equation (\ref{eqn-W2-dalphaoveralpha}), we get 
\begin{equation*}
\int_0^\eta K_{Cr} \frac{da}{a}
 \leq \int_{1/(\lambda\eta e)}^\infty F_0(r)^2 r^{n-3} dr.
\end{equation*}
\begin{equation}
\boxed{
0 \leq \int_0^\eta K_{Cr}K_{C1}K_2 \frac{da}{a^{n+1}}
 \leq \Bigl( \tfrac{2\pi^2}{3} (n-1) + \tfrac{\pi^2}{2} (n-2)^2 (1+\tfrac{5}{n-3}) \Bigr) \int_{1/(\lambda\eta e)}^\infty F_0(r)^2 r^{n-3} dr.
}
\label{eqn-K-C}
\end{equation}

}{}

%%%%%%%%%%%%%%%%%%%%%%%%%%%%%%%%%%%%%%%%%%%%%%%%%%%%%%%%%%%%%%%%%%%%%%%%%%%%%%%

\section{The Ball in $\RR^n$}

\ifthenelse{\boolean{sec4}}{

We prove Theorem \ref{thm-ball}.

\subsection{The low-frequency components}

First, we claim that almost all the power in $\hat{f}$ is located at frequencies above some threshold $1/\lambda$.  This justifies our use of the curvelet transform, and theorems \ref{thm-cct-1} and \ref{thm-cct-2}, for an appropriate choice of the parameter $\lambda$.

% NOTE:  see our notes from 5/16/08, 5/13/08

We start by proving an upper-bound on the integral $\int_0^z t^{-1} J_\nu(t)^2 dt$, for $\nu > 0$.  Note that (\cite{abram-stegun}, eqn. (9.1.62)) 
\begin{equation}
|J_\nu(t)| \leq \frac{(\tfrac{1}{2}t)^\nu}{\nu!} \quad (\nu \geq -\tfrac{1}{2}, \: t \geq 0).
\end{equation}
Also (\cite{abram-stegun}, eqn. (6.1.38)), 
\begin{equation}
\nu! > \sqrt{2\pi} \nu^{(\nu+\tfrac{1}{2})} e^{-\nu} \quad (\nu > 0).
\end{equation}
Hence 
\begin{equation}
|J_\nu(t)| < \frac{(\tfrac{1}{2}t)^\nu}{\sqrt{2\pi} \nu^{\nu+\tfrac{1}{2}} e^{-\nu}} = \frac{1}{\sqrt{2\pi\nu}} \Bigl(\frac{te}{2\nu}\Bigr)^\nu,
\label{eqn-J-closetoorigin-upperbound}
\end{equation}
and
\begin{equation}
\begin{split}
\int_0^z t^{-1} J_\nu(t)^2 dt
 < \int_0^z \frac{1}{t} \frac{1}{2\pi\nu} \Bigl(\frac{te}{2\nu}\Bigr)^{2\nu} dt
 = \frac{1}{2\pi\nu} \Bigl(\frac{e}{2\nu}\Bigr)^{2\nu} \frac{1}{2\nu} t^{2\nu} \Big|_0^z
 = \frac{1}{4\pi\nu^2} \Bigl(\frac{ez}{2\nu}\Bigr)^{2\nu}.
\end{split}
\end{equation}
This upper bound is useful when $z \leq 2\nu/e$.

We can now calculate the amount of power contained in the low-frequency components of $f$:
\begin{equation}
\begin{split}
\int_{|\veck| \leq z} |\hat{f}(\veck)|^2 d\veck
 &= \int_0^z \frac{n}{S_0} \frac{1}{\rho^n} J_{n/2}(2\pi\rho\beta)^2 \cdot S_0 \rho^{n-1} d\rho \\
 &= n \int_0^{2\pi\beta z} t^{-1} J_{n/2}(t)^2 \cdot dt \\
 &< n \frac{1}{\pi n^2} \Bigl(\frac{2\pi\beta ez}{n}\Bigr)^n.
\end{split}
\end{equation}
Setting $z = n/(2\pi\beta e)$, we get 
\begin{equation}
\int_{|\veck| \leq n/(2\pi\beta e)} |\hat{f}(\veck)|^2 d\veck < \frac{1}{\pi n}.
\end{equation}
Recall that we set the parameter $\lambda$ so that $\lambda \geq 2\pi\beta e/n$.  So the region $\set{|\veck| \leq 1/\lambda}$ contains at most a $1/(\pi n)$ fraction of the total power.  

\subsection{The decay of $J_\nu(x)$}

% NOTE:  See our notes from 8/5/08, 8/22/08

We now prove some technical lemmas on the decay of $J_\nu(x)$ for $x \geq 2\nu$, $\nu \geq 1/2$.  These follow from classical results on Bessel functions \cite{abram-stegun,watson}, though some care is required near the transition region at $x \approx \nu$.  In particular, the usual asymptotic expansions for $J_\nu(x)$ only work when $x \geq \nu^2$, or when $x = \alpha\nu$ for some fixed constant $\alpha$.  For our purposes, we use an asymptotic expansion of $J_\nu(x)^2 + Y_\nu(x)^2$, that behaves well when $x \geq \nu$.  

\subsubsection{}

We start by quoting the following result from (\cite{watson}, p.447).  Define 
\begin{equation}
M_\nu(x) = \sqrt{J_\nu(x)^2 + Y_\nu(x)^2}.
\end{equation}
Then for all $x \geq \nu \geq 1/2$, 
\begin{equation}
\frac{2}{\pi x} < M_\nu(x)^2 < \frac{2}{\pi\sqrt{x^2-\nu^2}}.
\label{eqn-Msquared-lowerbound}
\end{equation}

This immediately implies an upper bound on $J_\nu(x)^2$, for all $x \geq 2\nu$, $\nu \geq 1/2$:
\begin{equation}
J_\nu(x)^2 \leq M_\nu(x)^2 < \frac{2}{\pi\sqrt{x^2-\nu^2}} \leq \frac{2}{\pi x} \cdot \frac{2}{\sqrt{3}}.
\label{eqn-Jsquared-upperbound}
\end{equation}

\subsubsection{}

We next prove a lower bound on $|J_\nu(x)|$, for $x$ within certain intervals.  Note that $J_\nu(x)$ is large at a zero of $Y_\nu(x)$.  We will show that (1) the zeroes of $Y_\nu(x)$ are not too far apart, and (2) $J_\nu(x)$ is large in a neighborhood around each zero of $Y_\nu(x)$.

To see this, note that $J_\nu(x)$ and $Y_\nu(x)$ can be written in terms of a modulus and phase, 
\begin{align}
\label{eqn-J-modphase}
J_\nu(x) &= M_\nu(x) \cos \theta_\nu(x),\\
Y_\nu(x) &= M_\nu(x) \sin \theta_\nu(x),
\end{align}
where $M_\nu(x)$ is as defined above, and $\theta_\nu(x)$ satisfies the equation
\begin{equation}
\theta'_\nu(x) = \frac{2}{\pi x M_\nu(x)^2}
\end{equation}
(see \cite{abram-stegun}, eqn. 9.2.21, and \cite{watson}, p.514).  This implies lower and upper bounds on $\theta'_\nu(x)$, for all $x \geq 2\nu$:
\begin{equation}
\tfrac{\sqrt{3}}{2} < \theta'_\nu(x) < 1.
\label{eqn-thetaprime-bounds}
\end{equation}

First, we claim that for any $t \geq 2\nu$, the interval $[t, t+\tfrac{2\pi}{\sqrt{3}}]$ contains a zero of $Y_\nu(x)$.  To see this, write the following, for any $\delta \geq 0$:
\[
\theta_\nu(x+\delta) = \theta_\nu(x) + \int_x^{x+\delta} \theta'_\nu(y) dy \geq \theta_\nu(x) + \tfrac{\sqrt{3}}{2} \delta.
\]
So $\theta_\nu(t+\tfrac{2\pi}{\sqrt{3}}) \geq \theta_\nu(t) + \pi$.  So $\theta_\nu(x)$ must equal an integer multiple of $\pi$ for some $x \in [t, t+\tfrac{2\pi}{\sqrt{3}}]$; and $Y_\nu(x)$ must vanish at that point.  This proves our first claim.

Second, let $\phi$ be a zero of $Y_\nu(x)$, satisfying $\phi \geq 2\nu$.  We claim that, for any $\delta \in [-\pi/2, \pi/2]$, 
\begin{equation}
|J_\nu(\phi+\delta)| \geq M_\nu(\phi+\delta) \cos\delta.
\end{equation}
To see this, write 
\[
|J_\nu(\phi+\delta)|
 = M_\nu(\phi+\delta) |\cos\theta_\nu(\phi+\delta)|
 = M_\nu(\phi+\delta) \bigl| \cos |\theta_\nu(\phi+\delta)-\theta_\nu(\phi)| \bigr|.
\]
(The last step follows because $\theta_\nu(\phi)$ is an integer multiple of $\pi$.)  Then note that 
\[
|\theta_\nu(\phi+\delta) - \theta_\nu(\phi)|
 = \Bigl| \int_\phi^{\phi+\delta} \theta'_\nu(y) dy \Bigr|
 \leq |\delta|.
\]
Hence we have 
\[
\bigl| \cos |\theta_\nu(\phi+\delta)-\theta_\nu(\phi)| \bigr|
 \geq \bigl| \cos|\delta| \bigr|
 = \cos\delta.
\]
This proves our second claim.

\subsubsection{}

Finally, we prove the following lower bound on a sum of squares of Bessel functions:
\begin{quotation}
Let $\nu_1, \ldots, \nu_m \in [1/2, \nu_\text{max}]$.  Let $t \geq 2\nu_\text{max}$.  Then there exists some $t' \in [t - \tfrac{\pi}{2}, t + \tfrac{2\pi}{\sqrt{3}} + \tfrac{\pi}{2}]$ such that 
\begin{equation}
\sum_{k=1}^m J_{\nu_k}(t')^2 \geq \frac{m}{7t'}.
\end{equation}
\end{quotation}

Proof:  Essentially, we will show that there must exist a point $t'$ where a constant fraction of the functions $J_{\nu_k}(t')^2$ ($k = 1,\ldots,m$) are large simultaneously.

Define the interval $I = [t, t + \tfrac{2\pi}{\sqrt{3}}]$.  For each $k = 1,\ldots,m$, $I$ contains a zero of $Y_{\nu_k}(x)$, call it $\phi_k$.  Now define the function 
\[
\chi_k(x) = \begin{cases}
  \cos^2(x-\phi_k) & \text{if $\phi_k-\tfrac{\pi}{2} \leq x \leq \phi_k+\tfrac{\pi}{2}$}, \\
  0 & \text{otherwise}.
\end{cases}
\]
Note that 
\[
J_{\nu_k}(x)^2 \geq M_{\nu_k}(x)^2 \chi_k(x) > \frac{2}{\pi x} \chi_k(x).
\]
Furthermore, define the function 
\[
u(x) = \sum_{k=1}^m \chi_k(x), 
\]
and note that 
\[
\sum_{k=1}^m J_{\nu_k}(x)^2 \geq \frac{2}{\pi x} u(x).
\]

Define the interval $I' = [t - \tfrac{\pi}{2}, t + \tfrac{2\pi}{\sqrt{3}} + \tfrac{\pi}{2}]$; this interval contains the support of all of the functions $\chi_k(x)$ ($k = 1,\ldots,m$).  Then write 
\[
\int_{I'} u(x)dx
 = \sum_{k=1}^m \int_{\phi_k-(\pi/2)}^{\phi_k+(\pi/2)} \chi_k(x)dx
 = m \cdot \frac{\pi}{2}.
\]
So there must exist a point $t' \in I'$ such that 
\[
u(t') \geq \frac{1}{|I'|} \int_{I'} u(x)dx
 \geq \frac{1}{7} \cdot m \cdot \frac{\pi}{2}
 = \frac{\pi}{14} m, 
\]
and the claim follows.

\subsection{The probability of observing a scale $a$}

Next we claim that $\hat{f}$ has a heavy tail.  This implies that we will observe fine-scale elements ($a$ small) with significant probability.

% NOTE:  see our notes from 4/21/08, 5/15/08, 6/3/08, 6/4/08, 7/8/08, 8/7/08

Again, we start by proving a lower bound on $\int_z^\infty t^{-1} J_\nu(t)^2 dt$, when $\nu$ is of the form $m$ or $m + (1/2)$ (where $m$ is an integer), $\nu \geq 1$, and $z \geq 2\nu$.  We will show that:
\begin{equation}
\int_z^\infty t^{-1} J_\nu(t)^2 dt
 \geq \Bigl(1-\frac{1}{2\nu}\Bigr) \frac{1}{7(z+5.20)}.
\label{eqn-intzinfty-Jsquared}
\end{equation}

First, consider the case of $\nu = m$.  We assume $m \geq 1$ and $z \geq 2m$.  Using (\cite{abram-stegun}, eqn. 11.3.36), we can write 
\begin{equation*}
\begin{split}
\int_z^\infty t^{-1} J_\nu(t)^2 dt
 &= \int_z^\infty t^{-1} J_m(t)^2 dt \\
 &= -\frac{1}{2m} \Bigl( J_0(t)^2 + J_m(t)^2 + 2\sum_{k=1}^{m-1} J_k(t)^2 \Bigr) \Big|_z^\infty \\
 &= \frac{1}{2m} \Bigl( J_0(z)^2 + J_m(z)^2 + 2\sum_{k=1}^{m-1} J_k(z)^2 \Bigr) \\
 &\geq \frac{1}{2m} \Bigl( J_m(z)^2 + 2\sum_{k=1}^{m-1} J_k(z)^2 \Bigr).
\end{split}
\end{equation*}
Then, using the lemma from the previous section, we get the following, for some $z' \in [z,z+5.20]$:
\begin{equation}
\int_z^\infty t^{-1} J_\nu(t)^2 dt
 \geq \frac{1}{2m} \frac{2m-1}{7z'}
 \geq \Bigl(1-\frac{1}{2\nu}\Bigr) \frac{1}{7(z+5.20)}.
\end{equation}

Next, consider the case $\nu = m+(1/2)$.  We assume $m \geq 1$ and $z \geq 2m+1$.  Using (\cite{abram-stegun}, eqn. 11.3.36), we can write 
\begin{equation*}
\begin{split}
\int_z^\infty t^{-1} J_\nu(t)^2 dt
 &= \int_z^\infty t^{-1} J_{m+(1/2)}(t)^2 dt \\
 &= \frac{1}{2m+1} \int_z^\infty t^{-1} J_{1/2}(t)^2 dt - \frac{1}{2m+1} \Bigl( J_{1/2}(t)^2 + J_{m+(1/2)}(t)^2 + 2\sum_{k=1}^{m-1} J_{k+(1/2)}(t)^2 \Bigr) \Big|_z^\infty \\
 &\geq \frac{1}{2m+1} \Bigl( J_{1/2}(z)^2 + J_{m+(1/2)}(z)^2 + 2\sum_{k=1}^{m-1} J_{k+(1/2)}(z)^2 \Bigr). 
\end{split}
\end{equation*}
Then, using the lemma from the previous section, we get the following, for some $z' \in [z,z+5.20]$:
\begin{equation}
\int_z^\infty t^{-1} J_\nu(t)^2 dt
 \geq \frac{1}{2m+1} \frac{2m}{7z'}
 \geq \Bigl(1-\frac{1}{2\nu}\Bigr) \frac{1}{7(z+5.20)}. 
\end{equation}

\vskipline

We now proceed to lower-bound the probability of observing a fine-scale element $a$.  The following bound holds for any $n \geq 2$ and any $\eta \leq 1/e^2$.  
\begin{equation*}
\begin{split}
\Pr[a\leq\eta]
 &\geq \int_{|\veck| \geq 1/(\lambda\eta)} |\hat{f}(\veck)|^2 d\veck \\
 &= \int_{1/(\lambda\eta)}^\infty \frac{n}{S_0} \rho^{-n} J_{n/2}(2\pi\rho\beta)^2 \cdot S_0 \rho^{n-1} d\rho \\
 &= n \int_{2\pi\beta/(\lambda\eta)}^\infty t^{-1} J_{n/2}(t)^2 dt \\
 &\geq n \int_{n/(e\eta)}^\infty t^{-1} J_{n/2}(t)^2 dt, 
\end{split}
\end{equation*}
where in the last step we used the fact that $\lambda \geq 2\pi\beta e/n$.  Then, by equation (\ref{eqn-intzinfty-Jsquared}), and using the fact that $n/(e\eta) \geq 2e \geq 5.43$, we get 
\begin{equation}
\Pr[a\leq\eta]
 \geq n \Bigl(1-\frac{1}{n}\Bigr) \frac{1}{7(\frac{n}{e\eta}+5.20)}
 \geq n \Bigl(1-\frac{1}{n}\Bigr) \frac{1}{14\frac{n}{e\eta}}
 = \frac{e\eta}{14} \Bigl(1-\frac{1}{n}\Bigr).
\end{equation}
\begin{equation}
\boxed{
\Pr[a\leq\eta]
 \geq \frac{e\eta}{14} \Bigl(1-\frac{1}{n}\Bigr).
}
\label{eqn-praleqeta}
\end{equation}

\subsection{The variance of $\vecb$ orthogonal to $\vectheta$}

% NOTE:  see our notes from 6/4/08, 6/13/08, 6/25/08, 7/17/08, 8/7/08

First, we give a simple upper bound on integrals of the form 
\[
\int_z^\infty t^{-k} J_\nu(t)^2 dt, 
\]
for $k \geq 1$, $z \geq 2\nu$ and $\nu \geq 1/2$.  This follows from equation (\ref{eqn-Jsquared-upperbound}):
\begin{equation}
\begin{split}
\int_z^\infty t^{-k} J_\nu(t)^2 dt
 &\leq \int_z^\infty t^{-k} \frac{2}{\pi t} \frac{2}{\sqrt{3}} dt \\
 &= \frac{4}{\pi\sqrt{3}} \int_z^\infty t^{-k-1} dt
  = \frac{4}{\pi\sqrt{3}} (-1/k) t^{-k} \Big|_z^\infty
  = \frac{4}{\pi\sqrt{3}k} z^{-k}.
\end{split}
\end{equation}

\vskipline

We now use this to upper-bound the integral 
\[
\int_{1/(\lambda\eta e)}^\infty F_0(r)^2 r^{n-k} dr, 
\]
for $k \geq 1$ and $\eta \leq 1/2e^2$.  We write the following:  (for the last step, recall that $\lambda \leq 2\cdot 2\pi\beta e/n$, which implies $2\pi\beta/(\lambda\eta e) \geq n/(2\eta e^2)$)
\begin{equation}
\begin{split}
\int_{1/(\lambda\eta e)}^\infty F_0(r)^2 r^{n-k} dr
 &= \frac{n}{S_0} \int_{1/(\lambda\eta e)}^\infty J_{n/2}(2\pi\beta r)^2 r^{-k} dr \\
 &= \frac{n}{S_0} \int_{2\pi\beta/(\lambda\eta e)}^\infty J_{n/2}(t)^2 t^{-k} dt \cdot (2\pi\beta)^{k-1} \\
 &\leq \frac{n}{S_0} \frac{4}{\pi\sqrt{3} k} \Bigl(\frac{2\eta e^2}{n}\Bigr)^k \cdot (2\pi\beta)^{k-1}.
\end{split}
\end{equation}

\vskipline

In a similar way, we can upper-bound the integral 
\[
\int_{1/(\lambda\eta e)}^\infty F'_0(r)^2 r^{n-k} dr, 
\]
for $k \geq 1$ and $\eta \leq \tfrac{n}{n+2} (1/2e^2)$.  First, note that 
\begin{equation}
F'_0(r) = -\sqrt{\tfrac{n}{S_0}} r^{-n/2} J_{(n/2)+1}(2\pi\beta r) (2\pi\beta).
\end{equation}
(To see this, write $F_0(r) = \sqrt{\tfrac{n}{S_0}} (2\pi\beta)^{n/2} g(2\pi\beta r)$, where $g(x) = x^{-n/2} J_{n/2}(x)$.  Then $F'_0(r) = \sqrt{\tfrac{n}{S_0}} (2\pi\beta)^{n/2} g'(2\pi\beta r) (2\pi\beta)$, where $g'(x) = -x^{-n/2} J_{(n/2)+1}(x)$, see \cite{abram-stegun} eqn. 9.1.30.)  Then we write the following:  (for the last step, recall the fact that $\lambda \leq 2\cdot 2\pi\beta e/n$, which implies $2\pi\beta/(\lambda\eta e) \geq n/(2\eta e^2)$)
\begin{equation}
\begin{split}
\int_{1/(\lambda\eta e)}^\infty F'_0(r)^2 r^{n-k} dr
 &= \frac{n}{S_0} \int_{1/(\lambda\eta e)}^\infty J_{(n/2)+1}(2\pi\beta r)^2 r^{-k} dr \cdot (2\pi\beta)^2 \\
 &= \frac{n}{S_0} \int_{2\pi\beta/(\lambda\eta e)}^\infty J_{(n/2)+1}(t)^2 t^{-k} dt \cdot (2\pi\beta)^{k+1} \\
 &\leq \frac{n}{S_0} \frac{4}{\pi\sqrt{3} k} \Bigl(\frac{2\eta e^2}{n}\Bigr)^k \cdot (2\pi\beta)^{k+1}.
\end{split}
\end{equation}

\vskipline

We now combine this with the results of section 3, to show a bound on the variance of $\vecb$ perpendicular to $\vectheta$, conditioned on $a \leq \eta$.  

We can simplify equations (\ref{eqn-I-B}), (\ref{eqn-I-A}) and (\ref{eqn-I-C}) as follows:
\[
\Bigl| \int_0^\eta I_{Br} I_{B1} I_2 \frac{da}{a^{n+1}} \Bigr|
 \leq \tfrac{1}{4}(n-2) \int_{1/(\lambda\eta e)}^\infty F_0(r)^2 r^{n-3} dr, 
\]
\[
0 \leq \int_0^\eta I_{Ar} I_{A1} I_2 \frac{da}{a^{n+1}}
 \leq \frac{\pi^2}{2(n-1)} \cdot \frac{1}{\lambda} \int_{1/(\lambda\eta e)}^\infty F_0'(r)^2 r^{n-2} dr
  + \frac{\pi^2}{2(n-1)} \cdot \frac{17}{\lambda} \int_{1/(\lambda\eta e)}^\infty F_0(r)^2 r^{n-4} dr, 
\]
\[
0 \leq \int_0^\eta I_{Cr} I_{C1} I_2 \frac{da}{a^{n+1}}
 \leq (2n+9+\tfrac{10}{n-3}) e\lambda \int_{1/(\lambda\eta e)}^\infty F_0(r)^2 r^{n-2} dr.
\]
Plugging in our bounds for the integrals on the right hand side, we get:
\[
\Bigl| \int_0^\eta I_{Br} I_{B1} I_2 \frac{da}{a^{n+1}} \Bigr|
 \leq \frac{25}{S_0} \frac{(2\eta)^3}{n} (2\pi\beta)^2, 
\]
\[
0 \leq \int_0^\eta I_{Ar} I_{A1} I_2 \frac{da}{a^{n+1}}
 \leq \frac{5}{n-1} \Bigl( \frac{8}{S_0} (2\eta)^2 (2\pi\beta)^2 + \frac{3600}{S_0} \frac{(2\eta)^4}{n^2} (2\pi\beta)^2 \Bigr), 
\]
\[
0 \leq \int_0^\eta I_{Cr} I_{C1} I_2 \frac{da}{a^{n+1}}
 \leq \Bigl(1+\frac{5}{n-3}\Bigr) \frac{640}{S_0} \frac{(2\eta)^2}{n} (2\pi\beta)^2.
\]
Substituting into equation (\ref{eqn-varbperp-conda}), and using (\ref{eqn-praleqeta}), we get:
\begin{equation}
\begin{split}
E(\vecb^T&(I-\vectheta\vectheta^T)\vecb \:|\: a\leq\eta) \\
 &\leq \frac{5.20}{\eta} \Bigl(1+\frac{1}{n-1}\Bigr) \beta^2 \Bigl( 640(2\eta)^2 + 40(2\eta)^2 + \frac{3200}{n-3}(2\eta)^2 + 50(2\eta)^3 + \frac{18000}{n^2}(2\eta)^4 \Bigr) \\
 &\leq (1+O(\tfrac{1}{n})) \beta^2 \eta \Bigl( 3536 + 260(2\eta) + O(\tfrac{1}{n}) \Bigr) \cdot 4.
\end{split}
\end{equation}
Using our assumption that $2\eta \leq 1/e^2$, we can rewrite this as 
\begin{equation}
\boxed{
E(\vecb^T(I-\vectheta\vectheta^T)\vecb \:|\: a\leq\eta)
 \leq \eta \beta^2 ( 14300 + O(\tfrac{1}{n}) ).
}
\label{eqn-bl-varbperp-conda}
\end{equation}

\subsection{The variance of $\vecb$ parallel to $\vectheta$}

We also get a bound on the variance of $\vecb$ parallel to $\vectheta$, conditioned on $a \leq \eta$.  

Substituting into equations (\ref{eqn-K-B}), (\ref{eqn-K-A}) and (\ref{eqn-K-C}), we get:
\[
\Bigl| \int_0^\eta K_{Br} K_{B1} K_2 \frac{da}{a^{n+1}} \Bigr|
 \leq \frac{25}{S_0} (2\eta)^3 (2\pi\beta)^2, 
\]
\[
0 \leq \int_0^\eta K_{Ar} K_{A1} K_2 \frac{da}{a^{n+1}}
 \leq \frac{12}{S_0} (2\eta) (2\pi\beta)^2 + \frac{5300}{S_0} \frac{(2\eta)^3}{n^2} (2\pi\beta)^2, 
\]
\[
0 \leq \int_0^\eta K_{Cr} K_{C1} K_2 \frac{da}{a^{n+1}}
 \leq \frac{700}{S_0} \frac{(2\eta)^3}{n} (2\pi\beta)^2 + \frac{500}{S_0} (2\eta)^3 \Bigl(1+\frac{5}{n-3}\Bigr) (2\pi\beta)^2.
\]
Substituting into equation (\ref{eqn-varbpara-conda}), and using (\ref{eqn-praleqeta}), we get:
\begin{equation}
\begin{split}
E((\vecb\cdot\vectheta)^2 \:|\: a\leq\eta)
 &\leq \frac{5.20}{\eta} \Bigl(1+\frac{1}{n-1}\Bigr) \beta^2 \Bigl( 12(2\eta) + 100(2\eta)^3 (\tfrac{53}{n^2} + \tfrac{1}{2} + \tfrac{7}{n} + 5 + \tfrac{25}{n-3}) \Bigr) \\
 &\leq (1+O(\tfrac{1}{n})) \beta^2 \Bigl( 63 + 3120(2\eta)^2 + O(\tfrac{1}{n}) \Bigr) \cdot 2.
\end{split}
\end{equation}
Using our assumption that $2\eta \leq 1/e^2$, we can rewrite this as 
\begin{equation}
\boxed{
E((\vecb\cdot\vectheta)^2 \:|\: a\leq\eta)
 \leq \beta^2 ( 242 + O(\tfrac{1}{n}) ).
}
\label{eqn-bl-varbpara-conda}
\end{equation}

\vskipline

% FIXME:  assume $n \geq 5$ and $\eta \leq 5/(7e^2)$?

}{}

%%%%%%%%%%%%%%%%%%%%%%%%%%%%%%%%%%%%%%%%%%%%%%%%%%%%%%%%%%%%%%%%%%%%%%%%%%%%%%%

\section{Spherical Shells}

\ifthenelse{\boolean{sec5}}{

\subsection{Spherical shell with square cross-section}

In this section, we give a heuristic analysis of the spherical shell with square cross-section.  

% NOTE:  See our notes from 10/27/08

We can write $f = h-g$, where $h$ is $C$ times the indicator function of a ball of radius $\beta+\delta$ around the origin, and $g$ is $C$ times the indicator function of a ball of radius $\beta$ around the origin.  Then its Fourier transform is $\hat{f} = \hat{h}-\hat{g}$, where $\hat{h}$ and $\hat{g}$ are calculated as in Section 4.  We can write this as:  $\hat{f}(\veck) = F_0(|\veck|)$, 
\begin{equation}
F_0(\rho)
 = \frac{C}{\rho^{n/2}} (\beta+\delta)^{n/2} J_{n/2}(2\pi(\beta+\delta)\rho)
 - \frac{C}{\rho^{n/2}} \beta^{n/2} J_{n/2}(2\pi\beta\rho).
\end{equation}

We are interested in the case where $\delta \ll \beta$.  Note that interference between the two Bessel functions begins to play a major role when $\rho \gtrsim 1/(2\pi\delta)$.  We claim that $F_0(\rho)$ decays quite slowly, out to distance $\rho \sim 1/(2\pi\delta)$.  

It will be convenient to define 
\begin{equation}
K(\beta) = \beta^{n/2} J_{n/2}(2\pi\beta\rho), 
\end{equation}
so we have 
\begin{equation}
F_0(\rho) = \frac{C}{\rho^{n/2}} (K(\beta+\delta) - K(\beta)).
\end{equation}
We can approximate $F_0(\rho)$ by a simpler expression.  First, when $\delta \lesssim \beta/n$, we can write $C$ as follows:
\begin{equation}
C \approx \frac{1}{\sqrt{n\beta^{n-1}\delta B_0}}
 = \sqrt{\frac{\beta}{n\delta}} \frac{1}{\sqrt{\beta^n B_0}}.
\end{equation}
Also, when $\delta$ is sufficiently small (we will elaborate on this point later), 
\begin{equation}
F_0(\rho) \approx \frac{C}{\rho^{n/2}} \delta K'(\beta).
\end{equation}
A straightforward calculation (see \cite{abram-stegun}, equation 9.1.30) shows that 
\begin{equation}
K'(\beta) = (2\pi\rho) \beta^{n/2} J_{(n/2)-1}(2\pi\beta\rho).
\end{equation}
Note that $K'(\beta)$ is roughly $2\pi\rho$ times larger than $K(\beta)$, so we expect the approximation to be accurate when $\delta \lesssim 1/(2\pi\rho)$, or equivalently, when $\rho \lesssim 1/(2\pi\delta)$.  Combining the above equations, we get the following approximation for $F_0(\rho)$:
\begin{equation}
\begin{split}
F_0(\rho)
 &\approx \sqrt{\frac{\beta}{n\delta}} \frac{1}{\sqrt{\beta^n B_0}} \cdot \frac{1}{\rho^{n/2}} \delta \cdot (2\pi\rho) \beta^{n/2} J_{(n/2)-1}(2\pi\beta\rho) \\
 &= \sqrt{\frac{\beta\delta}{S_0}} \frac{2\pi}{\rho^{(n/2)-1}} J_{(n/2)-1}(2\pi\beta\rho) \text{ when } \rho \lesssim 1/(2\pi\delta), 
\end{split}
\label{eqn-ss-squarecs-F0}
\end{equation}
where $S_0$ is the surface area of the sphere in $\RR^n$ (note that $B_0 = S_0/n$).

Compared to the Fourier transform of the ball (Section 4), this function decays more slowly, out to distance $\rho \sim 1/(2\pi\delta)$.  Thus, when we apply the curvelet transform, with significant probability, we can observe fine-scale elements $a\leq\eta$, where $\eta$ shrinks proportional to $\delta$.  This suggests that a very thin spherical shell (i.e., $\delta$ very small) allows us to find the center with very high precision, proportional to $\delta$.

\subsection{Spherical shell with Gaussian cross-section}

In the next few sections, we will prove Theorem \ref{thm-sphericalshell}, for a spherical shell with Gaussian cross-section.

% NOTE:  see our notes beginning 9/22/08

We begin by proving upper and lower bounds on the normalization factor $C_f$.  The following identity will be useful:  (this follows from the definition of $F_0(\rho)$ and a change of variables)
\begin{equation}
\int_\alpha^{\alpha'} F_0(\rho)^2 \rho^{n-1} d\rho
 = C_f^2 \varepsilon^n \beta^{2n-2} \int_{2\pi\beta\alpha}^{2\pi\beta\alpha'} \exp(-\tfrac{1}{2\pi}\varepsilon^2t^2) J_{(n/2)-1}(t)^2 t dt.  
\label{eqn-ss-power}
\end{equation}
Then the $L^2$ norm of $\hat{f}(\veck)$ is given by:
\begin{equation}
\begin{split}
\int_{\RR^n} |\hat{f}(\veck)|^2 d\veck
 &= S_0 \int_0^\infty F_0(\rho)^2 \rho^{n-1} d\rho \\
 &= S_0 C_f^2 \varepsilon^n \beta^{2n-2} \int_0^\infty \exp(-\tfrac{1}{2\pi}\varepsilon^2t^2) J_{(n/2)-1}(t)^2 t dt \\
 &= S_0 C_f^2 \varepsilon^n \beta^{2n-2} (N_1+N_2), 
\end{split}
\label{eqn-ss-norm}
\end{equation}
where we split the integral into two parts, 
\begin{equation}
N_1 = \int_0^{n-2} \exp(-\tfrac{1}{2\pi}\varepsilon^2t^2) J_{(n/2)-1}(t)^2 t dt, 
\end{equation}
\begin{equation}
N_2 = \int_{n-2}^\infty \exp(-\tfrac{1}{2\pi}\varepsilon^2t^2) J_{(n/2)-1}(t)^2 t dt.  
\label{eqn-ss-N2-def}
\end{equation}

We now prove upper bounds on $N_1$ and $N_2$.  For $N_1$, using trivial upper bounds on $\exp(-x^2)$ and $J_\nu(x)^2$ (see \cite{abram-stegun}, eqn. 9.1.60), we get:
\begin{equation}
N_1 \leq \int_0^{n-2} \tfrac{1}{2} t dt
 = \tfrac{1}{4} (n-2)^2.  
\end{equation}
For $N_2$, using the upper bound on $J_\nu(x)^2$ from equation (\ref{eqn-Jsquared-upperbound}), we get:
\begin{equation}
\begin{split}
N_2 &\leq \int_{n-2}^\infty \exp(-\tfrac{1}{2\pi}\varepsilon^2t^2) \tfrac{4}{\pi\sqrt{3}} dt \\
 &= \tfrac{4}{\pi\sqrt{3}} \sqrt{2\pi} \tfrac{1}{\varepsilon} \int_{\frac{1}{\sqrt{2\pi}}\varepsilon(n-2)}^\infty \exp(-\tau^2) d\tau \\
 &\leq \tfrac{4}{\pi\sqrt{3}} \sqrt{2\pi} \tfrac{1}{\varepsilon} \tfrac{\sqrt{\pi}}{2}
 = 2 \sqrt{\tfrac{2}{3}} \tfrac{1}{\varepsilon}.
\end{split}
\end{equation}

Substituting into (\ref{eqn-ss-norm}), we get 
\begin{equation*}
\int_{\RR^n} |\hat{f}(\veck)|^2 d\veck
 \leq S_0 C_f^2 \varepsilon^n \beta^{2n-2} (\tfrac{1}{4} (n-2)^2 + 2 \sqrt{\tfrac{2}{3}} \tfrac{1}{\varepsilon}).  
\end{equation*}
We assumed $\varepsilon \leq \tfrac{6}{(n-2)^2}$, and it is easy to check that this implies $\tfrac{1}{4} (n-2)^2 \leq 2 \sqrt{\tfrac{2}{3}} \tfrac{1}{\varepsilon}$.  Thus 
\begin{equation*}
\int_{\RR^n} |\hat{f}(\veck)|^2 d\veck
 \leq S_0 C_f^2 \varepsilon^n \beta^{2n-2} \cdot 4 \sqrt{\tfrac{2}{3}} \tfrac{1}{\varepsilon}.  
\end{equation*}
Setting the left side equal to 1 implies a lower bound on $C_f^2$:
\begin{equation}
\boxed{
C_f^2 \geq \frac{1}{S_0 \varepsilon^{n-1} \beta^{2n-2}} \cdot \tfrac{1}{4} \sqrt{\tfrac{3}{2}}.  
}
\label{eqn-ss-cf2-lowerbound}
\end{equation}

Next we show lower bounds on $N_1$ and $N_2$.  For $N_1$ we have a trivial lower bound,
\begin{equation}
N_1 \geq 0.
\end{equation}
For $N_2$, we use the lower bound $J_\nu(x)^2 > \frac{2}{\pi x} \cos^2(\theta_\nu(x))$ (see equations (\ref{eqn-J-modphase}) and (\ref{eqn-Msquared-lowerbound})).  For convenience, we define $\theta(x) = \theta_\nu(x)$, suppressing the $\nu$ subscript.  We get:
\begin{equation}
N_2 \geq \frac{2}{\pi} \int_{n-2}^\infty \exp(-\tfrac{1}{2\pi}\varepsilon^2t^2) \cos^2(\theta(t)) dt.  
\label{eqn-ss-N2-lowerbound-begin}
\end{equation}

Note that $\theta(t)$ is a monotone increasing function (equation (\ref{eqn-thetaprime-bounds})), hence it is one-to-one and has a well-defined inverse.  We make a change of variables, $\tau = \theta(t)$, $t = \theta^{-1}(\tau)$:
\begin{equation}
N_2 \geq \frac{2}{\pi} \int_{\theta(n-2)}^\infty \exp(-\tfrac{1}{2\pi}\varepsilon^2(\theta^{-1}(\tau))^2) \cos^2\tau \cdot (\theta^{-1})'(\tau) d\tau.  
\end{equation}
Note that, whenever $\tau = \theta(t)$, we have $(\theta^{-1})'(\tau) = \frac{1}{\theta'(t)}$.  Hence, by equation (\ref{eqn-thetaprime-bounds}), 
\begin{equation}
1 < (\theta^{-1})'(\tau) < \tfrac{2}{\sqrt{3}}, 
\quad \text{for $\tau \geq \theta(n-2)$}.  
\end{equation}
Also note that 
\begin{equation}
\begin{split}
\theta^{-1}(\tau)
 &= \theta^{-1}(\theta(n-2)) + \int_{\theta(n-2)}^\tau (\theta^{-1})'(x) dx \\
 &\leq n-2 + \tfrac{2}{\sqrt{3}} (\tau-\theta(n-2)).  
\end{split}
\end{equation}
Substituting in, we get:
\begin{equation}
N_2 \geq \frac{2}{\pi} \int_{\theta(n-2)}^\infty \exp(-\tfrac{1}{2\pi}\varepsilon^2 \cdot (n-2 + \tfrac{2}{\sqrt{3}} (\tau-\theta(n-2)))^2) \cos^2\tau d\tau.  
\end{equation}

We will use the following simple fact:  if a function $f$ is nonnegative and monotone decreasing on the interval $[\alpha,\infty)$, then 
\begin{equation}
\int_\alpha^\infty f(x) \cos^2x dx
 \geq \tfrac{1}{2} \int_{\alpha+\pi}^\infty f(x) dx.
\end{equation}
This follows because 
\begin{equation}
\begin{split}
\int_\alpha^\infty f(x) \cos^2x dx
 &= \sum_{k=0}^\infty \int_{\alpha+k\pi}^{\alpha+(k+1)\pi} f(x) \cos^2x dx \\
 &\geq \sum_{k=0}^\infty f(\alpha+(k+1)\pi) \int_{\alpha+k\pi}^{\alpha+(k+1)\pi} \cos^2x dx \\
 &= \sum_{k=0}^\infty f(\alpha+(k+1)\pi) \tfrac{\pi}{2} \\
 &= \sum_{k=0}^\infty f(\alpha+(k+1)\pi) \tfrac{1}{2} \int_{\alpha+(k+1)\pi}^{\alpha+(k+2)\pi} dx \\
 &\geq \sum_{k=0}^\infty \tfrac{1}{2} \int_{\alpha+(k+1)\pi}^{\alpha+(k+2)\pi} f(x) dx \\
 &= \tfrac{1}{2} \int_{\alpha+\pi}^\infty f(x) dx.
\end{split}
\end{equation}

Using the above fact, and a change of variables, we get 
\begin{equation}
\begin{split}
N_2 &\geq \frac{1}{\pi} \int_{\theta(n-2)+\pi}^\infty \exp(-\tfrac{1}{2\pi}\varepsilon^2 \cdot (n-2 + \tfrac{2}{\sqrt{3}} (\tau-\theta(n-2)))^2) d\tau \\
 &= \frac{\sqrt{3}}{2\pi} \int_{n-2+\frac{2\pi}{\sqrt{3}}}^\infty \exp(-\tfrac{1}{2\pi}\varepsilon^2x^2) dx \\
 &= \frac{\sqrt{3}}{\sqrt{2\pi}} \frac{1}{\varepsilon} \int_{\frac{1}{\sqrt{2\pi}} \varepsilon (n-2+\frac{2\pi}{\sqrt{3}})}^\infty \exp(-y^2) dy.  
\end{split}
\label{eqn-ss-N2-lowerbound-end}
\end{equation}
Recall that we assumed $\varepsilon \leq \frac{1}{n+2}$.  This implies $\frac{1}{\sqrt{2\pi}} \varepsilon (n-2+\frac{2\pi}{\sqrt{3}}) \leq \frac{1}{\sqrt{2\pi}}$.  So, substituting in and integrating numerically, we get that 
\begin{equation}
N_2 \geq \frac{\sqrt{3}}{\sqrt{2\pi}} \frac{1}{\varepsilon} \int_{\frac{1}{\sqrt{2\pi}}}^\infty \exp(-y^2) dy
 \geq \frac{1}{4\varepsilon}.  
\end{equation}

Substituting into (\ref{eqn-ss-norm}), we get that 
\begin{equation}
\int_{\RR^n} |\hat{f}(\veck)|^2 d\veck
 \geq S_0 C_f^2 \varepsilon^n \beta^{2n-2} (0+\tfrac{1}{4\varepsilon})
 = S_0 C_f^2 \varepsilon^{n-1} \beta^{2n-2} \cdot \tfrac{1}{4}.
\end{equation}
Setting the left side equal to 1 implies an upper bound on $C_f^2$:
\begin{equation}
\boxed{
C_f^2 \leq \frac{4}{S_0 \varepsilon^{n-1} \beta^{2n-2}}.
}
\label{eqn-ss-cf2-upperbound}
\end{equation}

\subsection{The low-frequency components}

Next, we show that $\hat{f}(\veck)$ has very little power at low frequencies, corresponding to coarse scales $a \geq 1$.  This justifies our use of the curvelet transform, which effectively ignores these low-frequency components (recall Theorems \ref{thm-cct-1} and \ref{thm-cct-2}).  

The total amount of power at frequencies less than $z$ (for any $z \geq 0$) is given by:
\begin{equation}
\begin{split}
\int_{|\veck| \leq z} |\hat{f}(\veck)|^2 d\veck
 &= S_0 \int_0^z F_0(\rho)^2 \rho^{n-1} d\rho \\
 &= S_0 C_f^2 \varepsilon^n \beta^{2n-2} \int_0^{2\pi\beta z} \exp(-\tfrac{1}{2\pi}\varepsilon^2t^2) J_{(n/2)-1}(t)^2 t dt.  
\end{split}
\end{equation}
(We used equation (\ref{eqn-ss-power}).)  Using a trivial upper bound $\exp(-x^2) \leq 1$, and the upper bound for $J_\nu(x)$ (when $x$ is small) from equation (\ref{eqn-J-closetoorigin-upperbound}), we get:
\begin{equation}
\begin{split}
\int_{|\veck| \leq z} |\hat{f}(\veck)|^2 d\veck
 &\leq S_0 C_f^2 \varepsilon^n \beta^{2n-2} \int_0^{2\pi\beta z} \frac{1}{\pi(n-2)} \Bigl(\frac{te}{n-2}\Bigr)^{n-2} t dt \\
 &= S_0 C_f^2 \varepsilon^n \beta^{2n-2} \cdot \frac{e^{n-2}}{\pi(n-2)^{n-1}} \cdot \frac{1}{n} (2\pi\beta z)^n.  
\end{split}
\end{equation}
Using our upper bound on $C_f^2$ (equation (\ref{eqn-ss-cf2-upperbound})), we get 
\begin{equation}
\begin{split}
\int_{|\veck| \leq z} |\hat{f}(\veck)|^2 d\veck
 &\leq 4\varepsilon \cdot \frac{e^{n-2}}{\pi(n-2)^{n-1}} \cdot \frac{1}{n} (2\pi\beta z)^n \\
 &\leq \frac{4\varepsilon}{\pi e^2} \cdot \frac{e^n}{(n-2)^n} \cdot (2\pi\beta z)^n.  
\end{split}
\end{equation}

Now, fix $z = \frac{n-2}{2\pi\beta e}$.  Recall that $\lambda \geq \frac{2\pi\beta e}{n-2}$, hence $1/\lambda \leq z$.  Then we have 
\begin{equation}
\boxed{
\int_{|\veck| \leq 1/\lambda} |\hat{f}(\veck)|^2 d\veck
 \leq \frac{4\varepsilon}{\pi e^2} \leq \frac{\varepsilon}{5}.  
}
\end{equation}
Recall that we assumed $\varepsilon \leq \frac{1}{n+2}$.  So the frequencies below $1/\lambda$ only constitute a small fraction of the total probability mass.  This justifies our use of the curvelet transform, with this choice of the parameter $\lambda$.  

\subsection{The probability of measuring a fine-scale element}

We give a lower bound on the probability of measuring the scale variable to be small, $a \leq \eta_c$, where 
\begin{equation}
\eta_c = \frac{\delta}{\betatilde} \frac{(n-2)}{e}.  
\end{equation}
We will show that $a \leq \eta_c$ with at least constant probability.  

First, we write $\Pr[a \leq \eta_c]$ as follows, 
using Section 3.1 and equation (\ref{eqn-ss-power}):
% using equations (\ref{eqn-rf-praleqeta}) and (\ref{eqn-ss-power}):
\begin{equation}
\begin{split}
\Pr[a \leq \eta_c]
 &\geq S_0 \int_{1/(\lambda\eta_c)}^\infty F_0(\rho)^2 \rho^{n-1} d\rho \\
 &= S_0 C_f^2 \varepsilon^n \beta^{2n-2} \int_{2\pi\beta/(\lambda\eta_c)}^\infty \exp(-\tfrac{1}{2\pi}\varepsilon^2t^2) J_{(n/2)-1}(t)^2 t dt.  
\end{split}
\end{equation}
The lower limit of integration can be simplified, by substituting in the definitions of $\lambda$ and $\eta_c$, 
\begin{equation}
\frac{2\pi\beta}{\lambda\eta_c}
 = 2\pi\beta \Bigl(\frac{2\pi\betatilde e}{n-2}\Bigr)^{-1} \Bigl(\frac{\delta}{\betatilde}\frac{(n-2)}{e}\Bigr)^{-1}
 = 2\pi\beta (2\pi\delta)^{-1} = \frac{\beta}{\delta} = \frac{1}{\varepsilon}.
\end{equation}
This integral is similar to the integral $N_2$ which we encountered earlier (equation (\ref{eqn-ss-N2-def})).  We can get a lower bound using the same technique (equations (\ref{eqn-ss-N2-lowerbound-begin}) - (\ref{eqn-ss-N2-lowerbound-end})); in particular, note that $1/\varepsilon \geq n-2$, as required in that calculation (this holds because we assumed $\varepsilon \leq 1/(n+2)$).  This leads to:  
\begin{equation}
\begin{split}
\Pr[a \leq \eta_c]
 &\geq S_0 C_f^2 \varepsilon^n \beta^{2n-2} \cdot \frac{\sqrt{3}}{\sqrt{2\pi}} \frac{1}{\varepsilon} \int_{\frac{1}{\sqrt{2\pi}} \varepsilon (\frac{1}{\varepsilon} + \frac{2\pi}{\sqrt{3}})}^\infty \exp(-y^2) dy.  
\end{split}
\end{equation}

The lower limit of integration can be written as 
\begin{equation}
\begin{split}
\frac{1}{\sqrt{2\pi}} \varepsilon \Bigl( \frac{1}{\varepsilon} + \frac{2\pi}{\sqrt{3}} \Bigr)
  = \frac{1}{\sqrt{2\pi}} \Bigl( 1 + \frac{2\pi}{\sqrt{3}} \varepsilon \Bigr)
  \leq \frac{1}{\sqrt{2\pi}} \cdot 2 = \sqrt{\frac{2}{\pi}}, 
\end{split}
\end{equation}
where the last inequality follows because $\varepsilon \leq \frac{1}{5}$.  
Thus we can lower-bound our integral as follows:
\begin{equation}
\begin{split}
\Pr[a \leq \eta_c]
 &\geq S_0 C_f^2 \varepsilon^n \beta^{2n-2} \cdot \frac{\sqrt{3}}{\sqrt{2\pi}} \frac{1}{\varepsilon} \int_{\sqrt{2/\pi}}^\infty \exp(-y^2) dy \\
 &> S_0 C_f^2 \varepsilon^n \beta^{2n-2} \cdot \frac{0.15}{\varepsilon} \\
 &= (0.15) S_0 C_f^2 \varepsilon^{n-1} \beta^{2n-2}.  
\end{split}
\label{eqn-ss-praleqetac-raw}
\end{equation}
Now, using our lower bound for $C_f^2$ (equation (\ref{eqn-ss-cf2-lowerbound})), we get:  
\begin{equation}
\boxed{
\Pr[a \leq \eta_c]
 \geq (0.15) \cdot \tfrac{1}{4} \sqrt{\tfrac{3}{2}} > 0.045.
}
\label{eqn-ss-praleqetac}
\end{equation}

\subsection{Some more integrals}

Our next goal is to bound the variance of $\vecb$.  We begin by proving upper bounds on certain integrals involving $F_0(r)$ and $F_0'(r)$.  Then, in the following two sections, we will bound the variance of $\vecb$ in the directions orthogonal and parallel to $\vectheta$.

First, we consider integrals of the following form, where $k \geq 1$:
\begin{equation}
\int_\alpha^\infty F_0(r)^2 r^{n-k} dr.
\end{equation}
Using the definition of $F_0(r)$, and a change of variables, 
\begin{equation*}
\int_\alpha^\infty F_0(r)^2 r^{n-k} dr
 = C_f^2 \varepsilon^n (2\pi)^{k-1} \beta^{2n+k-3} \int_{2\pi\beta\alpha}^\infty \exp(-\tfrac{1}{2\pi}\varepsilon^2t^2) J_{(n/2)-1}(t)^2 \frac{dt}{t^{k-2}}.
\end{equation*}

We will upper-bound this integral, assuming that $\alpha \geq \frac{n-2}{2\pi\beta}$.  First, using equation (\ref{eqn-Jsquared-upperbound}), we have $J_{(n/2)-1}(t)^2 \leq \frac{4}{\pi\sqrt{3}} \frac{1}{t}$, and we get:  
\begin{equation*}
\int_\alpha^\infty F_0(r)^2 r^{n-k} dr
 \leq C_f^2 \varepsilon^n (2\pi)^{k-1} \beta^{2n+k-3} \cdot \frac{4}{\pi\sqrt{3}} \int_{2\pi\beta\alpha}^\infty \exp(-\tfrac{1}{2\pi}\varepsilon^2t^2) \frac{dt}{t^{k-1}}.
\end{equation*}
Next, we use a simple inequality:  $t^{k-1} \geq (2\pi\beta\alpha)^{k-1}$, whenever $t \geq 2\pi\beta\alpha$.  Thus, 
\begin{equation*}
\int_\alpha^\infty F_0(r)^2 r^{n-k} dr
 \leq C_f^2 \varepsilon^n \beta^{2n-2} \cdot \frac{4}{\pi\sqrt{3}} \frac{1}{\alpha^{k-1}} \int_{2\pi\beta\alpha}^\infty \exp(-\tfrac{1}{2\pi}\varepsilon^2t^2) dt.
\end{equation*}
The integral on the right hand side can be bounded as follows:
\begin{equation*}
\int_{2\pi\beta\alpha}^\infty \exp(-\tfrac{1}{2\pi}\varepsilon^2t^2) dt
 = \int_{\frac{1}{\sqrt{2\pi}}\varepsilon \cdot 2\pi\beta\alpha}^\infty \exp(-\tau^2) d\tau \cdot \sqrt{2\pi} \frac{1}{\varepsilon}
 \leq \frac{\sqrt{\pi}}{2} \cdot \sqrt{2\pi} \frac{1}{\varepsilon}
 = \frac{\pi}{\sqrt{2}} \frac{1}{\varepsilon}.
\end{equation*}
Substituting in, we get:
\begin{equation}
\boxed{
\int_\alpha^\infty F_0(r)^2 r^{n-k} dr
 \leq C_f^2 \varepsilon^{n-1} \beta^{2n-2} \cdot \frac{4}{\sqrt{6}} \frac{1}{\alpha^{k-1}}, 
 \qquad \text{for $\alpha \geq \frac{n-2}{2\pi\beta}$}.
}
\label{eqn-ss-intF0-upperbound}
\end{equation}

Next, we consider integrals of the following form, where $k = 1, 2$:
\begin{equation}
\int_\alpha^\infty F_0'(r)^2 r^{n-k} dr.
\end{equation}

In order to calculate $F_0'(r)$, we write $F_0(r)$ in the following form:
\begin{equation*}
F_0(r) = C_f \varepsilon^{n/2} (2\pi) \beta^n \cdot (2\pi\beta)^{(n/2)-1} \cdot H(r) \cdot K(2\pi\beta r), 
\end{equation*}
where 
\begin{equation*}
H(x) = \exp(-\pi\varepsilon^2\beta^2x^2), \quad
K(x) = \frac{1}{x^{(n/2)-1}} J_{(n/2)-1}(x).
\end{equation*}
Then $F_0'(r)$ can be written as 
\begin{equation*}
F_0'(r) = F_0^{(a)}(r) + F_0^{(b)}(r), 
\end{equation*}
where 
\begin{align*}
F_0^{(a)}(r) &= C_f \varepsilon^{n/2} (2\pi) \beta^n \cdot (2\pi\beta)^{(n/2)-1} \cdot H'(r) \cdot K(2\pi\beta r), \\
F_0^{(b)}(r) &= C_f \varepsilon^{n/2} (2\pi) \beta^n \cdot (2\pi\beta)^{(n/2)-1} \cdot H(r) \cdot K'(2\pi\beta r) (2\pi\beta).
\end{align*}
We can expand this out.  Note that (see \cite{abram-stegun}, equation (9.1.30))
\begin{equation*}
H'(x) = \exp(-\pi\varepsilon^2\beta^2x^2) (-2\pi\varepsilon^2\beta^2x), \quad
K'(x) = -\frac{1}{x^{(n/2)-1}} J_{n/2}(x).  
\end{equation*}
Substituting in, we get 
\begin{align*}
F_0^{(a)}(r) &= C_f \varepsilon^{n/2} (2\pi) \beta^n \frac{1}{r^{(n/2)-1}} \cdot \exp(-\pi\varepsilon^2\beta^2r^2) (-2\pi\varepsilon^2\beta^2r) J_{(n/2)-1}(2\pi\beta r), \\
F_0^{(b)}(r) &= C_f \varepsilon^{n/2} (2\pi) \beta^n \frac{1}{r^{(n/2)-1}} \cdot \exp(-\pi\varepsilon^2\beta^2r^2) (-1) J_{n/2}(2\pi\beta r) (2\pi\beta).
\end{align*}
Thus $F_0'(r)$ is given by:
\begin{equation}
F_0'(r) = C_f \varepsilon^{n/2} (2\pi) \beta^n \frac{1}{r^{(n/2)-1}} \exp(-\pi\varepsilon^2\beta^2r^2) \cdot \Bigl( -2\pi\varepsilon^2\beta^2r J_{(n/2)-1}(2\pi\beta r) - 2\pi\beta J_{n/2}(2\pi\beta r) \Bigr).
\end{equation}
Substituting into our integral, and performing a change of variables, we get:
\begin{equation}
\int_\alpha^\infty F_0'(r)^2 r^{n-k} dr
 = C_f^2 \varepsilon^n (2\pi)^{k-1} \beta^{2n+k-3} \int_{2\pi\beta\alpha}^\infty \exp(-\tfrac{1}{2\pi}\varepsilon^2t^2) \Bigl( \varepsilon^2\beta t J_{(n/2)-1}(t) + 2\pi\beta J_{n/2}(t) \Bigr)^2 \frac{dt}{t^{k-2}}.
\end{equation}

We will upper-bound this integral, assuming that $\alpha \geq \frac{n}{2\pi\beta}$.  First, using equation (\ref{eqn-Jsquared-upperbound}), we have $J_{n/2}(t)^2 \leq \frac{4}{\pi\sqrt{3}} \frac{1}{t}$, and similarly for $J_{(n/2)-1}(t)^2$.  So we get:  
\begin{equation}
\int_\alpha^\infty F_0'(r)^2 r^{n-k} dr
 \leq C_f^2 \varepsilon^n (2\pi)^{k-1} \beta^{2n+k-3} \int_{2\pi\beta\alpha}^\infty \exp(-\tfrac{1}{2\pi}\varepsilon^2t^2) \Bigl( \varepsilon^2\beta t + 2\pi\beta \Bigr)^2 \Bigl(\frac{4}{\pi\sqrt{3}} \frac{1}{t}\Bigr) \frac{dt}{t^{k-2}}.
\end{equation}
Changing variables and rearranging, we get:
\begin{equation}
\begin{split}
\int_\alpha^\infty & F_0'(r)^2 r^{n-k} dr \\
 &\leq C_f^2 \varepsilon^n (2\pi)^{k-1} \beta^{2n+k-3} \int_{\sqrt{2\pi}\varepsilon\beta\alpha}^\infty \exp(-\tau^2) (\sqrt{2\pi}\varepsilon\beta\tau + 2\pi\beta)^2 \frac{4}{\pi\sqrt{3}} \frac{d\tau}{\tau^{k-1}} \cdot (\tfrac{1}{\sqrt{2\pi}}\varepsilon)^{k-2} \\
 &= C_f^2 \varepsilon^n (2\pi)^{k-1} \beta^{2n+k-3} \cdot \tfrac{4}{\pi\sqrt{3}} (\tfrac{1}{\sqrt{2\pi}}\varepsilon)^{k-2} 
 \int_{\sqrt{2\pi}\varepsilon\beta\alpha}^\infty \exp(-\tau^2) (2\pi\varepsilon^2\beta^2\tau^2 + 2(2\pi)^{3/2}\varepsilon\beta^2\tau + (2\pi)^2\beta^2) \frac{d\tau}{\tau^{k-1}} \\
 &= C_f^2 \varepsilon^{n+k-2} \sqrt{2\pi}^k \beta^{2n+k-1} \cdot \tfrac{4}{\pi\sqrt{3}} 
 \int_{\sqrt{2\pi}\varepsilon\beta\alpha}^\infty \exp(-\tau^2) (2\pi\varepsilon^2\tau^2 + 2(2\pi)^{3/2}\varepsilon\tau + (2\pi)^2) \frac{d\tau}{\tau^{k-1}}.
\end{split}
\end{equation}

We can handle integrals of the form 
\begin{equation}
\int_a^\infty \exp(-\tau^2) \frac{d\tau}{\tau^\ell}
\end{equation}
as follows.  When $\ell \geq 0$, we have 
\begin{equation}
\int_a^\infty \exp(-\tau^2) \frac{d\tau}{\tau^\ell}
 \leq \frac{1}{a^\ell} \int_a^\infty \exp(-\tau^2) d\tau
 \leq \frac{1}{a^\ell} \cdot \frac{\sqrt{\pi}}{2}.
\end{equation}
When $\ell = -1$, we have 
\begin{equation}
\int_a^\infty \exp(-\tau^2) \tau d\tau
 \leq \int_0^\infty \exp(-\tau^2) \tau d\tau
 = -\tfrac{1}{2} \exp(-\tau^2) \Big|_0^\infty
 = \tfrac{1}{2}.
\end{equation}
When $\ell = -2$, we have 
\begin{equation}
\int_a^\infty \exp(-\tau^2) \tau^2 d\tau
 \leq \int_0^\infty \exp(-\tau^2) \tau^2 d\tau
 = -\tfrac{1}{2} \exp(-\tau^2) \tau \Big|_0^\infty - \int_0^\infty -\tfrac{1}{2} \exp(-\tau^2) d\tau
 = \frac{\sqrt{\pi}}{4} < 0.45.
\end{equation}

Now, we can upper-bound our integral in the $k=2$ case:
\begin{equation}
\begin{split}
\int_\alpha^\infty & F_0'(r)^2 r^{n-2} dr \\
 &\leq C_f^2 \varepsilon^n (2\pi) \beta^{2n+1} \cdot \tfrac{4}{\pi\sqrt{3}} 
 \int_{\sqrt{2\pi}\varepsilon\beta\alpha}^\infty \exp(-\tau^2) (2\pi\varepsilon^2\tau^2 + 2(2\pi)^{3/2}\varepsilon\tau + (2\pi)^2) \frac{d\tau}{\tau} \\
 &\leq C_f^2 \varepsilon^n (2\pi) \beta^{2n+1} \cdot \tfrac{4}{\pi\sqrt{3}} 
 \Bigl( 2\pi\varepsilon^2 \cdot \tfrac{1}{2} + 2(2\pi)^{3/2}\varepsilon \cdot \tfrac{\sqrt{\pi}}{2} + (2\pi)^2 \cdot \tfrac{1}{\sqrt{2\pi}\varepsilon\beta\alpha} \tfrac{\sqrt{\pi}}{2} \Bigr) \\
 &= C_f^2 \varepsilon^n (2\pi) \beta^{2n+1} \cdot \tfrac{8}{\sqrt{3}} 
 \Bigl( \tfrac{1}{2} \varepsilon^2 + \sqrt{2}\pi\varepsilon + \tfrac{\pi}{\sqrt{2}} \tfrac{1}{\varepsilon\beta\alpha} \Bigr), 
 \qquad \text{for $\alpha \geq \frac{n}{2\pi\beta}$}.
\end{split}
\label{eqn-ss-intF0prime-upperbound2}
\end{equation}

We can also upper-bound our integral in the $k=1$ case:
\begin{equation}
\begin{split}
\int_\alpha^\infty & F_0'(r)^2 r^{n-1} dr \\
 &\leq C_f^2 \varepsilon^{n-1} \sqrt{2\pi} \beta^{2n} \cdot \tfrac{4}{\pi\sqrt{3}} 
 \int_{\sqrt{2\pi}\varepsilon\beta\alpha}^\infty \exp(-\tau^2) (2\pi\varepsilon^2\tau^2 + 2(2\pi)^{3/2}\varepsilon\tau + (2\pi)^2) d\tau \\
 &\leq C_f^2 \varepsilon^{n-1} \sqrt{2\pi} \beta^{2n} \cdot \tfrac{4}{\pi\sqrt{3}} 
 \Bigl( 2\pi\varepsilon^2 \cdot \tfrac{\sqrt{\pi}}{4} + 2(2\pi)^{3/2}\varepsilon \cdot \tfrac{1}{2} + (2\pi)^2 \cdot \tfrac{\sqrt{\pi}}{2} \Bigr) \\
 &= C_f^2 \varepsilon^{n-1} \sqrt{2\pi} \beta^{2n} \cdot \tfrac{8}{\sqrt{3}} 
 \Bigl( \tfrac{\sqrt{\pi}}{4} \varepsilon^2 + \sqrt{2\pi} \varepsilon + \pi^{3/2} \Bigr), 
 \qquad \text{for $\alpha \geq \frac{n}{2\pi\beta}$}.
\end{split}
\label{eqn-ss-intF0prime-upperbound1}
\end{equation}

\subsection{The variance of $\vecb$ orthogonal to $\vectheta$}

We now bound the variance of $\vecb$ orthogonal to $\vectheta$, conditioned on observing $a \leq \eta_c$.  Recall that $\eta_c = \frac{\delta}{\betatilde} \frac{(n-2)}{e}$.  

We start with the results of Section 3.  From equation (\ref{eqn-I-B}), we get:
\begin{equation}
\Bigl| \int_0^\eta I_{Br} I_{B1} I_2 \frac{da}{a^{n+1}} \Bigr|
 \leq \tfrac{1}{4} (n-2) \lambda\eta e \int_{1/(\lambda\eta e)}^\infty F_0(r)^2 r^{n-2} dr.
\end{equation}
(We used the fact that $1 \leq r \cdot \lambda\eta e$, for all $r$ in this interval.)  From equations (\ref{eqn-I-A}), (\ref{eqn-I-A-G1}) and (\ref{eqn-I-A-G2}), we get:
\begin{equation}
\begin{split}
0 \leq \int_0^\eta I_{Ar} I_{A1} I_2 \frac{da}{a^{n+1}}
 &\leq \frac{\pi^2}{2(n-1)} \cdot \frac{1}{\lambda} \int_{1/(\lambda\eta e)}^\infty F_0'(r)^2 r^{n-2} dr + \\
 &\quad \frac{\pi^2}{2(n-1)} \cdot 24e^2\lambda\eta^2 \int_{1/(\lambda\eta e)}^\infty F_0(r)^2 r^{n-2} dr.
\end{split}
\end{equation}
(Again, we used the fact that $1 \leq r \cdot \lambda\eta e$, for all $r$ in this interval.)  From equation (\ref{eqn-I-C}), we get:
\begin{equation}
0 \leq \int_0^\eta I_{Cr} I_{C1} I_2 \frac{da}{a^{n+1}}
 \leq 2(n+10)e\lambda \int_{1/(\lambda\eta e)}^\infty F_0(r)^2 r^{n-2} dr.
\end{equation}
(We used the fact that $\frac{8}{3} + 2(n-2)(1+\frac{5}{n-3}) = \frac{8}{3} + 2(n+3+\frac{5}{n-3}) \leq 2(n+10)$, assuming $n \geq 4$.)

Now we fix $\eta = \eta_c$, and we upper-bound these integrals, using equations (\ref{eqn-ss-intF0-upperbound}) and (\ref{eqn-ss-intF0prime-upperbound2}).  (Note that, in order to apply these results, we must have $\frac{1}{\lambda\eta_c e} \geq \frac{n}{2\pi\beta}$.  Recall that $\lambda = \frac{2\pi\betatilde e}{n-2}$, and $\eta_c = \frac{\delta}{\betatilde} \frac{(n-2)}{e}$.  Also, recall that we assumed $\varepsilon \leq \frac{1}{en}$.  Then $\lambda\eta_c = 2\pi\delta = 2\pi\varepsilon\beta$, and $\frac{1}{\lambda\eta_c e} = \frac{1}{2\pi\beta\varepsilon e} \geq \frac{n}{2\pi\beta}$, as desired.)

After some tedious calculation, we get:
\begin{align}
\Bigl| \int_0^{\eta_c} I_{Br} I_{B1} I_2 \frac{da}{a^{n+1}} \Bigr|
 &\leq (n-2) \cdot C_f^2\varepsilon^{n-1}\beta^{2n-2} \cdot 120\varepsilon^2\beta^2, \\
0 \leq \int_0^{\eta_c} I_{Ar} I_{A1} I_2 \frac{da}{a^{n+1}}
 &\leq 5 \cdot C_f^2\varepsilon^{n-1}\beta^{2n-2} \cdot \varepsilon\beta^2 (12000\varepsilon^2 + 9\varepsilon + 80), \\
0 \leq \int_0^{\eta_c} I_{Cr} I_{C1} I_2 \frac{da}{a^{n+1}}
 &\leq C_f^2\varepsilon^{n-1}\beta^{2n-2} \cdot 2600\varepsilon\beta^2(1+\tfrac{12}{n-2}) \cdot S.
\end{align}
Then, by combining these equations and using our assumption that $\varepsilon \leq O(\frac{1}{n^2})$, we get:
\begin{equation}
\begin{split}
\int_0^{\eta_c} & \Bigl( I_{Ar}I_{A1}I_2 + 2 I_{Br}I_{B1}I_2 + I_{Cr}I_{C1}I_2 \Bigr) \frac{da}{a^{n+1}} \\
 &\leq C_f^2\varepsilon^{n-1}\beta^{2n-2} \cdot \varepsilon\beta^2 (60000\varepsilon^2 + 45\varepsilon + 400 + 2(n-2)120\varepsilon + 2600(1+\tfrac{12}{n-2})S) \\
 &\leq C_f^2\varepsilon^{n-1}\beta^{2n-2} \cdot \varepsilon\beta^2 (400 + 2600 + O(\tfrac{1}{n})) \cdot S.
\end{split}
\end{equation}

Next, recall from equation (\ref{eqn-ss-praleqetac-raw}) that:
\begin{equation}
\Pr[a \leq \eta_c] \geq (0.15) S_0 C_f^2 \varepsilon^{n-1} \beta^{2n-2}.
\end{equation}
Finally, by substituting into equation (\ref{eqn-varbperp-conda}) and simplifying, we get a bound on the variance of $\vecb$, in the subspace orthogonal to $\vectheta$, conditioned on observing $a \leq \eta_c$:
\begin{equation}
\boxed{
\text{E}(\vecb^T (I-\vectheta\vectheta^T) \vecb \;|\; a \leq \eta_c)
 \leq (n-1) \varepsilon \beta^2 (507 + O(\tfrac{1}{n})) \cdot S.
}
\label{eqn-ss-varbperp-conda}
\end{equation}

\subsection{The variance of $\vecb$ parallel to $\vectheta$}

We now bound the variance of $\vecb$ parallel to $\vectheta$, conditioned on observing $a \leq \eta_c$.  Recall that $\eta_c = \frac{\delta}{\betatilde} \frac{(n-2)}{e}$.  

We start with the results of Section 3.  From equation (\ref{eqn-K-B}), we get:
\begin{equation}
\Bigl| \int_0^\eta K_{Br} K_{B1} K_2 \frac{da}{a^{n+1}} \Bigr|
 \leq \tfrac{1}{4} (n-1) (n-2) \int_{1/(\lambda\eta e)}^\infty F_0(r)^2 r^{n-3} dr.
\end{equation}
From equations (\ref{eqn-K-A}), (\ref{eqn-K-A-G1}) and (\ref{eqn-K-A-G2}), we get:
\begin{equation}
\begin{split}
0 \leq \int_0^\eta K_{Ar} K_{A1} K_2 \frac{da}{a^{n+1}}
 &\leq 2 \int_{1/(\lambda\eta e)}^\infty F_0'(r)^2 r^{n-1} dr + \\
 &\quad 2 \cdot \tfrac{8}{3}\pi^2 \int_{1/(\lambda\eta e)}^\infty F_0(r)^2 r^{n-3} dr.
\end{split}
\end{equation}
From equation (\ref{eqn-K-C}), we get:
\begin{equation}
0 \leq \int_0^\eta K_{Cr} K_{C1} K_2 \frac{da}{a^{n+1}}
 \leq \tfrac{\pi^2}{2} (n^2+3n+3) \int_{1/(\lambda\eta e)}^\infty F_0(r)^2 r^{n-3} dr.
\end{equation}
(We used the fact that $\frac{2\pi^2}{3} (n-1) + \frac{\pi^2}{2} (n-2)^2 (1+\frac{5}{n-3}) = \frac{\pi^2}{2} (n^2 + \frac{7}{3}n - \frac{7}{3} + \frac{5}{n-3}) \leq \frac{\pi^2}{2} (n^2 + 3n + 3)$, assuming $n \geq 4$.)

Now we fix $\eta = \eta_c$, and we upper-bound these integrals, using equations (\ref{eqn-ss-intF0-upperbound}) and (\ref{eqn-ss-intF0prime-upperbound1}).  (Note that, in order to apply these results, we must have $\frac{1}{\lambda\eta_c e} \geq \frac{n}{2\pi\beta}$.  Recall that $\lambda = \frac{2\pi\betatilde e}{n-2}$, and $\eta_c = \frac{\delta}{\betatilde} \frac{(n-2)}{e}$.  Also, recall that we assumed $\varepsilon \leq \frac{1}{en}$.  Then $\lambda\eta_c = 2\pi\delta = 2\pi\varepsilon\beta$, and $\frac{1}{\lambda\eta_c e} = \frac{1}{2\pi\beta\varepsilon e} \geq \frac{n}{2\pi\beta}$, as desired.)

After some tedious calculation, we get:
\begin{align}
\Bigl| \int_0^{\eta_c} K_{Br} K_{B1} K_2 \frac{da}{a^{n+1}} \Bigr|
 &\leq C_f^2\varepsilon^{n-1}\beta^{2n-2} \cdot n^2 \cdot 120\varepsilon^2\beta^2, \\
0 \leq \int_0^{\eta_c} K_{Ar} K_{A1} K_2 \frac{da}{a^{n+1}}
 &\leq C_f^2\varepsilon^{n-1}\beta^{2n-2} \cdot \beta^2 (25200\varepsilon^2 + 60\varepsilon + 132), \\
0 \leq \int_0^{\eta_c} K_{Cr} K_{C1} K_2 \frac{da}{a^{n+1}}
 &\leq C_f^2\varepsilon^{n-1}\beta^{2n-2} \cdot 2400(n^2+3n+3)\varepsilon^2\beta^2.
\end{align}
Then, by combining these equations and using our assumption that $\varepsilon \leq O(\frac{1}{n^2})$, we get:
\begin{equation}
\begin{split}
\int_0^{\eta_c} & \Bigl( K_{Ar}K_{A1}K_2 - 2 K_{Br}K_{B1}K_2 + K_{Cr}K_{C1}K_2 \Bigr) \frac{da}{a^{n+1}} \\
 &\leq C_f^2\varepsilon^{n-1}\beta^{2n-2} \cdot \beta^2 (25200\varepsilon^2 + 60\varepsilon + 132 + 2 \cdot 120n^2\varepsilon^2 + 2400(n^2+3n+3)\varepsilon^2) \\
 &\leq C_f^2\varepsilon^{n-1}\beta^{2n-2} \cdot \beta^2 (132 + O(\tfrac{1}{n^2})).
\end{split}
\end{equation}

Next, recall from equation (\ref{eqn-ss-praleqetac-raw}) that:
\begin{equation}
\Pr[a \leq \eta_c] \geq (0.15) S_0 C_f^2 \varepsilon^{n-1} \beta^{2n-2}.
\end{equation}
Finally, by substituting into equation (\ref{eqn-varbpara-conda}) and simplifying, we get a bound on the variance of $\vecb$, in the direction $\vectheta$, conditioned on observing $a \leq \eta_c$:
\begin{equation}
\boxed{
\text{E}((\vecb \cdot \vectheta)^2 \;|\; a \leq \eta_c)
 \leq \beta^2 (23 + O(\tfrac{1}{n^2})).
}
\label{eqn-ss-varbpara-conda}
\end{equation}

}{}

%%%%%%%%%%%%%%%%%%%%%%%%%%%%%%%%%%%%%%%%%%%%%%%%%%%%%%%%%%%%%%%%%%%%%%%%%%%%%%%

\section{A Fast Quantum Curvelet Transform}

\ifthenelse{\boolean{sec6}}{

\subsection{The Discrete Curvelet Transform}

% NOTE:  for discussion of discrete vs continuous transforms, see 3/17/09, 10/13/08, 8/31/08

First, we argue that the discrete Fourier transform approximates the continuous Fourier transform, in the sense described in Section 6.1.  This follows from the definitions of the different transforms.  Recall the continuous Fourier transform that takes a function on $\RR^n$ to a function on $\RR^n$:  
\begin{align}
\mathcal{F}_{cont}(f)(\veck)
 &= \int_{\RR^n} f(\vecx) e^{-2\pi i \veck\cdot\vecx} d\vecx, \\
\mathcal{F}_{cont}^{-1}(g)(\vecx)
 &= \int_{\RR^n} g(\veck) e^{2\pi i \veck\cdot\vecx} d\veck.
\end{align}
Now consider the Fourier transform that takes a function on the cube $C = [-L,L)^n$ (or equivalently, a function on $\RR^n$ that is periodic with respect to the lattice $(2L\ZZ)^n$) to a function on the (dual) lattice $\hat{C} = (\tfrac{1}{2L}\ZZ)^n$.  We refer to this as the ``semi-discrete'' Fourier transform:
\begin{align}
\mathcal{F}_{semi}(f)(\veck)
 &= (\tfrac{1}{2L})^{n/2} \int_C f(\vecx) e^{-2\pi i \veck\cdot\vecx} d\vecx, \\
\mathcal{F}_{semi}^{-1}(g)(\vecx)
 &= (\tfrac{1}{2L})^{n/2} \sum_{\veck\in\hat{C}} g(\veck) e^{2\pi i \veck\cdot\vecx}.
\end{align}
Also recall the discrete Fourier transform, that takes a function on $Z = (\sigma\ZZ)^n \cap [-L,L)^n$ to a function on $\hat{Z} = (\tfrac{1}{2L}\ZZ)^n \cap [-\tfrac{1}{2\sigma}, \tfrac{1}{2\sigma})^n$:
\begin{align}
\mathcal{F}_{dis}(f)(\veck)
 &= (\tfrac{\sigma}{2L})^{n/2} \sum_{\vecx\in Z} f(\vecx) e^{-2\pi i \veck\cdot\vecx}, \\
\mathcal{F}_{dis}^{-1}(g)(\vecx)
 &= (\tfrac{\sigma}{2L})^{n/2} \sum_{\veck\in\hat{Z}} g(\veck) e^{2\pi i \veck\cdot\vecx}.
\end{align}

We are given a function $f_{cont}$ on $\RR^n$ that vanishes outside the cube $C$.  We define a function $f_{semi}$ on $C$ by restriction, $f_{semi} = f_{cont} |_C$.  Then it follows from the definitions that $\hat{f}_{semi} = (\tfrac{1}{2L})^{n/2} \hat{f}_{cont} |_{\hat{C}}$.  

Recall that $\hat{f}_{cont}$ has most of its probability mass inside the cube $[-\tfrac{1}{2\sigma},\tfrac{1}{2\sigma})^n$.  Then the same should be true for $\hat{f}_{semi}$.  Now define a function $\hat{f}_{dis}$ on $\hat{Z}$ by restriction, $\hat{f}_{dis} = \hat{f}_{semi} |_{\hat{Z}}$.  Then, using the definitions, we see that $f_{dis} \approx \sigma^{n/2} f_{semi} |_Z$.

\vskipline

Note that an example of a discrete curvelet transform (over $\RR^2$ or $\RR^3$) can be found in \cite{fast-curvelet-trans}.  There, the frequency space is partitioned into concentric cubes according to the scale $a$, and these are divided into wedges according to the direction $\vectheta$.  For our purposes, we will use a tiling based on concentric balls (in $\RR^n$), which more closely approximates the continuous curvelet transform defined in Section 2.

As a classical computation, the discrete curvelet transform can be implemented using the fast Fourier transform \cite{fast-curvelet-trans} (see in particular the ``wrapping'' method).  We will use these ideas to implement a quantum curvelet transform.  The following discussion will be self-contained; but for readers who are familiar with \cite{fast-curvelet-trans}, we mention that we omit the ``wrapping'' step.  Our transform produces curvelet coefficients that are somewhat oversampled, but this does not cause any problems in our situation.

\subsection{Constructing the Window Functions $\chi_{a,\vectheta}(\veck)$}

We will construct two families of window functions $\chi_{a,\vectheta}(\veck)$, for which the operation $\cal{X}$ can be performed efficiently.  

First, suppose we have some partition of the frequency domain into disjoint subsets, $(\ZZ_M)^n = \bigcup_{a,\vectheta} S_{a,\vectheta}$, such that given any point $\veck \in (\ZZ_M)^n$, we can efficiently compute which set $S_{a,\vectheta}$ contains $\veck$.  Define the window functions to be the indicator functions for these sets, 
\begin{equation}
\chi_{a,\vectheta}(\veck) = \begin{cases}
 1, & \text{if $\veck \in S_{a,\vectheta}$}, \\
 0, & \text{otherwise}.
\end{cases}
\end{equation}
Then the operation $\cal{X}$ can be implemented efficiently:  it simply maps $\ket{\veck} \ket{0,\vec{0}} \mapsto \ket{\veck} \ket{a,\vectheta}$, where $a$ and $\vectheta$ denote the set $S_{a,\vectheta}$ that contains $\veck$.  

Unfortunately, these window functions are sharply discontinuous, so the resulting curvelets are not very well-localized in space.  This makes them poorly suited for the applications proposed in this paper (recall that the results of Sections 3, 4 and 5 required window functions that were $C^1$-smooth).

Smooth window functions are more challenging to implement, because the supports of the functions $\chi_{a,\vectheta}(\veck)$ necessarily overlap.  Thus, at a given point $\veck$, the operation $\cal{X}$ must create a superposition of many values of $a$ and $\vectheta$.  These superpositions can be complicated:  for instance, if we imagine that the tiling of frequency space looks (locally) like an array of $n$-dimensional cubes, then a significant amount of volume lies near the corners of the cubes, and each corner point touches $2^n$ different cubes, so we would have to prepare superpositions of $2^n$ different values of $a$ and $\vectheta$.  This seems impossible for many choices of the window functions.

However, the above example also suggests a solution to the problem.  We can use spherical coordinates, which look locally like Cartesian coordinates (except at the poles).  If we define the window functions to be products of simpler functions, each depending on a single variable, then we can prepare these superpositions efficiently.  We now demonstrate this construction.

First, recall the definition of spherical coordinates in $\RR^n$:  we have $(r,\phi_1,\ldots,\phi_{n-1})$, where $r \in [0,\infty)$, $\phi_1,\ldots,\phi_{n-2} \in [0,\pi] \cup \set{\undef}$, and $\phi_{n-1} \in (-\pi,\pi] \cup \set{\undef}$.  We use the value $\undef$ to represent points on the ``poles'' of the sphere, e.g., if $\phi_j = 0$ or $\pi$, then $\phi_{j+1} = \cdots = \phi_{n-1} = \undef$ (a similar situation arises when $r = 0$).  

Cartesian coordinates are written in terms of spherical coordinates as follows:
\begin{align}
x_1 &= r \cos\phi_1 \quad \text{(or $0$ if undefined)}, \\
x_j &= r \sin\phi_1 \cdots \sin\phi_{j-1} \cos\phi_j 
\quad \text{(or $0$ if undefined)} \quad (j = 2,\ldots,n-1), \\
x_n &= r \sin\phi_1 \cdots \sin\phi_{n-1} \quad \text{(or $0$ if undefined)}.
\end{align}
The reverse mapping is given by:
\begin{align}
r &= \sqrt{x_1^2 + \cdots + x_n^2}, \\
\phi_1 &= \arccos(x_1 / r) \quad \text{(or $\undef$ if $r=0$)}, \\
\phi_j &= \arccos(x_j / (r \sin\phi_1 \cdots \sin\phi_{j-1})) 
\quad \text{(or $\undef$ if $\phi_{j-1} \in \set{0,\pi,\undef}$)} 
\quad (j = 2,\ldots,n-2), \\
\phi_{n-1} &= \text{sign}(x_n) \arccos(x_{n-1} / (r \sin\phi_1 \cdots \sin\phi_{n-2})) 
\quad \text{(or $\undef$ if $\phi_{n-2} \in \set{0,\pi,\undef}$)}.
\end{align}

Next we will define discrete values for the scale variable $a$ and the direction variable $\vectheta$.  In the notation, it will be convenient to represent the scale variable by $s$ instead of $a$, where 
\begin{equation}
a = 2^{-s}.  
\end{equation}
We will then define window functions $\chi_{s,\vectheta}(\veck)$.  These will be products of radial and angular components (we write $\veck = (r,\vecphi)$ using spherical coordinates):
\begin{equation}
\chi_{s,\vectheta}(\veck) = w_s(\lambda r) v_{s,\vectheta}(\vecphi).
\end{equation}
Here, $\lambda$ is a parameter that sets the radial scaling.  For future use, we define the function $c:\: [0,\infty) \rightarrow \RR$, 
\begin{equation}
c(x) = \begin{cases}
 \cos x, & 0 \leq x \leq \pi/2, \\
 0, & x > \pi/2.
\end{cases}
\end{equation}

We begin with the scale variable $a = 2^{-s}$.  We fix the cutoff values $s_{min}, s_{max} \in \ZZ$, where $1 \leq s_{min} \leq s_{max}$.  Then we let $s \in \set{s_{min}, s_{min}+1, \ldots, s_{max}} \cup \set{\coarse,\fine}$.  

We define radial window functions $w_s(r)$ as follows:
\begin{align}
w_s(r) &= \begin{cases}
 c(\tfrac{\pi}{2} (2^s-r)/2^{s-1}), & 2^{s-1} \leq r \leq 2^s, \\
 c(\tfrac{\pi}{2} (r-2^s)/2^s), & 2^s \leq r \leq 2^{s+1}, \\
 0, & \text{otherwise},
\end{cases} \\
w_\coarse(r) &= \begin{cases}
 1, & 0 \leq r \leq 2^{s_{min}-1}, \\
 c(\tfrac{\pi}{2} (r-2^{s_{min}-1})/2^{s_{min}-1}), & 2^{s_{min}-1} \leq r \leq 2^{s_{min}}, \\
 0, & r \geq 2^{s_{min}},
\end{cases} \\
w_\fine(r) &= \begin{cases}
 0, & 0 \leq r \leq 2^{s_{max}}, \\
 c(\tfrac{\pi}{2} (2^{s_{max}+1}-r)/2^{s_{max}}), & 2^{s_{max}} \leq r \leq 2^{s_{max}+1}, \\
 1, & r \geq 2^{s_{max}}.
\end{cases}
\end{align}
An example is shown in Figure \ref{fig-radialwindows}.  It is easy to check that 
\begin{equation}
\sum_s w_s(r)^2 = 1 \quad (\forall r \geq 0).
\end{equation}
(Note that at any given point $r$, at most two of the functions $w_s(r)$ are nonzero.)

\begin{figure}
\centering
\includegraphics[width=\textwidth]{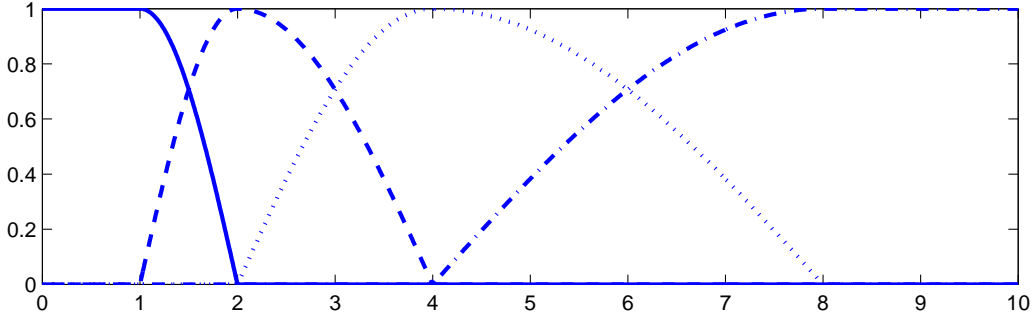}
\caption{A family of radial window functions:  $w_\coarse$, $w_1$, $w_2$ and $w_\fine$.}
\label{fig-radialwindows}
\end{figure}

We let the direction variable $\vectheta$ take on values in the set $G_s(S^{n-1})$.  (Assume for the time being that $s \notin \set{\coarse,\fine}$; we will handle those special cases later.)  The set $G_s(S^{n-1})$ contains grid points on the sphere $S^{n-1}$, defined using spherical coordinates, with angular spacing $\pi/2^{\ceil{s/2}} \approx \pi/\sqrt{2^s} = \pi\sqrt{a}$.  This set is defined recursively:
\begin{align}
G_s(S^1)
 &= \set{\pi t / 2^{\ceil{s/2}} \;|\; t \in \ZZ, \; 0 \leq t \leq 2\cdot 2^{\ceil{s/2}} - 1}, \\
G_s(S^k)
 &= \set{\pi t / 2^{\ceil{s/2}} \;|\; t \in \ZZ, \; 1 \leq t \leq 2^{\ceil{s/2}} - 1} \times G_s(S^{k-1})
  \cup \set{0,\pi} \times \set{\undef} \quad (k \geq 2).
\end{align}
(See Figure \ref{fig-sphere} for an example.)

\begin{figure}
\centering
\includegraphics[scale=0.8]{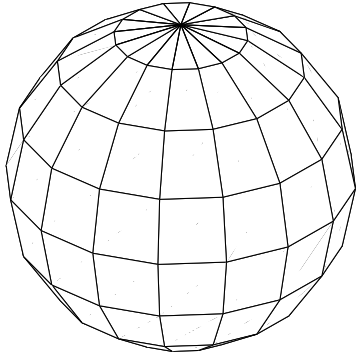}
\caption{$G_6(S^2)$, the set of grid points on the sphere $S^2 \subset \RR^3$, with angular spacing $\pi/8$.}
\label{fig-sphere}
\end{figure}

We define angular window functions $v_{s,\vectheta}(\vecphi)$ as follows:
\begin{equation}
v_{s,\vectheta}(\vecphi) = \prod_{j=1}^{n-1} u_{s,\theta_j}(\phi_j),
\end{equation}
where
\begin{equation}
u_{s,\theta_j}(\phi_j) = c(2^{\ceil{s/2}} |\phi_j-\theta_j| / 2).
\end{equation}
This requires some further explanation.  

Intuitively, $u_{s,\theta_j}(\phi_j)$ is a one-dimensional ``bump'' function centered around $\theta_j$, and $v_{s,\vectheta}(\vecphi)$ is a product of these functions.  

Note that, for the first $n-2$ coordinates $j = 1,\ldots,n-2$, $u_{s,\theta_j}(\phi_j)$ is defined on the interval $[0,\pi]$, whereas for the last coordinate $j = n-1$, $u_{s,\theta_j}(\phi_j)$ is defined on the circle $(-\pi,\pi]$; in this latter case, we interpret $|\phi_j-\theta_j|$ as the shortest-path distance around the circle.  

Also, in the definition of $v_{s,\vectheta}(\vecphi)$, we simply omit those factors that have $\theta_j = \undef$ or $\phi_j = \undef$.  We claim that this is a natural thing to do, in that it yields a simple geometric picture.  Intuitively, $\theta_j = \undef$ means that $\vectheta$ is located on a pole of the sphere, with some coordinate $\theta_i$ ($i<j$) equal to $0$ or $\pi$.  Then this construction produces a bump function that covers a circular region around the pole, and so does not depend on $\phi_j$.  On the other hand, if $\phi_j = \undef$, then $\vecphi$ is located on a pole of the sphere, with some coordinate $\phi_i$ ($i<j$) equal to $0$ or $\pi$.  If $\theta_i \neq \phi_i$, then $u_{s,\theta_i}(\phi_i) = 0$, hence $v_{s,\vectheta}(\vecphi) = 0$, independent of $\phi_j$.  If $\theta_i = \phi_i$, then $\vectheta$ is located on a pole, hence $v_{s,\vectheta}(\vecphi)$ does not depend on $\phi_j$.

Note that at least the first ($j=1$) factor will always be defined, since $\theta_1$ is always defined whenever $\vectheta \in G_s(S^{n-1})$, and $\phi_1$ is always defined whenever $\veck \neq \vec{0}$ (we can ignore the case of $\veck = \vec{0}$, because it is relevant only when $s = \coarse$, in which case we will not use these angular windows).  

Finally, in the special case where $s = \coarse$ or $\fine$, we do not resolve any directions $\vectheta$.  Instead, we fix $\vectheta = \undef$, and we define the angular window to be trivial, $v_{s,\vectheta}(\vecphi) = 1$.

We now show how to perform the operation $\cal{X}$ that maps 
\begin{equation}
\ket{\veck} \ket{0,\vec{0}} \mapsto \ket{\veck} \sum_{s,\vectheta} \chi_{s,\vectheta}(\veck) \ket{s,\vectheta}.  
\end{equation}
We will do this by converting $\veck$ to spherical coordinates $(r,\vecphi)$, performing an operation $\cal{X}'$ that creates the superposition over $s$ and $\vectheta$, then converting back to Cartesian coordinates:
\begin{equation}
\begin{split}
\ket{\veck} \ket{0,\vec{0}} \ket{0,\vec{0}}
 &\mapsto \ket{\veck} \ket{r,\vecphi} \ket{0,\vec{0}} \\
 &\mapsto \ket{\veck} \ket{r,\vecphi}
  \sum_{s,\vectheta} w_s(\lambda r) v_{s,\vectheta}(\vecphi) \ket{s,\vectheta} \\
 &= \ket{\veck} \ket{r,\vecphi}
  \sum_{s,\vectheta} \chi_{s,\vectheta}(\veck) \ket{s,\vectheta} \\
 &\mapsto \ket{\veck} \ket{0,\vec{0}}
  \sum_{s,\vectheta} \chi_{s,\vectheta}(\veck) \ket{s,\vectheta}.
\end{split}
\end{equation}
The operation $\cal{X}'$ is implemented recursively, acting on the variables $r,\phi_1,\ldots,\phi_{n-1}$ one at a time:
\begin{tabbing}
// The ``$s$'' register \\
If $\lambda r < 2^{s_{min}}$, then let $s_1 = \coarse$ and $s_2 = s_{min}$. \\
Else if $\lambda r > 2^{s_{max}}$, then let $s_1 = s_{max}$ and $s_2 = \fine$. \\
Else, let $s_1 = \floor{\lg (\lambda r)}$ and $s_2 = \ceil{\lg (\lambda r)}$. \\
If $s_1 = s_2$, then set the ``$s$'' register to $\ket{s_1}$. \\
Else, set the ``$s$'' register to $w_{s_1}(\lambda r) \ket{s_1} + w_{s_2}(\lambda r) \ket{s_2}$. \\
\\
// The ``$\theta_1$'' register \\
If $s \in \set{\coarse, \fine}$, then set the ``$\theta_1$'' register to $\ket{\undef}$. \\
Else, \= begin: \\
\> Let $\tau_1 = (\pi/2^s) \floor{\phi_1 (2^s/\pi)}$ and $\tau_2 = (\pi/2^s) \ceil{\phi_1 (2^s/\pi)}$. \\
\> If $\tau_1 = \tau_2$, then set the ``$\theta_1$'' register to $\ket{\tau_1}$. \\
\> Else, set the ``$\theta_1$'' register to $u_{s,\tau_1}(\phi_1) \ket{\tau_1} + u_{s,\tau_2}(\phi_1) \ket{\tau_2}$. \\
\> // Note, $\phi_1 \neq \undef$, since otherwise we would have $s = \coarse$ \\
\> // Note, if $\phi_1 \in \set{0,\pi}$, then $\tau_1 = \tau_2$, hence $\theta_1 \in \set{0,\pi}$ \\
End. \\
Recurse on the ``$\theta_2$'' register. \\
\\
// The ``$\theta_\ell$'' register, for $\ell = 2,\ldots,n-2$ \\
If $\theta_{\ell-1} \in \set{0,\pi,\undef}$, then set the ``$\theta_\ell$'' register to $\ket{\undef}$. \\
Else, begin: \\
\> Let $\tau_1 = (\pi/2^s) \floor{\phi_\ell (2^s/\pi)}$ and $\tau_2 = (\pi/2^s) \ceil{\phi_\ell (2^s/\pi)}$. \\
\> If $\tau_1 = \tau_2$, then set the ``$\theta_\ell$'' register to $\ket{\tau_1}$. \\
\> Else, set the ``$\theta_\ell$'' register to $u_{s,\tau_1}(\phi_\ell) \ket{\tau_1} + u_{s,\tau_2}(\phi_\ell) \ket{\tau_2}$. \\
\> // Note, $\phi_\ell \neq \undef$, since otherwise we would have, in some previous iteration $k$, \\
\> // $\phi_k \in \set{0,\pi}$, hence $\theta_k \in \set{0,\pi}$, and $\theta_{\ell-1} \in \set{0,\pi,\undef}$ \\
\> // Note, if $\phi_\ell \in \set{0,\pi}$, then $\tau_1 = \tau_2$, hence $\theta_\ell \in \set{0,\pi}$ \\
End. \\
Recurse on the ``$\theta_{\ell+1}$'' register. \\
\\
// The ``$\theta_{n-1}$'' register \\
If $\theta_{n-2} \in \set{0,\pi,\undef}$, then set the ``$\theta_{n-1}$'' register to $\ket{\undef}$. \\
Else, begin: \\
\> Let $\tau_1 = (\pi/2^s) \floor{\phi_{n-1} (2^s/\pi)}$ and $\tau_2 = (\pi/2^s) \ceil{\phi_{n-1} (2^s/\pi)}$. \\
\> If $\tau_1 = \tau_2$, then set the ``$\theta_{n-1}$'' register to $\ket{\tau_1}$. \\
\> Else, set the ``$\theta_{n-1}$'' register to $\tilde{u}_{s,\tau_1}(\phi_{n-1}) \ket{\tau_1} + \tilde{u}_{s,\tau_2}(\phi_{n-1}) \ket{\tau_2}$. \\
\> // Note, $\phi_{n-1} \neq \undef$, since otherwise we would have, in some previous iteration $k$, \\
\> // $\phi_k \in \set{0,\pi}$, hence $\theta_k \in \set{0,\pi}$, and $\theta_{n-2} \in \set{0,\pi,\undef}$ \\
End. \\
\end{tabbing}

This construction yields a fast quantum curvelet transform using smooth window functions.  Note that we can carry out this construction using other choices of the function $c(x)$, which lead to different window functions $\chi_{s,\vectheta}(\veck)$.  We only need $c(x)$ to satisfy the identity $c(x)^2 + c(\tfrac{\pi}{2}-x)^2 = 1$ (for all $0 \leq x \leq \pi/2$).  

For instance, we can define 
\begin{equation}
c(x) = \begin{cases}
 \cos(h(x)), & 0 \leq x \leq \pi/2, \\
 0, & x > \pi/2,
\end{cases}
\end{equation}
where $h(x)$ is any increasing function that satisfies $h(0) = 0$, $h(\tfrac{\pi}{2}) = \tfrac{\pi}{2}$, and $h(\tfrac{\pi}{2}-x) = \tfrac{\pi}{2}-h(x)$ (for all $0 \leq x \leq \pi/2$).  

In particular, if we set $h(x) = \tfrac{\pi}{2} \sin^2 x$, then the resulting function $c(x)$ is $C^1$-smooth.  Thus we get window functions $\chi_{s,\vectheta}(\veck)$ that are $C^1$-smooth, and are qualitatively similar to the ones used in Sections 3-5 of this paper.  % FIXME:  more details?

}{}

%%%%%%%%%%%%%%%%%%%%%%%%%%%%%%%%%%%%%%%%%%%%%%%%%%%%%%%%%%%%%%%%%%%%%%%%%%%%%%%

\section{Quantum Algorithms using the Curvelet Transform}

\ifthenelse{\boolean{sec7}}{

\subsection{Single-shot measurement of a quantum-sample state}

Here we analyze a ``continuous'' analogue of our single-shot measurement procedure for finding the center of a ball.

First, we claim that $a \leq \eta$ with probability $\geq \Omega(\nu^2)$.  By Theorem \ref{thm-ball}, we have 
\begin{equation}
\Pr[a \leq \eta] \geq (0.19) \eta (1-\tfrac{1}{n}).
\end{equation}
Note that $\eta \geq \Omega(\nu^2)$, which follows from the definition of $\eta$ and the fact that $\mu = \nu\beta \geq \nu\betatilde/2$.  This shows the claim.

From this point on, all probabilities are conditioned on having $a \leq \eta$.

Without loss of generality, assume $\vecc = \vec{0}$.  The algorithm succeeds when it outputs a point close to $\vec{0}$.  Let $\Pi_1$ be the projector onto the subspace orthogonal to $\vectheta$, and let $\Pi_2$ be the projector onto the direction $\vectheta$.  We will show that, with constant probability, $|\Pi_1\vecb|$ is small and $|\Pi_2\vecb|$ is not too large.

Let $X = |\Pi_1\vecb|^2$, $\mu_X = \text{E}(X)$, and let $Y = |\Pi_2\vecb|^2$, $\mu_Y = \text{E}(Y)$.  By Markov's inequality, 
\begin{equation*}
\Pr[X \geq 3\mu_X] \leq \tfrac{1}{3}, \quad
\Pr[Y \geq 3\mu_Y] \leq \tfrac{1}{3}.
\end{equation*}
Then, the union bound implies:
\begin{equation*}
\Pr[X \leq 3\mu_X \text{ and } Y \leq 3\mu_Y] \geq 1 - \Pr[X \geq 3\mu_X] - \Pr[Y \geq 3\mu_Y] \geq \tfrac{1}{3}.
\end{equation*}
So, with probability $\geq 1/3$, we have $X \leq 3\mu_X$ and $Y \leq 3\mu_Y$.  

We now rewrite this in terms of $|\Pi_1\vecb|$ and $|\Pi_2\vecb|$.  From Theorem \ref{thm-ball}, we know that 
\begin{equation*}
\mu_X \leq \eta\beta^2 (14300+(Q_1/n)), \quad
\mu_Y \leq \beta^2 (242+(Q_2/n)), 
\end{equation*}
for some constants $Q_1$ and $Q_2$.  So, we have 
\begin{equation}
|\Pi_1\vecb| \leq \sqrt{3} \sqrt{\eta} \beta \sqrt{14300+(Q_1/n)}, \quad
|\Pi_2\vecb| \leq \sqrt{3} \beta \sqrt{242+(Q_2/n)}.
\end{equation}

This shows that $|\Pi_1\vecb|$ (the error orthogonal to $\vectheta$) is small.  Indeed, substituting in our choice of $\eta$, and using the fact that $\beta/\betatilde \leq 1$, we see that 
\begin{equation}
|\Pi_1\vecb| \leq \tfrac{1}{\sqrt{2}} \mu.
\end{equation}

However, $|\Pi_2\vecb|$ (the error parallel to $\vectheta$) is not so small.  So the algorithm tries to guess this error and output a corrected point.  It succeeds when 
\begin{equation}
\bigl| \Pi_2\vecb - u (\sqrt{3}\betatilde\sqrt{242+(Q_2/n)}) \vectheta \bigr|
 \leq \tfrac{1}{\sqrt{2}} \mu.
\end{equation}
Call this event $E$.  The probability of $E$ is the probability that a random point chosen uniformly from the interval $\sqrt{3}\betatilde\sqrt{242+(Q_2/n)} \cdot [-1,1]$ lies within distance $\tfrac{1}{\sqrt{2}} \mu$ of some fixed point in the (possibly smaller) interval $\sqrt{3}\beta\sqrt{242+(Q_2/n)} \cdot [-1,1]$.  This probability is lower-bounded by 
\begin{equation}
\Pr[E]
 \geq \frac{\frac{1}{\sqrt{2}} \mu}{2\sqrt{3}\betatilde\sqrt{242+(Q_2/n)}}
 = \frac{1}{2\sqrt{6}\sqrt{242+(Q_2/n)}} \frac{\mu}{\betatilde}, 
\end{equation}
which is $\geq \Omega(\nu)$, since $\mu = \nu\beta \geq \nu\betatilde/2$.

So the algorithm succeeds with probability $\geq \Omega(\nu^3)$.  This proves Theorem \ref{thm-ball-alg}.  $\square$

\vskipline

Finally, we argue that, when the grid $G = (\sigma\ZZ)^n \cap [-L,L)^n$ is chosen properly, the discrete algorithm will behave like the continuous one.  Recall from Section 6 that we can approximate a function $f$ on $\RR^n$ with a function $f_2$ on $G$, provided that most of the probability mass of $f$ lies within distance $L$ of the origin, and most of the probability mass of $\hat{f}$ lies within distance $\frac{1}{2\sigma}$ of the origin.  This holds for our algorithm, provided that:
\begin{equation}
R+\beta \leq L, \quad
\frac{100}{\lambda\eta} \leq \frac{1}{2\sigma}.
\end{equation}
(The first condition follows immediately, since $f$ is supported on a ball.  The second condition follows from the decay of $\hat{f}$; details omitted.)  These conditions hold whenever 
\begin{equation}
L \geq R+\betatilde, \quad
\sigma \leq \frac{\pi e}{600} \frac{\mu^2}{\betatilde} \frac{1}{14300n+Q_1}.
\end{equation}

Also, we argue that the discretization of the ``direction'' variable $\vectheta$ will not introduce too much error in the output of the algorithm.  Our algorithm constructs a line $\ell = \set{\vecb+\lambda\vectheta \;|\; \lambda\in\RR}$, and if $\vectheta$ were a continuous variable, this line would pass within distance $O(\sqrt{a}\beta)$ of the center $\vecc$.  Recall from Section 6 that the discrete curvelet transform resolves $\vectheta$ within error $\pm \sqrt{a}$ in angular distance.  Since $\vecb$ lies at distance $O(\beta)$ from the center $\vecc$, the error in $\vectheta$ can increase the distance from $\ell$ to $\vecc$ by at most $O(\sqrt{a}\beta)$.  Thus the error in the output of the algorithm increases by at most a constant factor.

We can bound the running time of our algorithm as follows.  Let $M$ be the number of grid points along one direction.  Then $M = 2L/\sigma$, and the running time is $\leq \poly(n, \log M)$.  Say we choose $L$ and $\sigma$ so that the above inequalities are tight (up to constant factors).  Then $M \leq O(R\betatilde n / \mu^2)$, and the running time is $\leq \poly(n, \log R, \log\betatilde, \log\frac{1}{\mu})$.  

\subsection{Quantum algorithm for finding the center of a radial function}

Here we analyze the ``continuous'' analogue of our algorithm for finding the center of a radial function.

First, consider what happens for each $i \in \set{1,2}$.

We quantum-sample over a ball of radius $R'$ around $\vec{0}$, then measure the value of $f$, and get a superposition over a shell of radius $\beta^{(i)}$ around $\vecc$.  If the shell is too large, it will not lie completely within our original ball, so we only get a fragment of the shell.  However, if $\beta^{(i)} \leq R'-R$, then we are guaranteed to get a complete shell.  

We claim that we observe $\beta^{(i)} \leq R'-R$ with constant probability.  To see this, write:
\begin{equation}
\begin{split}
\Pr[\beta^{(i)} \leq R'-R]
 &= \frac{\text{volume of ball of radius $R'-R$ around $\vecc$}}{\text{volume of ball of radius $R'$ around $\vec{0}$}} \\
 &= \frac{(R'-R)^n}{(R')^n} = (1-\tfrac{1}{n})^n \geq e^{-1.2} > 0.30, 
\end{split}
\end{equation}
using the fact that $1-x \geq e^{-(1.2)x}$ for all $0 \leq x \leq 1/4$ (recall that we assumed $n \geq 4$).  

Our algorithm does not know the shell's true radius $\beta^{(i)}$, so it uses $\betatilde^{(i)} = R'-R$ as an estimate.  We claim that $\beta^{(i)} \leq \betatilde^{(i)} \leq (3/2)\beta^{(i)}$, with constant probability.  Observe that 
\begin{equation}
\begin{split}
\Pr[\beta^{(i)} < (2/3)(R'-R)]
 &= \frac{\text{volume of ball of radius $(2/3)(R'-R)$ around $\vecc$}}{\text{volume of ball of radius $R'$ around $\vec{0}$}} \\
 &= \frac{((2/3)(R'-R))^n}{(R')^n} = (2/3)^n (1-\tfrac{1}{n})^n.
\end{split}
\end{equation}
So 
\begin{equation}
\begin{split}
\Pr[(2/3)(R'-R) \leq \beta^{(i)} \leq R'-R]
 &= (1-(\tfrac{2}{3})^n) (1-\tfrac{1}{n})^n \\
 &\geq (1-0.20) (0.30) = 0.24, 
\end{split}
\end{equation}
using the fact that $n \geq 4$.

Next, we claim that we observe $a^{(i)} \leq \eta^{(i)}$, with constant probability.  This follows from Theorem \ref{thm-sphericalshell}:
\begin{equation}
\Pr[a^{(i)} \leq \eta^{(i)}] > 0.045.
\end{equation}

From this point on, we take probabilities conditioned on $a^{(i)} \leq \eta^{(i)}$.

Without loss of generality, assume $\vecc = \vec{0}$.  Let $\Pi_1^{(i)}$ be the projector onto the subspace orthogonal to $\vectheta^{(i)}$, and let $\Pi_2^{(i)}$ be the projector onto the direction $\vectheta^{(i)}$.  

We claim that $|\Pi_1^{(i)} \vecb^{(i)}|$ is small, and $|\Pi_2^{(i)} \vecb^{(i)}|$ is of order $\beta$, with constant probability.  We use the same argument as in the previous section, together with Theorem \ref{thm-sphericalshell}.  We define $\varepsilon^{(i)} = \delta/\beta^{(i)}$.  We get that, with probability $\geq 1/3$, 
\begin{equation}
|\Pi_1^{(i)} \vecb^{(i)}| \leq \sqrt{3} \sqrt{(n-1) \varepsilon^{(i)}} \beta^{(i)} \sqrt{761+(Q_1/n)}, \quad
|\Pi_2^{(i)} \vecb^{(i)}| \leq \sqrt{3} \beta^{(i)} \sqrt{23+(Q_2/n^2)}, 
\end{equation}
for some constants $Q_1$ and $Q_2$.

Using the definition of $\varepsilon^{(i)}$, and the fact that $\beta^{(i)} \leq R'-R = (n-1)R$, this implies that 
\begin{equation}
|\Pi_1^{(i)} \vecb^{(i)}| \leq \sqrt{3} (n-1)\sqrt{\delta R} \sqrt{761+(Q_1/n)}, \quad
|\Pi_2^{(i)} \vecb^{(i)}| \leq \sqrt{3} (n-1)R \sqrt{23+(Q_2/n^2)}.
\end{equation}

The algorithm carries out this procedure twice, for $i = 1$ and $2$.  With constant probability, this produces two lines, $\ell_1 = \set{\vecb^{(1)} + \lambda \vectheta^{(1)} \;|\; \lambda \in \RR}$ and $\ell_2 = \set{\vecb^{(2)} + \lambda \vectheta^{(2)} \;|\; \lambda \in \RR}$, which both pass near the point $\vecc$.  The algorithm then checks that these lines are nearly orthogonal, and if they are, it returns the point on $\ell_1$ closest to $\ell_2$.  A straightforward calculation shows that this point is given by $\frac{-s+rt}{1-r^2} \vectheta^{(1)} + \vecb^{(1)}$.

First, we claim that the lines $\ell_1$ and $\ell_2$ are nearly orthogonal ($|\vectheta^{(1)} \cdot \vectheta^{(2)}| \leq 3/4$) with at least constant probability.  

We want to upper-bound the probability that $|\vectheta^{(1)} \cdot \vectheta^{(2)}| > 3/4$.  Recall that these are independent random vectors, chosen uniformly from the unit sphere $S^{n-1}$ in $\RR^n$.  It follows that 
\begin{equation}
\Pr[|\vectheta^{(1)} \cdot \vectheta^{(2)}| > 3/4] = \Pr[|x_1| > 3/4], 
\end{equation}
where $\vecx = (x_1,\ldots,x_n)$ is a random vector chosen uniformly from $S^{n-1}$.  Note that $\text{E}(\vecx) = \vec{0}$, hence $\text{E}(x_1) = 0$; also, $\text{E}(|\vecx|^2) = 1$, hence $\text{E}(x_1^2) = 1/n$.  Then by Markov's inequality, 
\begin{equation}
\Pr[|x_1| > 3/4] = \Pr[x_1^2 > 9/16] \leq \tfrac{16}{9} \tfrac{1}{n} \leq \tfrac{4}{9} \text{ (for $n \geq 4$)}.
\end{equation}
Thus, we observe $|\vectheta^{(1)} \cdot \vectheta^{(2)}| \leq 3/4$, with probability $\geq 5/9$.  (This is a rather weak bound, especially when $n$ is large, but it is adequate for our purposes.  Actually, it is the case that $|\vectheta^{(1)} \cdot \vectheta^{(2)}| \leq O(1/\sqrt{n})$, with probability $\geq \Omega(1)$.)  

Next, we claim that when $\ell_1$ and $\ell_2$ are nearly orthogonal ($|\vectheta^{(1)} \cdot \vectheta^{(2)}| \leq 3/4$), the point on $\ell_1$ closest to $\ell_2$ (call it $p_1$) is close to $\vecc$.  

Let $q_1$ be the point on $\ell_1$ closest to $\vecc$, and let $p_1$ be the point on $\ell_1$ closest to $\ell_2$.  Similarly, let $q_2$ be the point on $\ell_2$ closest to $\vecc$, and let $p_2$ be the point on $\ell_2$ closest to $\ell_1$.  (See Figure \ref{fig-two-lines}.)  
\begin{figure}
\input{two-lines.pst}
\caption{}
\label{fig-two-lines}
\end{figure}

We know that $q_1$ and $q_2$ are both close to $\vecc$:  $|q_1-\vecc| \leq \Delta$, and $|q_2-\vecc| \leq \Delta$, where $\Delta = \sqrt{3} (n-1)\sqrt{\delta R} \sqrt{761+(Q_1/n)}$.  Furthermore, $p_1$ and $p_2$ are close together:  $|p_1-p_2| \leq 2\Delta$.

Now suppose that $p_1$ is far from $\vecc$:
\begin{equation}
|p_1-\vecc| \geq 8\Delta.
\label{eqn-triangle-suppose}
\end{equation}
From this we will derive a contradiction.  

First, note that $p_1$ is far from $q_1$:
\begin{equation}
|p_1-q_1| \geq 7\Delta.
\label{eqn-triangle-1}
\end{equation}

Consider the line $\ell_2' = \ell_2 + (p_1-p_2)$.  Define another point $q_2' = q_2 + (p_1-p_2)$.  The line $\ell_2'$ is parallel to $\ell_2$, it intersects $\ell_1$ at $p_1$, and it passes through $q_2'$.  Note that $q_2'$ is close to $\vecc$:  $|q_2'-\vecc| \leq 3\Delta$.

Note that $p_1$ is far from $q_2'$:
\begin{equation}
|p_1-q_2'| \geq 5\Delta.
\label{eqn-triangle-2}
\end{equation}

Note that $q_1$ and $q_2'$ are close to each other:
\begin{equation}
|q_1-q_2'| \leq 4\Delta.
\label{eqn-triangle-3}
\end{equation}

We will use equations (\ref{eqn-triangle-1}), (\ref{eqn-triangle-2}) and (\ref{eqn-triangle-3}) to show that the angle between $\ell_1$ and $\ell_2'$ is small.  (See Figure \ref{fig-two-lines}.)  We have $a \leq 4\Delta$, $b \geq 7\Delta$ and $c \geq 5\Delta$.

Using the law of cosines, and the fact that $x + \frac{1}{x} \geq 2$ for all $x>0$, we get:
\begin{equation}
\begin{split}
|\vectheta^{(1)} \cdot \vectheta^{(2)}|
 &= \cos \varphi = \frac{c^2+b^2-a^2}{2bc} \\
 &= \frac{1}{2} \Bigl( \frac{c}{b} + \frac{b}{c} - \frac{a^2}{bc} \Bigr) \\
 &\geq 1 - \frac{a^2}{2bc} \\
 &\geq 1 - \frac{(4\Delta)^2}{2(7\Delta)(5\Delta)} = 1 - \tfrac{8}{35} > \tfrac{3}{4}.
\end{split}
\end{equation}
This contradicts our assumption that $\ell_1$ and $\ell_2$ are nearly orthogonal.  

So we conclude that $p_1$ is close to $\vecc$, as desired:
\begin{equation}
|p_1-\vecc| \leq 8\Delta = 8 \sqrt{3} (n-1)\sqrt{\delta R} \sqrt{761+(Q_1/n)}.
\end{equation}

Finally, using our assumed upper bound on $\delta$, we get that:
\begin{equation}
|p_1-\vecc| \leq \mu.
\end{equation}
So the algorithm succeeds with constant probability.  This proves Theorem \ref{thm-crfn-alg}.  $\square$

\vskipline

Finally, we argue that, when the grid $G = (\sigma\ZZ)^n \cap [-L,L)^n$ is chosen properly, the discrete algorithm will behave like the continuous one.  Recall from Section 6 that we can approximate a function $f$ on $\RR^n$ with a function $f_2$ on $G$, provided that most of the probability mass of $f$ lies within distance $L$ of the origin, and most of the probability mass of $\hat{f}$ lies within distance $\frac{1}{2\sigma}$ of the origin.  This holds for our algorithm, provided that:
\begin{equation}
R' \leq L, \quad
\frac{100}{\delta} \leq \frac{1}{2\sigma}.
\end{equation}
(The first condition follows immediately, since $f$ is supported on a ball.  The second condition follows from the decay of $\hat{f}$; details omitted.)

Also, we argue that the discretization of the ``direction'' variable $\vectheta$ will not introduce too much error in the output of the algorithm.  The key part of our algorithm involves constructing a line $\ell = \set{\vecb+\lambda\vectheta \;|\; \lambda\in\RR}$; and if $\vectheta$ were a continuous variable, this line would pass within distance $O(\sqrt{a}\beta)$ of the center $\vecc$.  Recall from Section 6 that the discrete curvelet transform resolves $\vectheta$ within error $\pm \sqrt{a}$ in angular distance.  Since $\vecb$ lies at distance $O(\beta)$ from the center $\vecc$, the error in $\vectheta$ can increase the distance from $\ell$ to $\vecc$ by at most $O(\sqrt{a}\beta)$.  Thus the error in the output of the algorithm increases by at most a constant factor.

We can bound the running time of our algorithm as follows.  Say we choose $L$ and $\sigma$ so that the above inequalities are tight (up to constant factors).  Also, suppose $\delta$ satisfies equation (\ref{eqn-alg2-delta}) exactly up to constant factors.  Let $M$ be the number of grid points along one direction.  Then $M = 2L/\sigma \leq O(R^2 n^3 / \mu^2)$, and the running time is $\leq \poly(n, \log M) \leq \poly(n, \log R, \log\frac{1}{\mu})$.  

\subsection{Classical lower bound}

We will prove a lower bound on classical randomized algorithms for finding the center of a radial function.  Recall that an instance of the problem is specified by an oracle $f$ and parameters $n$, $R$, $\delta$ and $\mu$.  Let us define $m = \lg(R/\mu)$.  Consider algorithms that query points within some finite subset $D \subset \RR^n$.  We will show that $\Omega(nm/\lg(nm))$ queries are needed to solve this problem.
\begin{thm}
The following holds for any parameters $n$, $R$, $\delta$ and $\mu$, and for any finite subset $D \subset \RR^n$.  Let us define $m = \lg(R/\mu)$.  For any classical randomized algorithm that queries points within the set $D$ and uses at most $(\tfrac{1}{2}nm)/\lg(\tfrac{1}{2}nm)$ oracle queries, there exists a problem instance (depending on $D$, and having the specified parameters $n$, $R$, $\delta$ and $\mu$) that causes the algorithm to fail with probability at least $1-2^{-nm/2}$.
\end{thm}

The assumption that the algorithm queries points within the set $D$ can be understood intuitively as follows.  We are assuming that the algorithm follows some (arbitrary) convention for how it describes points in $\RR^n$ when it queries the oracle.  We let $D \subset \RR^n$ be the set of points that can be queried, and note that $D$ must be finite:  if the algorithm runs in time $T$, then clearly $|D| \leq 2^T$.

Note that this assumption does not weaken our lower bound.  The assumption is needed because the construction of the hard instance depends on $D$; however, the actual lower bound (i.e., the number of oracle queries and the resulting probability of success) is independent of $D$.  

\vskipline

\noindent
Proof:  We will use the following version of Yao's minimax lemma, due to Rademacher and Vempala \cite{Rademacher-dispersion-06}:
\begin{lem}
Let $\calI$ be a set of problem instances and $\calA$ be a set of deterministic algorithms.  For any probability measure $\pi$ over $\calI$, and any probability measure $\nu$ over $\calA$, we have that 
\begin{equation}
\inf_{A\in\calA} \Pr_{I\sim\pi(\calI)} 
[\text{algorithm $A$ fails on instance $I$}] 
\leq \sup_{I\in\calI} \Pr_{A\sim\nu(\calA)} 
[\text{algorithm $A$ fails on instance $I$}].
\end{equation}
\end{lem}
We will use the following approach.  First we will fix a probability distribution $\pi$ over instances, and prove that the best deterministic algorithm still fails with high probability.  Note that a randomized algorithm is simply a probability distribution over deterministic algorithms; so the minimax lemma implies that for any randomized algorithm $\nu$, there exists an instance that causes the algorithm to fail with high probability.

We let $\calA$ be the set of deterministic algorithms that query points within the subset $D$ and make at most $\ell$ queries, for some $\ell$ to be specified later.  

We let $\calI$ be the set of problem instances, such that $n$, $R$, $\delta$ and $\mu$ are fixed, but $f$ and $\vecc$ can vary.  That is, we fix the dimension $n$, the radius $R$ of the ball in which the center lies, the thickness $\delta$ of the spherical shells, and the desired accuracy $\mu$.  We define $m = \lg(R/\mu)$, which is fixed.  However, the values returned by the radial function $f$, and the location of its center $\vecc$, are arbitrary.  

We now fix a distribution $\pi$ on problem instances.  Random instances according to this distribution are constructed as follows.  First, we choose the center point $\vecc$ uniformly at random from the ball of radius $R$ around the origin.  Let $f$ be of the form 
\begin{equation}
f(\vecx) = [s_k \text{ if } \vecx \in A_k], 
\text{ where } A_k = \set{\vecx \;|\; k\delta \leq |\vecx-\vecc| < (k+1)\delta}, 
\text{ for } k = 0,1,2,\ldots.  
\end{equation}
Next, we choose the values of $f$ on the points in the set $D$ (these are the points that the algorithm can query).  Equivalently, we will choose the values of $s_k$, for those $k$ such that $A_k \cap D \neq \emptyset$.  Let $K$ be the set of those $k$, and note that $|K| \leq |D|$.  Let $S = \set{1,2,\ldots,|D|}$.  Choose a random injective map $\sigma:\: K \hookrightarrow S$, and then set $s_k = \sigma(k)$.  

Consider any deterministic algorithm $A \in \calA$, and let $U$ be the set of possible outputs of the algorithm.  We claim that $|U| \leq 2^{\ell\lg\ell}$.  To see this, note that the algorithm $A$ can be described as a decision tree, where each node represents a query to the oracle $f$, and the algorithm chooses which branch to follow depending on the oracle's answer.  We claim that after seeing the answer to its $k$'th query, the algorithm can have at most $k$ distinct branches.  This is because, while the oracle can return many different values, they are meaningless except in cases where the oracle returns the same value as it did for a previous query.  (Note that, for any permutation $\sigma$ on the range of $f$, we can replace the oracle $f$ with $\sigma \circ f$, to get a new instance of the problem that has the same desired solution.  These two instances occur with equal probability under the distribution $\pi$.)  So, when the algorithm receives the answer to its $k$'th query, all it can do is compare that value to the answers to its previous queries.  The number of branches is at most the number of distinct values that have been seen previously (which is at most $k-1$), plus 1 (if the new value does not match any of the previous ones); thus the number of branches is at most $k$.  Finally, note that the size of $U$ is at most the number of leaves at the final level of the decision tree, i.e., after the $\ell$'th query.  So $|U| \leq 1\cdot 2\cdot 3\cdots \ell \leq \ell^\ell = 2^{\ell\lg\ell}$.

Thus we can upper-bound the probability that algorithm $A$ succeeds on a random instance $I \sim \pi(\calI)$:
\begin{equation}
\begin{split}
\Pr_{I \sim \pi(\calI)} &[\text{algorithm $A$ succeeds on instance $I$}] \\
 &\leq \Pr_{\vecc} [\exists \vec{z} \in U \text{ s.t. } |\vec{z}-\vecc| \leq \mu]
  \leq \sum_{\vec{z} \in U} \Pr_{\vecc} [|\vec{z}-\vecc| \leq \mu] \\
 &\leq \sum_{\vec{z} \in U} \frac{\text{volume of ball of radius $\mu$}}{\text{volume of ball of radius $R$}}
  = |U| \Bigl(\frac{\mu}{R}\Bigr)^n \\
 &\leq 2^{\ell\lg\ell} 2^{-mn}.
\end{split}
\end{equation}
Now suppose that $\ell$, the number of queries, is at most $(\tfrac{1}{2}nm)/\lg(\tfrac{1}{2}nm)$.  Then 
\begin{equation}
\ell\lg\ell
 \leq \frac{\tfrac{1}{2}nm}{\lg(\tfrac{1}{2}nm)} (\lg(\tfrac{1}{2}nm) - \lg\lg(\tfrac{1}{2}nm))
 \leq \tfrac{1}{2}nm \text{ (assuming $nm \geq 4$)}.
\end{equation}
So we have 
\begin{equation}
\Pr_{I \sim \pi(\calI)} [\text{algorithm $A$ succeeds on instance $I$}]
 \leq 2^{-nm/2}, 
\end{equation}
and this holds for all algorithms $A \in \calA$.  So 
\begin{equation}
\inf_{A \in \calA} \Pr_{I \sim \pi(\calI)} [\text{algorithm $A$ fails on instance $I$}]
 \geq 1 - 2^{-nm/2}.
\end{equation}
Now plug into the minimax lemma and the result follows.  $\square$

\subsection{Finding the center through multiple iterations}

Here we analyze the ``continuous'' analogue of our algorithm for finding the center of a radial function.

First, we analyze the procedure OneRound().  Consider what happens for each $i \in \set{1,2}$.

We quantum-sample over a ball of radius $R'$ around $\vec{0}$, then measure the value of $f$, and get a superposition over a shell of radius $\beta^{(i)}$ around $\vecc$.  If the shell is too large, it will not lie completely within our original ball, so we only get a fragment of the shell.  However, if $\beta^{(i)} \leq R'-R$, then we are guaranteed to get a complete shell.  

We claim that we observe $\beta^{(i)} \leq R'-R$ with probability $\geq 1-O(\frac{1}{S})$.  To see this, write:
\begin{equation}
\begin{split}
\Pr[\beta^{(i)} \leq R'-R]
 &= \frac{\text{volume of ball of radius $R'-R$ around $\vecc$}}{\text{volume of ball of radius $R'$ around $\vec{0}$}} \\
 &= \frac{(R'-R)^n}{(R')^n} = (1-\tfrac{1}{nS})^n \geq e^{-(1.2)/S} \geq 1-\tfrac{1.2}{S}, 
\end{split}
\end{equation}
using the fact that $1-x \geq e^{-(1.2)x}$ for all $0 \leq x \leq 1/4$ (recall that we assumed $n \geq 4$).  

Our algorithm does not know the shell's true radius $\beta^{(i)}$, so it uses $\betatilde^{(i)} = R'-R$ as an estimate.  We claim that $\beta^{(i)} \leq \betatilde^{(i)} \leq S\beta^{(i)}$, with probability $\geq 1-O(\frac{1}{S})$.  Observe that 
\begin{equation}
\begin{split}
\Pr[\beta^{(i)} < (1/S)(R'-R)]
 &= \frac{\text{volume of ball of radius $(1/S)(R'-R)$ around $\vecc$}}{\text{volume of ball of radius $R'$ around $\vec{0}$}} \\
 &= \frac{((1/S)(R'-R))^n}{(R')^n} = (1/S)^n (1-\tfrac{1}{nS})^n.
\end{split}
\end{equation}
So 
\begin{equation}
\begin{split}
\Pr[(1/S)(R'-R) \leq \beta^{(i)} \leq R'-R]
 &= (1-(\tfrac{1}{S})^n) (1-\tfrac{1}{nS})^n \\
 &\geq (1-\tfrac{1}{S}) (1-\tfrac{1.2}{S}) > 1-\tfrac{2.2}{S}.
\end{split}
\end{equation}

Next, we claim that we observe $a^{(i)} \leq \eta^{(i)}$, with constant probability.  This follows from Theorem \ref{thm-sphericalshell}:
\begin{equation}
\Pr[a^{(i)} \leq \eta^{(i)}] > 0.045.
\end{equation}

From this point on, we take probabilities conditioned on $a^{(i)} \leq \eta^{(i)}$.

Without loss of generality, assume $\vecc = \vec{0}$.  Let $\Pi_1^{(i)}$ be the projector onto the subspace orthogonal to $\vectheta^{(i)}$, and let $\Pi_2^{(i)}$ be the projector onto the direction $\vectheta^{(i)}$.  

We claim that $|\Pi_1^{(i)} \vecb^{(i)}|$ is small, and $|\Pi_2^{(i)} \vecb^{(i)}|$ is of order $\beta$, with probability $\geq 1-O(\tfrac{1}{S})$.  We use a similar argument as in the previous section, together with Theorem \ref{thm-sphericalshell}.  We define $\varepsilon^{(i)} = \delta/\beta^{(i)}$.  We get that, with probability $\geq 1-\tfrac{2}{S}$, 
\begin{equation}
|\Pi_1^{(i)} \vecb^{(i)}| \leq \sqrt{S} \sqrt{(n-1) \varepsilon^{(i)}} \beta^{(i)} \sqrt{(507+\tfrac{Q_1}{n}) \cdot S}, \quad
|\Pi_2^{(i)} \vecb^{(i)}| \leq \sqrt{S} \beta^{(i)} \sqrt{23+\tfrac{Q_2}{n^2}}, 
\end{equation}
for some constants $Q_1$ and $Q_2$.

Using the definition of $\varepsilon^{(i)}$, and the fact that $\beta^{(i)} \leq R'-R = (nS-1)R < nSR$, this implies that 
\begin{equation}
|\Pi_1^{(i)} \vecb^{(i)}| \leq S^{3/2} n \sqrt{\delta R} \sqrt{507+\tfrac{Q_1}{n}}, \quad
|\Pi_2^{(i)} \vecb^{(i)}| \leq S^{3/2} n R \sqrt{23+\tfrac{Q_2}{n^2}}.
\end{equation}

The algorithm carries out this procedure twice, for $i = 1$ and $2$.  With probability $\geq 1-O(\tfrac{1}{S})$, this produces two lines, $\ell_1 = \set{\vecb^{(1)} + \lambda \vectheta^{(1)} \;|\; \lambda \in \RR}$ and $\ell_2 = \set{\vecb^{(2)} + \lambda \vectheta^{(2)} \;|\; \lambda \in \RR}$, which both pass near the point $\vecc$.  The algorithm then checks that these lines are nearly orthogonal, and if they are, it returns the point on $\ell_1$ closest to $\ell_2$.  A straightforward calculation shows that this point is given by $\frac{-s+rt}{1-r^2} \vectheta^{(1)} + \vecb^{(1)}$.

First, we claim that the lines $\ell_1$ and $\ell_2$ are nearly orthogonal ($|\vectheta^{(1)} \cdot \vectheta^{(2)}| \leq 3/4$) with probability $\geq 5/9$.  This follows from the same argument as in the previous section.

Next, we claim that when $\ell_1$ and $\ell_2$ are nearly orthogonal ($|\vectheta^{(1)} \cdot \vectheta^{(2)}| \leq 3/4$), the point on $\ell_1$ closest to $\ell_2$ (call it $p_1$) is close to $\vecc$.  Using the same argument as in the previous section, we conclude that:
\begin{equation}
|p_1-\vecc| \leq 8 S^{3/2} n \sqrt{\delta R} \sqrt{507+\tfrac{Q_1}{n}}.
\end{equation}

Finally, using our assumed upper bound on $\delta$, we get that:
\begin{equation}
|p_1-\vecc| \leq \sqrt{R} \sqrt{\mu/2}.
\end{equation}

In summary, we have shown that the procedure OneRound() has the following two properties:  (1) OneRound() returns a point $\vec{q}$ (rather than ``no answer'') with probability 
\begin{equation}
  \geq (0.045)^2 \cdot \tfrac{5}{9}
  > 0.0011,
\end{equation}
and (2) when OneRound() returns a point $\vec{q}$, that point $\vec{q}$ lies within distance $\sqrt{R} \sqrt{\mu/2}$ of the center point $\vecc$, with probability 
\begin{equation}
  \geq \Bigl( (1-\tfrac{2.2}{S}) (1-\tfrac{2}{S}) \Bigr)^2
  > (1-\tfrac{4.2}{S})^2
  > 1-\tfrac{8.4}{S}.
\end{equation}

\vskipline

We now analyze the complete algorithm, consisting of multiple iterations.  

First, let us upper-bound the number of iterations, assuming that every iteration is successful.  Let $k$ denote the number of iterations.  We claim that $k \leq \lceil \lg\lg \tfrac{2R}{\mu} \rceil$.  

Let $R_i$ denote the value of $R_{cur}$ following the $i$'th iteration, so we have $R_0 = R$ (when the algorithm starts), $R_{i+1} = \sqrt{R_i} \sqrt{\mu/2}$ (the recurrence relation), and $R_k \leq \mu$ (when the algorithm finishes).  

Let us define $\tilde{R}_i = \frac{R_i}{\mu/2}$.  Then we have $\tilde{R}_0 = \frac{R}{\mu/2}$ (when the algorithm starts), $\tilde{R}_{i+1} = \sqrt{\tilde{R}_i}$  (the recurrence relation), and $\tilde{R}_k \leq 2$ (when the algorithm finishes).  

It is easy to see that $\tilde{R}_k = (\tilde{R}_0)^{(1/2)^k}$, and a straightforward calculation shows that it suffices to set $k = \lceil \lg\lg \frac{2R}{\mu} \rceil$.

Next, we show that the algorithm succeeds (i.e., every iteration is successful) with constant probability.  First, consider a single iteration.  The algorithm makes $n_{tries}$ attempts to run OneRound(), and it succeeds if at least one of those attempts returns a point $\vec{q}$ that lies near the center.  The probability that OneRound() returns ``no answer'' every time is $\leq (0.9989)^{n_{tries}} = (0.9989)^{910 \log S} \leq 1/S$.  So the probability that OneRound() returns a point $\vec{q}$ at least once is $\geq 1-\frac{1}{S}$.  When this happens, the point $\vec{q}$ is near the center with probability $\geq 1-\frac{8.4}{S}$.  So the iteration succeeds with overall probability $\geq (1-\frac{1}{S}) (1-\frac{8.4}{S}) > 1-\frac{9.4}{S} = 1-\frac{1}{n_{iter}}$.  

The algorithm makes $n_{iter}$ iterations, and it succeeds if all of the iterations succeed.  This occurs with probability $\geq (1-\frac{1}{n_{iter}})^{n_{iter}} \geq 1/e^2$.  (We assumed that $R \geq 8\mu$, so $n_{iter} \geq 2$, and we used the fact that $1-x \geq e^{-2x}$ for all $0 \leq x \leq 1/2$.)  So the algorithm succeeds with constant probability.

Finally, note that the number of oracle queries is $2 n_{tries} n_{iter} \leq O(\lg\lg \frac{2R}{\mu} \lg\lg\lg \frac{2R}{\mu})$.  This proves Theorem \ref{thm-crfn-alg2}.  $\square$

\vskipline

Finally, we argue that, when the grid $G = (\sigma\ZZ)^n \cap [-L,L)^n$ is chosen properly, the discrete algorithm will behave like the continuous one.  Recall from Section 6 that we can approximate a function $f$ on $\RR^n$ with a function $f_2$ on $G$, provided that most of the probability mass of $f$ lies within distance $L$ of the origin, and most of the probability mass of $\hat{f}$ lies within distance $\frac{1}{2\sigma}$ of the origin.  This holds for our algorithm, provided that:
\begin{equation}
R' \leq L, \quad
\frac{100}{\delta} \leq \frac{1}{2\sigma}.
\end{equation}
(The first condition follows immediately, since $f$ is supported on a ball.  The second condition follows from the decay of $\hat{f}$; details omitted.)

Also, we argue that the discretization of the ``direction'' variable $\vectheta$ will not introduce too much error in the output of the algorithm.  The key part of our algorithm involves constructing a line $\ell = \set{\vecb+\lambda\vectheta \;|\; \lambda\in\RR}$; and if $\vectheta$ were a continuous variable, this line would pass within distance $O(\sqrt{a}\beta)$ of the center $\vecc$.  Recall from Section 6 that the discrete curvelet transform resolves $\vectheta$ within error $\pm \sqrt{a}$ in angular distance.  Since $\vecb$ lies at distance $O(\beta)$ from the center $\vecc$, the error in $\vectheta$ can increase the distance from $\ell$ to $\vecc$ by at most $O(\sqrt{a}\beta)$.  Thus the error in the output of the algorithm increases by at most a constant factor.

We can bound the running time of our algorithm as follows.  Say we choose $L$ and $\sigma$ so that the above inequalities are tight (up to constant factors).  Also, suppose $\delta$ satisfies equation (\ref{eqn-alg2-delta}) exactly up to constant factors.  Let $M$ be the number of grid points along one direction.  Then $M = 2L/\sigma \leq O(R^2 n^3 / \mu^2)$, and the running time is $\leq \poly(n, \log M) \leq \poly(n, \log R, \log\frac{1}{\mu})$.  

}{}

}{}

\end{document}